\documentclass[a4paper,fleqn,usenatbib]{mnras}
\usepackage[T1]{fontenc}
\usepackage{ae,aecompl}
\usepackage{amsmath}
\usepackage{amsfonts}
\usepackage{amssymb}
\usepackage{graphicx}
\usepackage{color,xcolor}
\def\mnras{MNRAS}
\def\aap{A$\&$A}
\def\nat{Nature}
\def\apj{ApJ}
\def\apjl{ApJ Letters}
\def\apjs{ApJS}

\def\aj{AJ}
\def\icarus{Icarus}

\newcommand{\me}{\, {\rm M}_{\oplus}}
\newcommand{\msun}{\, {\rm M}_{\odot}}

\newcommand{\rsun}{\, {\rm R}_{\odot}}
\newcommand{\au}{\, {\rm au}}
\newcommand{\ab}{\, a_{\rm b}}
\newcommand{\rgeuv}{\, {r_{\rm g,euv}}}
\newcommand{\rgfuv}{{\, {r_{\rm g, fuv}}}}
\newcommand{\acav}{\, {r_{\rm c}}}
\newcommand{\ecav}{\, {e_{\rm c}}}
\newcommand{\rcavapo}{\, {r_{\rm c,a}}}

\newcommand{\Stokes}{{\, {\rm St}}}

\title[Circumbinary Planet Formation With Pebbles]{Global N-body simulations of circumbinary planet formation around Kepler-16 and -34 analogues I: Exploring the pebble accretion scenario}
\author[G. A. L. Coleman et al]{Gavin A. L. Coleman$^1$\thanks{Email: gavin.coleman@qmul.ac.uk}, Richard. P. Nelson$^1$, Amaury H. M. J. Triaud$^2$\\
1. Astronomy Unit, Queen Mary University of London, Mile End Road, London, E1 4NS, UK\\
2. School of Physics and Astronomy, University of Birmingham, Edgbaston, Birmingham B15 2TT, UK}
\date{Accepted 2023 March 16; Received 2023 March 12; in original form 2023 January 27}
\pubyear{2023}
\begin{document}
\label{firstpage}
\pagerange{\pageref{firstpage}--\pageref{lastpage}}
\maketitle
\begin{abstract}
Numerous circumbinary planets have been discovered in surveys of transiting planets. Often, these planets are found to orbit near to the zone of dynamical instability, close to the central binary. The existence of these planets has been explained by hydrodynamical simulations that show that migrating circumbinary planets, embedded in circumbinary discs, halt at the central cavity that is formed by the central binary. Transit surveys are naturally most sensitive to finding circumbinary planets with the shortest orbital periods. The future promise of detecting longer period systems using radial-velocity searches, combined with the anticipated detection of numerous circumbinary planets by ESA's PLATO mission, points to the need to model and understand the formation and evolution of circumbinary planets in a more general sense than has been considered before. With this goal in mind, we present a newly developed global model of circumbinary planet formation that is based on the \textsc{mercury6} symplectic N-body integrator, combined with a model for the circumbinary disc and prescriptions for a range of processes involved in planet formation such as pebble accretion, gas envelope accretion and migration.
Our results show that under reasonable assumptions, the pebble accretion scenario can produce circumbinary systems that are similar to those observed, and in particular is able to produce planets akin to Kepler-16b and Kepler-34b. Comparing our results to other systems, we find that our models also adequately reproduce such systems, including multi-planet systems. Resonances between neighbouring planets are frequently obtained, whilst ejections of planets by the central binary acts as an effective source of free floating planets.

\end{abstract}

\section{Introduction}
\label{sec:intro}

Fourteen circumbinary planets have been discovered via transit surveys since the first was announced in 2011.
Kepler-16b, the first such planet to be discovered \citep{Doyle11}, is a $\sim$ Saturn-mass planet \citep{Triaud22} with an orbital period of 229 days orbiting a pair of sub-Solar mass stars that are on moderately eccentric orbits. The discoveries of Kepler-34b and -35b \citep{Welsh12} followed closely afterwards, with Kepler-34b being a $\sim 70 \me$ planet orbiting a pair of Solar-mass stars on significantly eccentric orbits. The discovery of the Kepler-47 system of three circumbinary planets \citep{Orosz12_k47} holds the current record for the largest known multiplicity  within a circumbinary system.
More recently {\textit{TESS}} found a transiting planet orbiting TOI-1338/BEBOP-1 \citep{Kostov20}, whilst radial velocity observations of the system have found an additional planet orbiting further out in the system, whilst also placing an upper limit on the mass of the transiting planet \citep{Standing23}.
One interesting feature shared by the majority of the circumbinary planetary systems is that they host a body that orbits near to the zone of dynamical instability \citep{Holman99}, inside of which a planet would be dynamically unstable due to gravitational perturbations from the central binary \citep[see][for a review of discovered circumbinary planets]{Martin18}.
More recent work, however, has shown that the instability region is more complex, with the outer edge of the zone of dynamical instability actually being the outer edge of an exclusion zone associated with the 3:1 mean-motion resonance with the central binary, accompanied by stable trajectories closer to the binary linked to resonant geometries and bifurcating limit cycles \citep{Langford23}.

In situ formation is one of the possible pathways that has been explored when attempting to explain the origins of these circumbinary planets. This scenario suffers from a number of issues that likely hinder the formation of planets near to the instability zone, including: gravitational interactions with non-axisymmetric features within circumbinary discs leading to large impact velocities between planetesimals \citep{Marzari08,Kley10}; differential pericentre alignment of eccentric planetesimals of different sizes that leads to corrosive collisions \citep{Scholl07}; excitation of planetesimal eccentricities through N-body interactions resulting in large relative velocities, which are disruptive for accretion onto planetary bodies \citep{Meschiari12a,Meschiari12b,Paardekooper12,Lines14,Bromley15}.
Ways to overcome the problems with in situ formation have been explored, including having extremely massive protoplanetary discs \citep{Marzari00,Martin13,Meschiari14,Rafikov15}, or if the fragments are reaccreted and form second or later generations of planetesimals \citep{Paardekooper10}.
More recently, it has been shown that in situ pebble accretion scenarios also suffer from difficulties because a parametric instability can generate hydrodynamical turbulence that stirs up pebbles, rendering pebble accretion onto planetary embryos inefficient \citep{Pierens20,Pierens21}. 

An alternative model is that the planets formed at a larger distance from the binary and then migrated to their observed orbits \citep{Nelson03,Pierens07,Pierens08a,Pierens08b,Thun18}. Numerous works have shown that migrating planets in circumbinary discs stall when they reach the central cavity. The precise stopping location depends on parameters such as the planet mass. Giant planets open a gap in the disc and circularise the eccentric inner cavity that is created by the binary. These planets tend to park closer to the binary \citep[although this increases the probability they may be ejected; ][]{Nelson03,Pierens08a,Thun18}.
Lower mass planets migrate to the edge of the inner cavity and their migration ceases there due to a strong corotation torque that counteracts the Lindblad torque \citep{Pierens07,Pierens08b}. The orbits of these low mass planets align with the inner eccentric disc and precess with it in a state of apsidal corotation \citep{Thun18}.
However, it has also been shown that low mass planets migrating slowly through mean-motion resonances with the binary can be ejected from the systems under certain conditions \citep{Martin22}.
Other works have found that the stopping locations and planet eccentricities are influenced by the mass of the discs that they form in \citep{Dunhill13}, by the effects of gas self-gravity on the disc structure \citep{Mutter17P}, or by the local dust-to-gas ratio close to the inner cavity \citep{Coleman22b}.
In a recent study, \citet{Penzlin21} found that disc parameters including the viscosity parameter $\alpha$ and the disc aspect ratio $H/R$ also affect the stopping locations of migrating planets, mainly by permitting the planets to open partial gaps in the disc, forcing the disc to become more circularised and allowing the planets to migrate closer to central binary. In attempting to match the locations of known circumbinary planets using hydrodynamical simulations, one outstanding problem has been to match systems such as Kepler-34b because the central cavity that forms is very large and eccentric in this case, causing simulated planets to park too far away from the central binary compared to what is observed \citep[e.g.][]{Pierens13, Penzlin21}.

In this work, we investigate whether a comprehensive model of circumbinary planet formation that includes pebble accretion and planet migration is able to form planets that are similar to Kepler-16b and Kepler-34b. We use an updated version of the \textsc{mercury6} symplectic integrator, taking into account the gravitational interactions between forming planets and the central binary \citep{Chambers,ChambersBinary}. This is combined with a 1D viscous disc model that incorporates thermal evolution through stellar irradiation from both central stars, viscous heating and blackbody cooling. We also include prescriptions pertaining specifically to circumbinary discs, including: an eccentric cavity formed through tidal torques from the central binary \citep{Artymowicz94,Dutrey94,Pierens13,Mutter17D}; gravitational torques arising through interactions between planets and material concentrated at the apocentres of the eccentric cavity; and approximating a two-dimensional disc when calculating gas and pebble velocities for pebble accretion rates. The simulations also incorporate up-to-date prescriptions for planet migration \citep{pdk10,pdk11,LinPapaloizou86}, gas accretion onto planetary cores \citep{Poon21}, and gas disc dispersal through photoevaporation on million-year time-scales \citep{Dullemond,Matsuyama03}. To account for pebbles in the disc, we use the evolution models of \citep{Lambrechts14} in which a pebble production forms, leading to a flux of inwardly drifting pebbles that can be accreted by planetary embryos \citep{Lambrechts12}. Our models successfully produce Kepler-16b and Kepler-34b analogues via the accretion of pebbles far from the central binary, before the planets migrate in towards the central cavity, and demonstrate more generally that circumbinary planetary systems similar to those that have been discovered can form through a combination of pebble accretion and migration.

This paper is organised as follows. Section~\ref{sec:base_model} outlines the basic physical model, while Section~\ref{sec:binary_addons} details the additions to the model due to the inclusion of a central binary system. In Section~\ref{sec:example_disc}, we describe the evolution of an example circumbinary disc.
We present our results in Section~\ref{sec:kep16} for Kepler-16 and Section~\ref{sec:kep34} for Kepler-34.
Finally, we discuss interesting outcomes of our results as a whole in Section~\ref{sec:discussions} before we draw our conclusions in Section~\ref{sec:conclusions}.

\section{Physical Model}
\label{sec:base_model}

In the following sections, we provide details of the physical model we adopt and the numerical scheme used to undertake the simulations.
We only model the circumbinary systems of Kepler-16 and Kepler-34, since they were amongst the first circumbinary planets to be discovered, and as such their formation processes and the evolution of their circumbinary discs are the most studied and well understood.
In future work we will expand our studies to explore circumbinary planet formation around a diverse population of binary stars, containing different binary parameters.
Initially we will present the general model that is similar to what we have previously used for simulations around single stars \citep[e.g.][]{ColemanNelson14,ColemanNelson16,ColemanNelson16b}, and then we will detail the additions to the model that take into account the effects of the binary stars on the inner regions of the circumbinary discs.

The N-body simulations presented here were performed using the {\textsc{mercury6}} symplectic N-body integrator \citep{Chambers}, updated to accurately model planetary orbits around a pair of binary stars \citep{ChambersBinary}.
We utilise the `close-binary' algorithm described in \citet{ChambersBinary} that calculates the temporal evolution of the positions and velocities of each body in the simulations with respect to the centre of mass of the binary stars, subject to gravitational perturbations from both stars and other large bodies.
Note that in this work we are interested in exploring the formation of circumbinary planets around systems that mimic Kepler-16 and -34, which have well-defined orbital parameters. Hence, we do not include the gravitational forces from planets and planetary embryos on to the central binary stars since this can lead to changes in their orbital elements away from the observed values. In future work, we will explore the consequences of this choice and examine the influence of the planets on the orbital evolution of the central binaries. We also do not include disc--binary interactions \citep[e.g.][]{Penzlin22} since this can also drive evolution of the binary.

Our model includes prescriptions for the evolution of protoplanetary discs and disc-planet interactions, which we describe below.

\subsection{Gas disc}

We adopt a 1D viscous disc model where the equilibrium temperature is calculated by balancing irradiation heating from the central stars, background heating from the residual molecular cloud, viscous heating and blackbody cooling.
The surface density, $\Sigma$, is evolved by solving the standard diffusion equation
\begin{equation}
\label{eq:diffusion}
\dfrac{d\Sigma}{dt}=\frac{1}{r}\dfrac{d}{dr}\left[3r^{1/2}\dfrac{d}{dr}\left(\nu\Sigma r^{1/2}\right)-\dfrac{2\Lambda\Sigma r^{3/2}}{GM_{\rm bin}}\right]-\dfrac{d\Sigma_{\rm pe}}{dt},
\end{equation}
where $\dfrac{d\Sigma_{\rm pe}}{dt} = \dfrac{d\Sigma_{\rm pe,int}}{dt} + \dfrac{d\Sigma_{\rm pe,ext}}{dt}$ is the rate change in surface density due to internally and externally driven photoevaporative winds, $\Lambda$ is the disc-planet torque that operates when a planet becomes massive enough to open a gap in the disc, $M_{\rm bin}$ is the combined stellar mass, and $\nu$ is the disc viscosity \citep{Shak}
\begin{equation}
\label{eq:viscosity}
\nu=\alpha c_{\rm s}^2/\Omega,
\end{equation}
where $c_{\rm s}$ is the local isothermal sound speed, $\Omega = \sqrt{\frac{GM_{\rm bin}}{r^3}}$ is the Keplerian frequency and $\alpha$ is the viscosity parameter.
The disc-planet torque per unit mass that applies for planets whose masses are large enough to open gaps is given by \cite{LinPapaloizou86}
\begin{equation}
\label{eq:t2torque}
    \Lambda = {\rm sign}(r-r_{\rm p})q^2\dfrac{GM_{\rm bin}}{2r}\left(\dfrac{r}{|\Delta_{\rm p}|}\right)^4
\end{equation}
where $q$ is the planet/star mass ratio, $r_{\rm p}$ is the planet orbital radius, and $|\Delta_{\rm p}| = \max (H\text{, }|r-r_{\rm p}|)$, where $H$ is the local disc scale height.

As the disc should be in thermal equilibrium, we use an iterative method to solve the following equation \citep{Dangelo12}
\begin{equation}
\label{eq:temperature}
Q_{\rm irr,A} + Q_{\rm irr,B} + Q_{\nu} + Q_{\rm cloud} - Q_{\rm cool} = 0,
\end{equation}
where $Q_{\rm irr,A}$ and $Q_{\rm irr,B}$ are the radiative heating rates due to the binary stars, $Q_{\nu}$ is the viscous heating rate per unit area of the disc, $Q_{\rm cloud}$ is the radiative heating due to the residual molecular cloud, and $Q_{\rm cool}$ is the radiative cooling rate.   
For a Keplerian disc, the energy flux due to dissipation is given by \citet{Mihalas} as
\begin{equation}
Q_{\nu} = \frac{9}{4}\nu\Sigma\Omega^2.
\end{equation}
The heating rate due to stellar irradiation from the $i$th star is given by \citet{Menou}
\begin{equation}
Q_{{\rm irr},i}=2\sigma T_{{\rm irr},i}^4/\tau_{\rm eff}, 
\end{equation}
where
\begin{equation}
\tau_{\rm eff} = \frac{3\tau_{\rm R}}{8}+\frac{1}{2}+\frac{1}{4\tau_{\rm P}},
\end{equation}
and $\tau_{\rm R}$ and $\tau_{\rm P}$ are the optical depths due to the Rosseland and Planck mean opacities respectively (assumed to be equivalent in this work).
For the irradiation temperature we follow \citet{Dangelo12} and take
\begin{equation}
T_{{\rm irr},i}^4=(T_{i}^4+T_{{\rm acc},i}^4)(1-\epsilon_{\rm alb})\left(\frac{R_i}{r}\right)^2 W_{\rm G}.
\end{equation}
Here $\epsilon_{\rm alb}$ is the disc albedo (estimated to be 0.5 in agreement with \citet{Dangelo12}), $T_{i}$ and $R_{i}$ are the effective temperature and radii of the $i$th star, $T_{{\rm acc},i}$ is the contribution made to the irradiation temperature by the accretion of gas on to the stars
\begin{equation}
    T_{{\rm acc},i}^4 = \dfrac{GM_{i}\dot{M}_{\rm disc}}{16\pi\sigma R_{i}^3}
\end{equation}
where $\dot{M}_{\rm disc}$ is the accretion rate recorded at the disc inner edge, and we assume that half of the gas accretes onto each star.
The geometrical factor $W_{\rm G}$ determines the flux of radiation that is intercepted by the disc surface, and approximates to
\begin{equation}
W_{\rm G} = 0.4 \left(\frac{R_*}{r}\right) +\frac{2}{7}\frac{H}{r},
\end{equation}
as given by \citet{Dangelo12}.
The scale height of the disc is denoted by $H$ in the equation above and is equal to $c_{\rm s}/\Omega$. We note that because a 1D disc model is adopted, the actual locations of the stars are not taken into account in the irradiation prescription, and instead the stars are both assumed to sit at the centre of mass of the system.
For $Q_{\rm cloud}$ we have
\begin{equation}
Q_{\rm cloud} = 2\sigma T_{\rm cloud}^4/\tau_{\rm eff}
\end{equation}
where we take $T_{\rm cloud}$ as being equal to 10 K, consistent with observed temperatures of molecular clouds \citep{Wilson97}.
For the cooling of the disc we have
\begin{equation}
Q_{\rm cool} = 2\sigma T_{\rm mid}^4/\tau_{\rm eff}
\end{equation}
with $T_{\rm mid}$ being the disc midplane temperature and is found iteratively using Brent's method \citep{Press07}.

We take the opacity, $\kappa$, to be equal to the Rosseland mean opacity, with the temperature and density dependencies calculated using the formulae in \citet{Bell97} for temperatures below 3730 K, and by \citet{Bell94} above 3730 K.
To account for changes in the disc metallicity, we multiply the opacity by the dust contribution to the metallicity relative to solar. The full opacity table can be found in Appendix \ref{sec:opacity}.

\subsection{Photoevaporation}

The absorption of UV radiation by the disc can heat the gas above the local escape velocity, and hence drive photoevaporative winds.
For extreme ultra-violet radiation (EUV), this creates a layer of ionised hydrogen with temperature $\sim$10,000~K \citep{Clarke2001}, whereas for far ultra-violet radiation (FUV), this creates a neutral layer of dissociated hydrogen with temperature of roughly 1000~K \citep{Matsuyama03}.
We incorporate both EUV radiation from the central stars (internal photoevaporation) and FUV radiation from other nearby stars (external photoevaporation).
We do not include here X-ray induced internal photoevaporation \citep[e.g.][]{Owen12, Picogna19}, since this operates in the outer regions of the disc where external photoevaporation operates, and the interplay between internal and external photoevaporation is poorly understood. FUV radiation from the central stars is also neglected, since it also operates in a similar location to FUV external photoevaporation, which we assume dominates the evolution of the disc in this region.
Whilst the internally originating FUV radiation is an important process, models for this are strongly dependent on the local disc properties, e.g. the size of dust in the penetrated region of the disc \citep{Gorti15}, as well as complex photochemistry, including the photo- and chromo-spheres of the central stars \citep{Gorti09a,Gorti09}.

\subsubsection{Internal photoevaporation}
To account for the radiation from the central stars we adopt the formula provided by \citet{Dullemond} to calculate the rate at which the surface density decreases due to this wind
\begin{equation}
\dfrac{d\Sigma_{\rm pe,int}}{dt} = 1.16\times10^{-11}G_{\rm fact}\sqrt{f_{41}}\left(\dfrac{1}{\rgeuv}\right)^{3/2}
\left(\dfrac{M_{\bigodot}}{\au^2 \, {\rm yr}}\right)
\end{equation}
where $G_{\rm fact}$ is a scaling factor defined as
\begin{equation}
G_{\rm fact} = \left\{ \begin{array}{ll}
\left(\dfrac{\rgeuv}{r}\right)^2 e^{\frac{1}{2}\left(1-\dfrac{\rgeuv}{r}\right)} 
& r\le \rgeuv, \\
\\
\left(\dfrac{\rgeuv}{r}\right)^{5/2} & r>\rgeuv.
\end{array} \right.
\end{equation}
Here, $\rgeuv$ is the characteristic radius beyond which gas becomes unbound from the system as a result of the EUV radiation launching a wind with a temperature of $10^4$ K, which is set to $7.6\au$ and $17.7\au$ for Kepler-16 and -34 respectively, and $f_{41}$ is the combined rate at which extreme UV ionising photons are emitted by the central stars in units of $10^{41}$ s$^{-1}$.

When the inner region of the disc becomes optically thin (i.e. when the gas surface density drops below a critical value within $0.2\times \rgeuv$, which we take as $10^{-5}\rm gcm^{-2}$), ionising photons can launch a wind off the inner edge of the disc, enhancing the photoevaporation rate.
The direct photoevaporation prescription that we adopt is taken from \citet{Alexander07} and \citet{Alexander09}, where the photoevaporative mass loss rate is given by 
\begin{equation}
\dfrac{d\Sigma_{\rm pe,int}}{dt}=2C_2\mu m_Hc_s\left(\dfrac{f_{41}}{4\pi\alpha_Bhr^3_{\rm in}}\right)^{1/2}
\left(\dfrac{r}{r_{\rm in}}\right)^{-2.42}.
\end{equation}
Here, $C_2=0.235$, $\alpha_{B}$ is the Case B recombination coefficient for atomic hydrogen at $10^4$K, having a value of $\alpha_B=2.6\times10^{-19}\text{m}^3\text{s}^{-1}$ \citep{Cox}, and $r_{\rm in}$ is the radial location of the inner disc edge.
We note that whilst the inner cavity region is optically thin in our models, and the inner edge of the disc is further out than occurs because of the magnetospheres of single stars, the cavity edge is always closer to the central stars than the location where a photoevaporative wind can be launched  ($\sim0.2\times \rgeuv$).
Therefore in our models, the only way direct photoevaporation can be triggered is if a giant planet removes material from this region by either accretion or through tidal torques, thus rendering the whole inner disc region optically thin, or by the inner disc around the cavity accreting on to the central stars.

\subsubsection{External photoevaporation}
In addition to EUV radiation from the central stars photoevaporating the protoplanetary disc, there is also a contribution from the discs external environment.
This is typically considered to be the radiation that is emanating from newly formed stars, in particular young, hot, massive stars that release vast amounts of high-energy radiation.
Here we include the effects of external photoevaporation due to far-ultraviolet (FUV) radiation emanating from massive stars in the vicinity of the discs \citep{Matsuyama03}.
This drives a wind outside of the gravitational radius where the sound speed in the heated layer is $T\sim$1000~K, denoted $\rgfuv$.
This leads to a reduction in the gas surface density as follows \citep{Matsuyama03}
\begin{equation}
\dfrac{d\Sigma_{\rm pe,ext}}{dt} =  \left\{ \begin{array}{ll}
0 & r\le \beta \rgfuv , \\
\\
\dfrac{\dot{M}_{\rm pe,ext}}{\pi(r_{\rm max}^2-\beta^2 \rgfuv^2)}& r>\beta \rgfuv.
\end{array} \right.
\end{equation}
where $\beta = 0.14$ \citep[similar to][]{AlexanderPascucci12} gives the effective gravitational radius that external photoevaporation operates above.
To ensure realistic disc lifetimes for both systems, we take the total rate $\dot{M}_{\rm pe,ext}$ to be equal to $3\times10^{-8} M_{\odot}/yr$, consistent with the rates found in \citet{Haworth18} for protoplanetary discs in low $G_0$ environments, where $G_0$ is the flux integral over 912--2400\AA, normalised to the value in the solar neighbourhood \citep{Habing68}.
Note that by also modifying the viscous alpha parameter, as well as the internal photoevaporation rate, realistic disc lifetimes can be obtained with weaker/stronger external photoevaporation rates.

\subsection{Planet Migration}
\subsubsection{Type I migration}
Planets with masses that significantly exceed a Lunar-mass undergo substantial migration through gravitational interactions with the surrounding disc for discs with masses similar to the Minimum Mass Solar Nebula (MMSN) model such as those studied here.
In our simulations we implement the torque formulae presented by \citet{pdk10,pdk11}.
These formulae take into account how planet masses, and changes in local disc conditions, modify the various torque contributions for the planet.
Corotation torques are especially sensitive to the ratio of the horseshoe libration time-scale to either the viscous or thermal diffusion time-scales across the horseshoe region.

In using equations 50-53 in \citet{pdk11}, we obtain an expression giving the 
total type I torque acting on a planet,
\begin{equation}
\label{eq:typeItorque}
\begin{split}
\Gamma_{\rm I,tot}&=F_L\Gamma_{\rm LR}+\left\lbrace\Gamma_{\rm VHS}F_{p_v}G_{p_v}\right.\\
&\left.+\Gamma_{\rm EHS}F_{p_v}F_{p_{\chi}}\sqrt{G_{p_v}G{p_{\chi}}}+\Gamma_{\rm LVCT}(1-K_{p_v})\right.\\
&\left.+\Gamma_{\rm LECT}\sqrt{(1-K_{p_v})(1-K_{p_{\chi}})}\right\rbrace F_e F_i
\end{split}
\end{equation}
where $\Gamma_{\rm LR}$, $\Gamma_{\rm VHS}$, $\Gamma_{\rm EHS}$, $\Gamma_{\rm LVCT}$ and $\Gamma_{\rm LECT}$, 
are the Lindblad torque, vorticity and entropy related horseshoe drag torques, and linear vorticity and 
entropy related corotation torques, respectively, as given by equations 3-7 in \citet{pdk11}. 
The functions $F_{p_v}$, $F_{p_{\chi}}$, $G_{p_v}$, $G_{p_{\chi}}$, $K_{p_v}$ and $K_{p_{\chi}}$ are 
related to the ratio between viscous/thermal diffusion time scales and horseshoe libration/horseshoe U-turn 
time-scales, as given by equations 23, 30 and 31 in \citet{pdk11}.
Changes in local disc conditions brought about by changes in temperature,
surface density, and metallicity/opacity, can alter the magnitude of the functions given in \citet{pdk11},
and thus the magnitude and possibly the direction of the torque calculated in Equation~\ref{eq:typeItorque}.
The factors $F_e$ and $F_i$, multiplying 
all terms relating to the corotation torque, allow for the fact that a planet's eccentricity and 
inclination can attenuate the corotation torque \citep{Bitsch}. To account for the effect of eccentricity,
we use the formula suggested by \citet{Fendyke}
\begin{equation}
F_e=\exp{\left(-\dfrac{e}{e_f}\right)},
\label{eqn:fendyke}
\end{equation}
where $e$ is the planet's eccentricity and $e_f$ is defined as
\begin{equation}
e_f=h/2+0.01
\end{equation}
where $h$ is the disc aspect ratio at the planet's location.
To account for the effect of orbital inclination we define $F_i$ as
\begin{equation}
F_i=1-\tanh(i/h),
\end{equation}
where $i$ is the inclination of the planet. 

The factor $F_{\rm L}$ in Equation~\ref{eq:typeItorque} accounts for the reduction in Lindblad 
torques when planets are on eccentric or inclined orbits, and is given by \citet{cressnels}
\begin{equation}
\begin{split}
F_L&=\left[P_e+\left(\dfrac{P_e}{|P_e|}\right)\times\left\lbrace0.07\left(\frac{i}{h}\right)+\right.\right.\\
&\left.\left.0.085\left(\frac{i}{h}\right)^4-0.08\left(\frac{e}{h}\right)\left(\frac{i}{h}\right)^2\right\rbrace\right]^{-1}
\end{split}
\end{equation}
where $P_e$ is defined as 
\begin{equation}
P_e=\dfrac{1+\left(\dfrac{e}{2.25h}\right)^{1.2}+\left(\dfrac{e}{2.84h}\right)^6}{1-\left(\dfrac{e}{2.02h}\right)^4}.
\end{equation}

To damp planet eccentricities and inclinations we follow the damping formulae given by \citet{Papaloizou2000},
\begin{equation}
F_{\rm damp,e}=-\dfrac{2v_{\rm r}}{t_{\rm edamp}},\,\,F_{\rm damp,i}=-\dfrac{v_{\rm z}}{t_{\rm idamp}}
\end{equation}
where the damping time-scales follow \citet{cressnels}
\begin{equation}
\begin{split}
t_{\rm edamp}&=\dfrac{t_{\rm wave}}{0.78}\\
&\times\left[1-0.14\left(\dfrac{e}{h}\right)^2+0.06\left(\dfrac{e}{h}\right)^3+
0.18\left(\dfrac{e}{h}\right)\left(\dfrac{i}{h}\right)^2\right],\\
\rm{and}\\
t_{\rm idamp}&=\dfrac{t_{\rm wave}}{0.544}\\
&\times\left[1-0.3\left(\dfrac{i}{h}\right)^2+0.24\left(\dfrac{i}{h}\right)^3+
0.14\left(\dfrac{i}{h}\right)\left(\dfrac{e}{h}\right)^2\right]
\end{split}
\end{equation}
where $t_{\rm wave}$ is specified as
\begin{equation}
t_{\rm wave}=\left(\dfrac{m_{\rm p}}{M_{\rm bin}}\right)^{-1}\left(\dfrac{a_{\rm p}\Omega_{\rm p}}{c_s}\right)^{-4}
\left(\dfrac{\Sigma_{\rm p}a^2_{\rm p}}{M_{\rm bin}}\right)^{-1}\Omega^{-1}_{\rm p}.
\end{equation}
with the subscript 'p' denoting the values taken at the planet's location \citep{Tanaka04}.

\subsubsection{Type II migration}
Once a planet becomes massive enough to form a gap in a disc, its migration changes from type I to type II.
This is where more massive planets begin to carve annular gaps centered on their orbits, until such a point that the viscous forces balance planetary torques, and the gaps reach an equilibrium state.
More recent work by \citet{Crida}, showed that not only viscous forces worked to balance planetary torques, but pressure forces arising from density waves launched by the planet assisted by transporting some of the gravitational torque away from the planet.
In balancing viscous and pressure forces with gravitational torques, \citet{Crida} showed that a gap can be opened in the disc when the following condition is satisfied
\begin{equation}
\label{eq:gapopening}
    \dfrac{3}{4}\dfrac{H}{r_{\rm H}}+\dfrac{50}{q{\it{Re}}} \le 1,
\end{equation}
where $r_{\rm H}$ is the planet Hill radius, $q$ is the planet to binary mass ratio, and ${\it{Re}}=r_{\rm p}^2\Omega_{\rm p}/\nu$ is the Reynolds number of the disc at the planet's location.

When the planet has opened a gap in the disc, the type II migration torque per unit mass is then given by
\begin{equation}
\label{eq:GammaII}  
\Gamma_{{\rm II}} = - \frac{2 \pi}{m_{\rm p}} 
\int_{r_{\rm in}}^{r_{\rm out}} r \Lambda \Sigma_{\rm g} dr.
\end{equation}
where $\Lambda$ is the disc-planet torque per unit mass as given by eq. \ref{eq:t2torque}.

\subsection{Gas Accretion}
Once a planet has significantly increased its mass through mutual collisions with other planets and via pebble accretion, it is able to accrete a gaseous envelope from the surrounding disc.
Ideally we would incorporate 1D envelope structure models \citep[e.g.][]{CPN17} into our simulations.
However these calculations are computationally expensive and would considerably increase simulation run times, therefore we opted to instead include fits to gas accretion rates obtained from 1D structure models.
Recently, \citet{Poon21} presented fits to gas accretion rates obtained using a 1D envelope structure model \citep{Pap-Terquem-envelopes,PapNelson2005,CPN17}.
To calculate these fits, \citet{Poon21} performed numerous simulations, embedding planets with initial core masses between 2--15 $\me$ at orbital radii spanning 0.2--50 $\au$, within gas discs of different masses.
This allowed for the effects of varying local disc properties to be taken into account when calculating the fits, which is a significant improvement on fits from earlier works \citep[e.g.][]{Hellary,ColemanNelson16}.
Using the 1D envelope structure model of \citet{CPN17}, the embedded planets were then able to accrete gas from the surrounding gas disc until either the protoplanetary disc dispersed, or the planets reached a critical state where they would then undergo runaway gas accretion.
With the results of these growing planets, \citet{Poon21} calculated fits to the gas accretion rates taking into account properties of both the planet and the local disc.
The gas accretion rate we adopt is
\begin{align}\label{eq:gasenvelope-gc}
\left (\dfrac{d M_{\mathrm{ge}}}{d t} \right )_{\mathrm{local}}=& 10^{-10.199} \left( \dfrac{\mathrm{M_{\oplus}}}{\mathrm{yr}}\right) f_{\mathrm{opa}}^{-0.963} \left ( \dfrac{T_{\mathrm{local}}}{\mathrm{1\,K}}\right )^{-0.7049}\nonumber \\
&\times \left (\dfrac{M_{\mathrm{core}}}{\mathrm{M}_{\oplus}} \right )^{5.6549}  \left (\dfrac{M_{\mathrm{ge}}}{\mathrm{M}_{\oplus}}  \right )^{-1.159}\nonumber \\
&\times\left [ \exp{\left ( \dfrac{M_{\mathrm{ge}}}{M_{\mathrm{core}}} \right )} \right ]^{3.6334}.
\end{align}
where $T_{\rm local}$ is the local disc temperature, $f_{\rm opa}$ is an envelope opacity reduction factor and $M_{\rm core}$ and $M_{\rm ge}$ are the planet's core and envelope masses, respectively.
When comparing the masses of gas accreting planets calculated through eq. \ref{eq:gasenvelope-gc} to the actual masses obtained using the 1D envelope structure model of \citet{CPN17}, \citet{Poon21} found excellent agreement.

In our simulations, we allow planets to start accreting a gaseous envelope once their mass exceeds an Earth mass.
The gas accretion rate given by eq.~\ref{eq:gasenvelope-gc} then applies until either the planet opens a gap in the disc (i.e. when eq.~\ref{eq:gapopening} is satisfied), or until the planet undergoes runaway gas accretion.
Once a planet undergoes runaway gas accretion, it can rapidly accrete material from its feeding zone until it reaches its `gas isolation mass', where the feeding zone has emptied and a gap has formed in the disc.
To calculate the gas isolation mass we follow the steps outlined in \citet{ColemanNelson16b}:\\
(i) Calculate the gas isolation mass, $m_{\rm iso}$, according to:
\begin{equation}
m_{\rm iso} = 2\pi r_{\rm p} \Sigma_g(r_{\rm p})\Delta r
\end{equation}
where $\Sigma_g (r_{\rm p})$ is the gas surface density taken at the planet's location, and $\Delta r$ is given by
\begin{equation}
\Delta r = 6\sqrt{3}R_{\rm H}
\end{equation}
where $R_{\rm H}$ is the planet's Hill radius.\\
(ii) Recalculate $m_{\rm iso}$ at each time step to account for the drop in $\Sigma_{\rm g}$ as the material in the planet's feeding zone diminishes.\\
(iii) Allow the planet to grow rapidly to $m_{\rm iso}$ by removing gas from the disc around the planet and adding it to the planet, using eq.~\ref{eq:gasenvelope-gc}.
Once the planet reaches $m_{\rm iso}$, it transitions to type II migration and begins accreting at the minimum between eq.~\ref{eq:gasenvelope-gc} and the viscous supply rate that is calculated at a distance of 10 planetary Hill radii exterior to the planet's orbit.
\begin{equation}
\label{eq:gasenvelope-gap}
    \frac{dM_{\rm ge}}{dt} = \min\left[\left(\frac{dM_{\rm ge}}{dt}\right)_{\rm local},3\pi\nu\Sigma \right].
\end{equation}
We define the point at which a planet undergoes runaway gas accretion when the gas accretion rate $\frac{dM_{\rm ge}}{dt} \ge 2\me$ per 1000 yr.
When the gas isolation mass is calculated we assume a maximum gas isolation mass of $400\sqrt{M_{\rm bin}}\me$ which accounts for when a planet enters the runaway gas accretion phase in a massive disc, where tidal torques from the planet would evacuate the feeding zone before the gas isolation mass was reached.
We note that a planet that does not reach the runaway gas accretion mass prior to reaching the local gap-forming mass would instead transition directly to type II migration without accreting the material within its feeding zone, and will begin accreting using eq. \ref{eq:gasenvelope-gap}.

\subsection{Pebble Accretion}
\label{sec:pebbles_model}
To account for the pebbles in the disc, we implement the pebble models of \citet{Lambrechts12,Lambrechts14} into our disc model.
As a protoplanetary disc evolves, a pebble production front extends outwards from the centre of the system as small pebbles and dust grains fall towards the disc midplane, gradually growing in size.
Once the pebbles that form reach a sufficient size they begin to migrate inwards through the disc due to aerodynamic drag.
The location of this pebble production front is defined as:
\begin{equation}
r_{\rm g}(t) = \left(\frac{3}{16}\right)^{1/3}(GM_{\rm bin})^{1/3}(\epsilon_{\rm d}Z_{0})^{2/3}t^{2/3},
\end{equation}
where $\epsilon_{\rm d} = 0.05$ is a free parameter that depends on the growth efficiency of pebbles, whilst $Z_{0}$ is the solids-to-gas ratio.
Since this front moves outwards over time, this provides a constant mass flux of inwardly drifting pebbles equal to:
\begin{equation}
\label{eq:massflux}
\dot{M}_{\rm flux} = 2\pi r_{\rm g}\dfrac{dr_{\rm g}}{dt}Z_{\rm peb}(r_{\rm g})\Sigma_{\rm gas}(r_{\rm g}),
\end{equation}
where $Z_{\rm peb}$ denotes the metallicity that is comprised solely of pebbles.
Combining the metallicity locked within pebbles with that to which contributes to the remaining dust in the disc, gives the total metallicity of the system:
\begin{equation}
Z_{0} = Z_{\rm peb} + Z_{\rm dust}.
\end{equation}
Here, we assume that 90 per cent of the total metallicity is converted into pebbles, and that this ratio remains constant throughout the entire disc lifetime.
The remaining metallicity is locked up within small dust grains that contribute to the opacity of the disc when calculating its thermal structure, and again we assume this remains constant over time.
Assuming that the mass flux of pebbles originating from $r_{\rm g}$ is constant throughout the disc, we follow \citet{Lambrechts14} in defining the pebble surface density, $\Sigma_{\rm peb}$, as the following:
\begin{equation}
\label{eq:sigma_peb}
\Sigma_{\rm peb} = \dfrac{\dot{M}_{\rm flux}}{2\pi r v_r},
\end{equation}
where $v_r$ is the radial velocity of the pebbles equal to
\begin{equation}
\label{eq:vr}
v_r = 2\frac{\Stokes}{\Stokes^2+1}\eta v_{\rm K}-\frac{v_{\rm r,gas}}{1+\Stokes^2}
\end{equation}
\citep{Weidenschilling_77,Nakagawa86}, where $\Stokes$ is the Stokes number of the pebbles, $v_K$ is the local Keplerian velocity, $v_{\rm r,gas}$ is the gas radial velocity, and $\eta$ is the dimensionless measure of gas pressure support \citep{Nakagawa86},
\begin{equation}
\eta = -\frac{1}{2}h^2\dfrac{\partial ~{\rm ln} P}{\partial ~{\rm ln} r}.
\end{equation}
For the Stokes number, we assume it is equal to:
\begin{equation}
    \Stokes = \min(\Stokes_{\rm drift}, \Stokes_{\rm frag})
\end{equation}
where $\Stokes_{\rm drift}$ is the drift-limited Stokes number that is obtained through an equilibrium between the drift and growth of pebbles to fit constraints of observations of pebbles in protoplanetary discs and from advanced coagulation models \citep{Birnstiel12}
\begin{equation}
    \Stokes_{\rm drift} = \dfrac{\sqrt{3}}{8}\dfrac{\epsilon_{\rm p}}{\eta}\dfrac{\Sigma_{\rm peb}}{\Sigma_{\rm gas}},
\end{equation}
where $\epsilon_{\rm p}$ is the coagulation efficiency between pebbles.
As well as the drift-limited Stokes number, we also include the fragmentation-limited Stokes number, ($\Stokes_{\rm frag}$) which we follow \citet{Ormel07} and is equal to
\begin{equation}
    \Stokes_{\rm frag} = \dfrac{v_{\rm frag}^2}{3\alpha c_{\rm s}^2}
\end{equation}
where $v_{\rm frag}$ is the impact velocity required for fragmentation, which we model as the smoothed function
\begin{equation}
    \dfrac{v_{\rm frag}}{1 \rm m\,s^{-1}} = 10^{0.5+0.5\tanh{((r-r_{\rm snow})/5H)}}.
\end{equation}
The fragmentation velocity therefore varies between 1$\rm m\,s^{-1}$ for rocky pebbles \citep{Guttler10}, to $10 \rm m\,s^{-1}$ for icy pebbles, consistent with some results in the literature \citep[though this is still an area of open research][]{Gundlach15,Musiolik19}.

As pebbles drift inwards, eventually they cross the water iceline, which we take as being where the local disc temperature is equal to 170 K.
Since pebbles are mostly comprised of ice and silicates, when they cross the iceline, the ices sublimate releasing trapped silicates, reducing the mass and size of the remaining silicate pebbles.
To account for the sublimation of ices, of which we assume comprise $50\%$ of the pebble mass, we multiply the mass flux of pebbles drifting through the disc at radial locations interior to the iceline by a factor of 0.5 \citep{Lambrechts14}.

As the pebbles drift through the disc, they can encounter planetary embryos and, given the right conditions, they can be accreted by the embryos.
This is due to the increased gas drag forces that allows them to become captured by the planet's gravity \citep{Lambrechts12}.
To calculate this accretion rate, we follow \citet{Johansen17} by distinguishing between the Bondi regime (small bodies) and the Hill regime (massive bodies).
The Bondi accretion regime occurs for low mass bodies where they do not accrete all of the pebbles that pass through their Hill sphere, i.e. the body's Bondi radius is smaller than the Hill radius.
Once the Bondi radius becomes comparable to the Hill radius, the accretion rate becomes Hill sphere limited, and so the body accretes in the Hill regime.
Normally, planets begin accreting in the Bondi regime before transitioning to the Hill regime when they reach the transition mass where the Bondi radius is equal to the Hill radius,
\begin{equation}
\label{eq:mtrans}
    M_{\rm trans} = \eta^3 M_{\rm bin}.
\end{equation}
A further distinction within the two regimes, is whether the body is accreting in a 2D or a 3D mode.
This is dependent on the relation between the Hill radius of the body and the scale height of the pebbles in the disc.
For bodies with a Hill radius smaller than the scale height of pebbles, the accretion is in the 3D mode since pebbles are passing through the entire Hill sphere, whilst for bodies with a Hill radius larger than the pebble scale height, regions of the Hill sphere remain empty of pebbles and as such the accretion rate becomes 2D as the body's mass increases.
Following \citet{Johansen17} the equations for the 2D and 3D accretion rates are
\begin{equation}
\label{eq:mdot2d}
\dot{M}_{\rm 2D} = 2 R_{\rm acc} \Sigma_{\rm peb} \delta v,
\end{equation}
and
\begin{equation}
\label{eq:mdot3d}
\dot{M}_{\rm 3D} = \pi R_{\rm acc}^2 \rho_{\rm peb} \delta v,
\end{equation}
where $\Sigma_{\rm peb}$ is the azimuthally averaged pebble surface density\footnote{We use the azimuthally averaged surface density instead of the instantaneous to be consistent with other accretion routines that are azimuthally averaged. In the area where there are significant deviations in surface density with azimuth (i.e. near the cavity, see sect. \ref{sec:binary_addons}), we tested that this was an adequate assumption. In those regions the changes in relative velocity due to eccentric planets and/or pebble orbits, sufficiently reduced accretion rates, resulting in the differences in choice of values for $\Sigma_{\rm peb}$ being negligible.}, while $\rho_{\rm peb}$ is the midplane pebble density.
Here $\delta v = \Delta v + \Omega R_{\rm acc}$ is the approach speed, with $\Delta v$ being the difference in velocity between the pebbles and accreting planets. 
The accretion radius $R_{\rm acc}$ depends on whether the accreting object is in the Hill or Bondi regime, and also on the friction time of the pebbles.
In order for pebbles to be accreted they must be able to significantly change direction on time-scales shorter than the friction time.
This inputs a dependence of the friction time onto the accretion radius, forming a criterion accretion radius $\hat{R}_{\rm acc}$ which is equal to 
\begin{equation}
 \hat{R}_{\rm acc} = \left( \frac{4 t_{\rm f}}{t_{\rm B}} \right)^{1/2} R_{\rm B},
\end{equation}
for the Bondi regime, and:
\begin{equation}
 \hat{R}_{\rm acc} = \left(  \frac{\Omega t_{\rm f}}{0.1} \right)^{1/3} R_{\rm H},
 \label{Racc_Hill}
\end{equation}
for the Hill regime.
Here $R_{\rm B}$ is the Bondi radius, while $R_{\rm H}$ is the Hill radius, $t_{\rm B} = R_{\rm B}/\Delta v$ is the Bondi sphere crossing time, and $t_{\rm f} = \Stokes/\Omega$ is the friction time.
The accretion radius is then equal to
\begin{equation}
    R_{\rm acc} = \hat{R}_{\rm acc} \exp[-\chi(t_{\rm f}/t_{\rm p})^{\gamma}]
\end{equation}
where $t_{\rm p} = G m_{\rm p}/(\Delta v + \Omega R_{H})^3$ is the characteristic passing time-scale, $\chi = 0.4$ and $\gamma = 0.65$ \citep{OrmelKlahr2010}.
Since some of the pebbles are being accreted by the planets, they can no longer drift further inwards. This alters the mass flux of pebbles defined in eq. \ref{eq:massflux} to the following:
\begin{equation}
    \dot{M}_{\rm flux}(r) = \dot{M}_{\rm flux} - \dot{M}_{\rm c}(r_{\rm p}>r)
\end{equation}
where $\dot{M}_{\rm c}(r_{\rm p}>r)$ sums up the mass flux of pebbles accreted by embryos exterior to an orbital radius r.

The planet then grows by accreting pebbles until it reaches the so-called pebble isolation mass, that is the mass required to perturb the gas pressure gradient in the disc: i.e. the gas velocity becomes super-Keplerian in a narrow ring outside the planet's orbit reversing the action of the gas drag.
The pebbles are therefore pushed outwards rather than inwards and accumulate at the outer edge of this ring stopping the core from accreting solids \citep{PaardekooperMellema06,Rice06}.
We follow \cite{Lambrechts14} and define the pebble isolation mass as:
\begin{equation}
\label{eq:peb_iso_mass}
q_{\rm iso} = \frac{h^3}{2},
\end{equation}
where $q_{\rm iso} = m_{\rm iso}/M_{\rm bin}$.
Once the pebble isolation mass is reached, we follow previous works \citep[e.g.][]{ColemanProxima17,Coleman19,Coleman21} and halt pebble accretion for planets downstream of the isolating planet, since the drifting pebbles will be trapped at the exterior pressure bump.
More recent calculations of the pebble isolation mass have included dependencies on local disc properties, such as the viscosity parameter $\alpha$ and the Stokes number of the pebbles \citep{Bitsch18,Ataiee18} and also planet eccentricities \citep{Chametla22}.
Note that whilst we did not use the more complex prescriptions for the pebble isolation mass, the simpler version from \citet{Lambrechts14} only differs in mass by at most being factor two smaller, and was typically within $\sim20\%$ of the other prescribed masses, mainly due to the Stokes numbers being $\sim0.1$.
We also note that even with the smaller pebble isolation masses, when planets reached this mass ($\sim10$--15$\me$), they had typically migrated in near the cavity region, where the eccentric orbits of the pebbles relative to the planets would have resulted in substantially reduced pebble accretion rates.
Even with the more massive cores with a more massive pebble isolation mass, the accretion rates would have been severely hindered by the eccentric orbits, and so we do not expect our results to change depending on which pebble isolation mass prescription is used.
In future work, We will include the more advanced prescriptions for the pebble isolation mass, taking into account the local disc properties.

\subsubsection{Eccentric pebble orbits}
In eqs \ref{eq:mdot2d} and \ref{eq:mdot3d}, the relative velocity between the pebbles drifting through the disc and the accreting planets is used to calculate $\delta v$.
Around single stars, where the discs can be assumed to be axisymmetric, pebbles are assumed to be moving at a sub-Keplerian velocity, and as such the relative velocity between pebbles drifting past an accreting planet follows \citep{Johansen17},
\begin{equation}
\label{eq:peb_ecc_inc}
\begin{split}
    \Delta v&=\{[e v_{\rm K}\cos(\theta)]^2+[-(1/2)e v_{\rm K}\sin(\theta)+\eta v_{\rm K}]^2  \\
    &+[i v_{\rm K}\cos(\theta)]^2\}\end{split}
\end{equation}
where $e$ and $i$ are the planet eccentricities and inclinations, $v_{\rm K}$ is the Keplerian velocity, and $\theta$ is the mean longitude of the planet.
For single stars, and for the regions of circumbinary discs where the gas pressure support dominates the effects of the eccentric binary, we calculate and average the relative velocity around the entire planet's orbit, taking into account the planet's eccentricity and inclination that also increases the relative velocity between a planet and pebbles drift through the planet's orbital plane.
However, circumbinary discs contain an inner region where the binary stars can perturb the disc, making it eccentric.
In this region of the disc, where the disc eccentricity is dominated by the binary effects, we assume that the pebbles are following the gas streamlines (appropriate for particles with low Stokes numbers or low dust-to-gas ratios \citep{Coleman22b}), and as such the relative velocities between pebbles and planets is now a function of azimuth.
With this being the case, we directly calculate the relative velocities between a planet's instantaneous velocity and the gas/pebble velocity at the planet's location.
In the next section, we discuss how we calculate the disc eccentricity as a function of radius, as well as the implementation of an inner cavity.

\section{Modelling the inner disc regions}
\label{sec:binary_addons}
The model described in Sect.~\ref{sec:base_model} is mostly appropriate for axisymmetric protoplanetary discs around single stars. However the inner regions of circumbinary discs are not axisymmetric, since tidal torques from the central binary lead to the formation of an eccentric inner cavity, through which gas accretes onto the stars via gas streamers. The eccentricity of the inner disc creates azimuthal asymmetries in the surface density and velocities that affect the interactions between gas and pebbles/planets. In the following sections, we outline the prescriptions we include to model these effects in the inner regions of circumbinary discs, and their subsequent effects on the forming planets.

\begin{figure}
\includegraphics[scale=0.5]{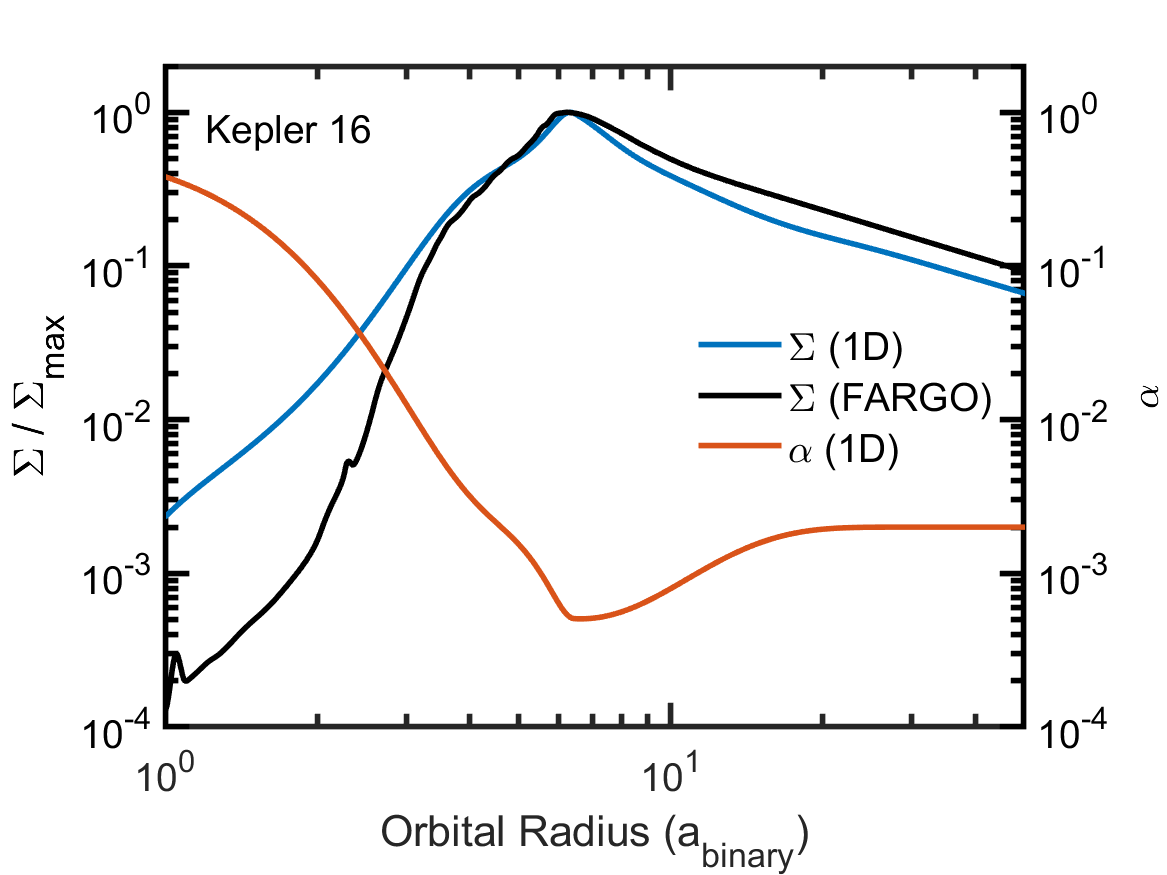}\\
\\
\includegraphics[scale=0.5]{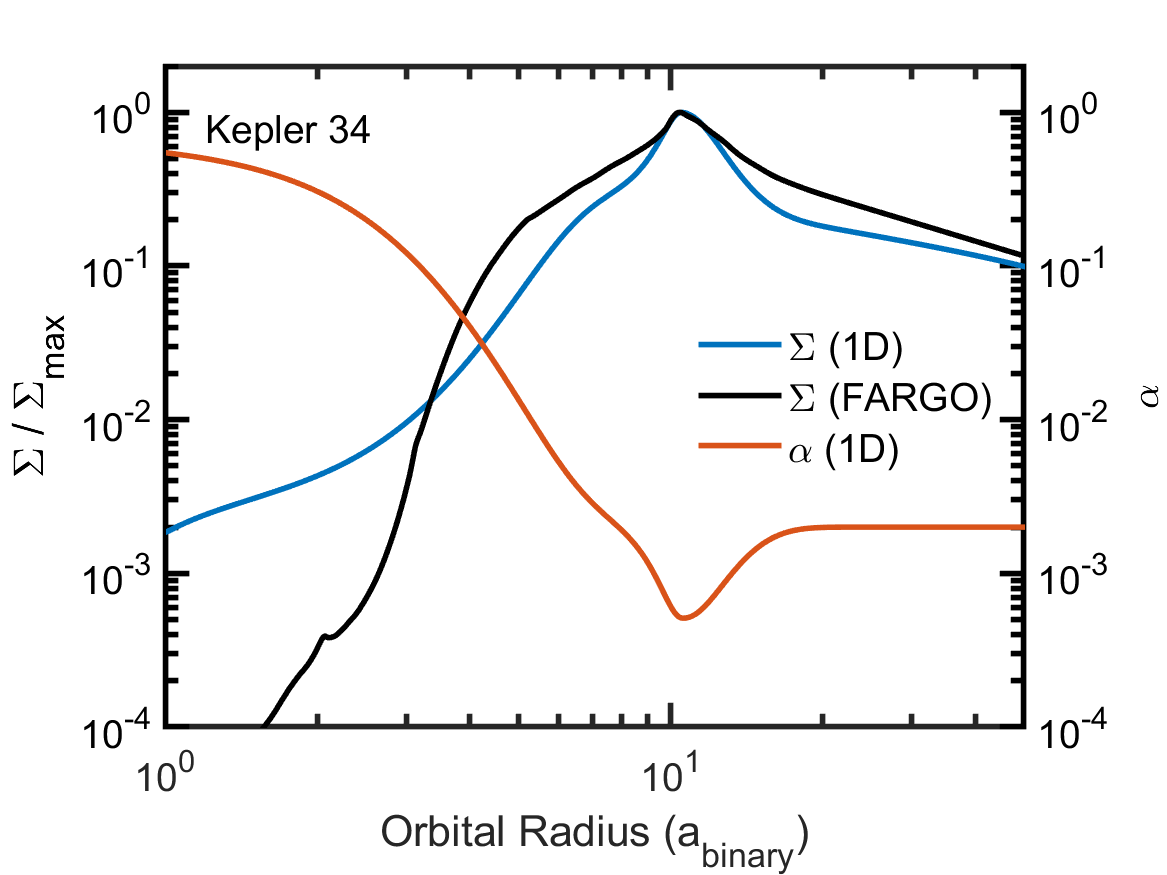}
\caption{Comparison of the surface density profile for a 1D model with varying $\alpha$ (blue line), and an azimuthally averaged profile from hydrodynamic simulations (black line). The varying $\alpha$ profile is shown as the red line. The top panel shows the profiles for Kepler-16 with the bottom panel showing Kepler-34.}
\label{fig:cavity_alpha}
\end{figure}

\subsection{Obtaining an inner cavity}
To mimic the decrease in the gas surface density inside the inner cavity, we adjust the viscous parameter $\alpha$ (see eq.~\ref{eq:viscosity}) in the regions of the disc close to binary stars. In a steady 1D disc with constant mass flow, there is a constant value of $\nu\Sigma$ where $\nu$ is the kinematic viscosity and $\Sigma$ is the gas surface density.
Therefore, by increasing the $\alpha$ parameter within the viscosity component, the surface density will decrease to maintain a constant flow rate.
The decreases in surface density can then be calibrated against azimuthally averaged values from hydrodynamical simulations using \textsc{fargo3d} \citep{FARGO-3D-2016}, to obtain a cavity of approximately the correct form in the gas surrounding the central binary stars.
As a function of radius, our prescription for $\alpha$ becomes:

\begin{equation}
\label{eq:alpha}
    \alpha(r) = \alpha_{\rm b} + \alpha_{\rm c} - \alpha_{\rm apo}
\end{equation}
where $\alpha_{\rm b}$ is equal to $2\times 10^{-3}$, $\alpha_{\rm c}$ is equal to
\begin{equation}
    \alpha_{\rm c} = 200\alpha_{\rm b}\times \left(\tanh{\left(\dfrac{3(\acav/4-r)}{3\acav/4}\right)}+1\right)
\end{equation}
and $\alpha_{\rm apo}$ equals
\begin{equation}
\label{eq:alpha_apo}
\alpha_{\rm apo} =  \left\{ \begin{array}{ll}
\dfrac{3\alpha_{\rm b}}{4}\times\exp{\left(\dfrac{-(r-C_1\,\rcavapo)^2}{C_2\rcavapo^2}\right)} ,& \dfrac{r}{\rcavapo}\le C_1 \\
\\
\dfrac{3\alpha_{\rm b}}{4}\times\exp{\left(\dfrac{-(r-C_1,\rcavapo)^2}{C_3\rcavapo^2}\right)} ,& C_1 < \dfrac{r}{\rcavapo} .\\ \end{array} \right.
\end{equation}
In the above equations, $\acav$ is the cavity radius, and $\rcavapo$ is the apocentre of the cavity ($\acav(1+e_{\rm c})$) where $e_{\rm c}$ is the cavity eccentricity.

In eq.~\ref{eq:alpha}, the first component of the right hand side denotes our nominal $\alpha$ value for the majority of the disc, far from the central binary where the effects of the varying gravitational potential are negligible.
This can be seen in the outer regions of the disc in the red  profiles in fig.~\ref{fig:cavity_alpha} that shows $\alpha$ as a function of radius in units of binary semi-major axes.
The second component of eq.~\ref{eq:alpha}, $\alpha_{\rm c}$, shows the significant increase in $\alpha$ that carves out the cavity region close to the central binary as seen by the sharp increase in $\alpha$ as the distance to the binary stars decreases in the inner region of fig.~\ref{fig:cavity_alpha}.
The third component of eq.~\ref{eq:alpha}, $\alpha_{\rm apo}$, represents a decrease and then increase in $\alpha$ that allows for an increase in surface density to arise around the apocentre of the eccentric cavity, as has been seen in multiple works \citep{Mutter17D,Thun17}.
This decrease and then increase in $\alpha$ can also be seen in the central regions of fig.~\ref{fig:cavity_alpha}.
The two factors, $C_2$ and $C_3$, in eq. \ref{eq:alpha_apo} allow for the slopes of the cavity apocentre to be easily adjusted, since the scale of the concentration of material at the cavity apocentre can change depending on the binary parameters.
Indeed the differences in concentrations can be seen in the black and blue profiles in fig. \ref{fig:cavity_alpha}, that compares the surface densities as a function of orbital distance of our 1D discs to their 2D azimuthally averaged counterparts derived from \textsc{fargo3d} simulations \citep{FARGO-3D-2016} for Kepler-16 (top panel) and Kepler-34 (bottom panel).
It can be seen that the agreement in the profiles is good when comparing the 1D to the 2D discs, except for the innermost regions close to the central binary, where the 2D discs tend to have significantly lower surface densities.
In order to attain such low surface densities in the 1D disc, $\alpha$ would have to be extremely high in that region (>1) and as such the time-step required to evolve the disc would be very short which would result in extremely long and unfeasible simulation run times. However, given that the surface density we obtain in that region is extremely low (at least two orders of magnitude lower than the cavity edge), and given that N-body interactions between the binary stars and individual planets will dominate their evolution, the inaccuracy in surface density compared to the 2D simulations should have negligible effects and is retained to allow for feasible simulation run times.

We note that an alternative way to treat the interaction between the central binary and the disc would have been to include the torque from the binary stars using an impulse approximation, similar to the way in which the torque on the disc due to a giant planet is included (see eq.~\ref{eq:GammaII} or the work of \citet{Alexander12}). The problem with this approach is that it would result in complete tidal truncation of the disc, without mass flow into the cavity and onto the stars occurring, because of the dominance of the tidal torque, and this behaviour is different to that which is observed in multi-dimensional hydrodynamical simulations of circumbinary discs, where significant accretion through the cavity is in fact observed.

\begin{figure}
\centering
\includegraphics[scale=0.6]{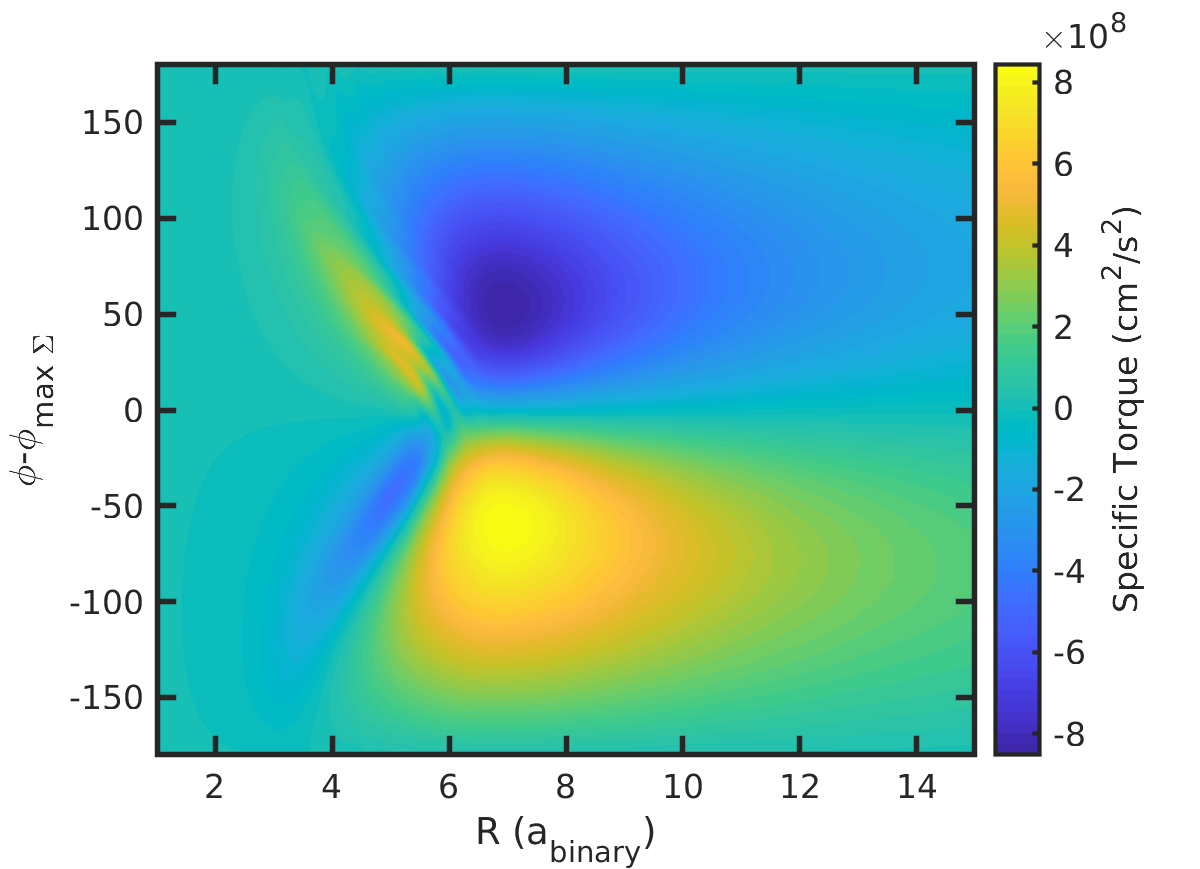}
\caption{Gravitational torque map from the eccentric cavity acting on solid objects as a function of azimuth in respect to the cavity apocentre and distance from the central binary.}
\label{fig:torque_map}
\end{figure}

\subsection{Gravitational potential of a 2D-disc}
Whilst the cavities now simulated in the 1D model are axisymmetric and match the azimuthally averaged 2D profiles, those in the {\textsc{fargo3d}} simulations are typically eccentric.
This eccentricity results in the concentration of gas at the apocentre of the cavity.
With such an imbalance of mass at a specific location of the disc (up to an order of magnitude), the gravitational potential of the disc can no longer be considered axisymmetric as would be the case in typical 1D discs. This azimuthally asymmetric potential induces significant perturbations in the orbits of planets and planetesimals, as seen in \citet{Marzari13}.

To account for the potential from the non-axisymmetric disc, we compute the average surface densities that arise in 2D {\textsc{fargo3d}} simulations of both Kepler-16 and Kepler-34.
These simulations have been allowed to reach a steady state, and we average the surface densities over the final 5000 binary orbits.\footnote{Other numbers of orbits were also checked to ensure consistency in calculating an average surface density, and thus gravitational potential.}
Using the average surface densities, a map of the gravitational potential was then generated as a function of the distance to the centre of the system and the azimuth in respect to the azimuth of the cavity apocentre.
Figure \ref{fig:torque_map} shows the generated map for the Kepler-16 system, with the Kepler-34 map containing similar characteristics.
The colour scale in fig.~\ref{fig:torque_map} shows the specific torque that would act on a planet/planetesimal located at a specific $R$ and $\phi$ in the disc, with zero indicating no torque acting on a planet, as it would be in an axisymmetric disc. We include the forces arising from such a torque map on planets/planetesimals in our simulations.

\begin{figure}
\centering
\includegraphics[scale=0.6]{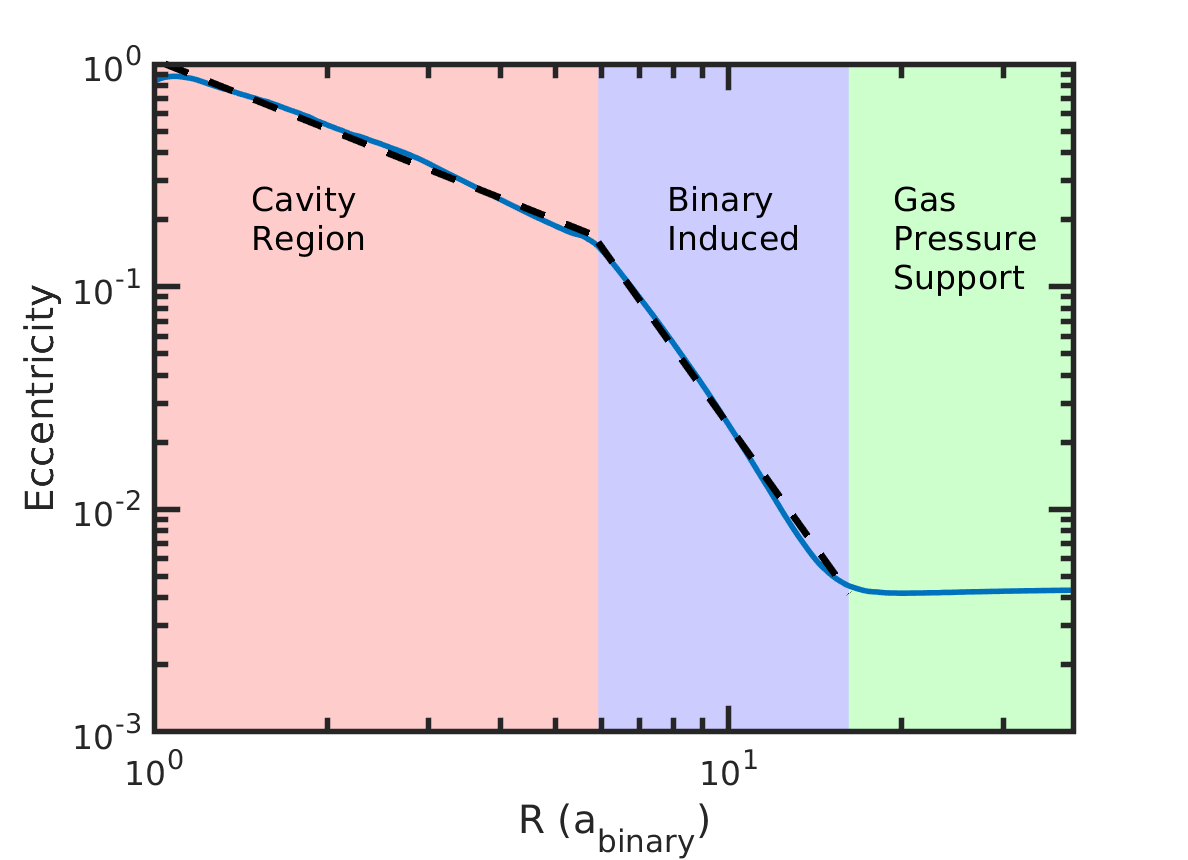}
\caption{Azimuthally averaged eccentricity profile of a disc around Kepler-16 (blue line), with coloured patches denoting different regions. The black dashed lines show the fits to the eccentricity from eq. \ref{eq:ecc_power_k16}.}
\label{fig:ecc_r}
\end{figure}

\subsection{Two dimensional gas velocities}
The eccentric inner region of the disc not only produces an asymmetric gravitational potential, it will also significantly affect the gas velocities that are used to calculate the pebble velocities when planets in the region are accreting pebbles that drift past their orbits.
The extent to which the gas velocities deviate from axisymmetry is also dependent on the distance to the central binary, with the strongest deviations occurring close to the central binary, in and around the cavity region.

Figure \ref{fig:ecc_r} shows the azimuthally averaged eccentricity profile arising from a 2D {\textsc{fargo3D}} simulation of a disc in the Kepler-16 system (blue line).
As can be seen by the profile of the eccentricity and the shaded regions, the eccentricity of the disc as a function of radius can be split into three components.
The eccentricity in the outer region of the disc at $r>15 \ab$, shown by the green shaded area, is dominated by the pressure support in the disc, $\eta$, and as such remains at a low value.
Since this region of the disc is far from the central binary, the time-varying gravitational potential is barely felt by the gas, i.e. the gas acts as if orbiting a single star, and as such the excitation of the gas orbits is diminished compared the effects of gas pressure support.
Though the eccentricity seems to be constant here, this is due to the simulation parameters with a constant aspect ratio, and similar surface density profiles in the outer disc region.
In the 1D simulations presented later in this work, the discs will have a varying $\eta$ across the disc, since the temperature and therefore the scale height of the disc are calculated to be in thermal equilibrium (eq. \ref{eq:temperature}).

Moving inwards closer to the stars, it is clear that the eccentricity profile can be fit by two power laws, shown by the dashed black lines and the red and blue shaded regions in fig. \ref{fig:ecc_r}.
The inner power law fit extends from the central binary out to the vicinity of the cavity apocentre ($5.6 \ab$ for Kepler-16), with the other extending further out towards the outer regions of the disc.
The two power laws correspond to
\begin{equation}
\label{eq:ecc_power_k16}
e(r) =  \left\{ \begin{array}{ll}
\ecav\times 10^{(-0.34959 - 1.0315 \times\log_{10}(r))} ,& r \le 1.1\times\rcavapo \\
\\
\ecav\times 10^{(-0.26591 - 3.622 \times\log_{10}(r))},& r > 1.1\times\rcavapo ,\\ \end{array} \right.
\end{equation}
for Kepler-16, and
\begin{equation}
\label{eq:ecc_power_k34}
e(r) =  \left\{ \begin{array}{ll}
\ecav\times 10^{(-0.38874 - 0.81896 \times\log_{10}(r))} ,& r \le 1.05\times\rcavapo \\
\\
\ecav\times 10^{(-0.32367 - 4.0975 \times\log_{10}(r))},& r > 1.05\times\rcavapo ,\\ \end{array} \right.
\end{equation}
for Kepler-34.
With the stellar masses, orbital separations and eccentricities being different for Kepler-16 and -34, the perturbations induced by the binaries on their respective discs are quantitatively different, which causes different profiles in the disc eccentricities.
This causes the fitting values in eqs. \ref{eq:ecc_power_k16} and \ref{eq:ecc_power_k34} to have different values to accurately fit the disc eccentricity.

With equations~\ref{eq:ecc_power_k16} and \ref{eq:ecc_power_k34}, we now have a function for the eccentricity as a function of orbital distance.
This does not however give us a value for the velocity of the gas as a function of azimuth.
Therefore using the equations for eccentricity, we numerically calculate the semi-major axis of gas at a given azimuth, assuming that the orientation of gas is the same as that of the precessing inner disc, that we assume has a precession period equal to 3,000 binary orbits, consistent with values typically found in hydrodynamical simulations \citep[e.g.][]{Thun17,Kley19,Coleman22b}.
Once the semi-major axis is known, we can therefore then calculate the velocities of the gas and pebbles that are included in the pebble accretion rates where necessary.

\begin{table}
\centering
\begin{tabular}{l|ccc}
Parameter & Kepler-16 & Kepler-34 & References\\
\hline
$M_{\rm A}\ (\msun)$ & 0.6897 & 1.0479 & $^{1,2}$\\
$M_{\rm B}\ (\msun)$ & 0.2025 & 1.0208 & $^{1,2}$\\
$M_{\rm bin}\ (\msun)$ & 0.8922 & 2.0687 & $^{1,2}$\\
$T_{\rm A}\ ({\rm K})$ & 4000 & 4300 & \\
$T_{\rm B}\ ({\rm K})$ & 3200 & 4300 & \\
$R_{\rm A}\ (\rsun)$ & 2 & 2 & \\
$R_{\rm B}\ (\rsun)$ & 1.5 & 2 & \\
$a_{\rm b}\ (\au)$ & 0.224 & 0.228 & $^{1,2}$\\
$e_{\rm b}$ & 0.15944 & 0.52087 & $^{1,2}$\\
Metallicity [m/H] (dex) & -0.3 & -0.1& $^{1,3}$\\
& & & \\
$\acav\ (\ab)$ & 3.8135 & 6.424 & \\
$\ecav$ & 0.4124 & 0.5598 &\\
$\rcavapo\ (\ab)$ & 5.3862 & 10.02 & \\
$C_1$ & 1.2 & 1.05 & \\
$C_2$ & 0.08 & 0.045 & \\
$C_3$ & 2.0 & 0.18 & \\
$\alpha_{\rm b}$ & $2\times 10^{-3}$ & $2\times 10^{-3}$ &\\
R$_{\rm in}\ (\au)$ & 0.224 & 0.228 & \\
R$_{\rm out}\ (\au)$ & 100 & 100 & \\
$f_{41}$ & 10 & 10 & \\
$r_{\rm g} (\au)$ & 7.66 & 17.76 & \\
$\dot{M}_{\rm pe,ext}\ (M_{\odot}/{\rm yr})$& $3\times 10^{-8}$ & $3\times 10^{-8}$ \\
$f_{\rm opa}$ & 1 & 1 & \\
$Z_{\rm peb}/Z_0$ & 0.9 & 0.9 & \\
$Z_{\rm dust}/Z_0$ & 0.1 & 0.1 & \\
$\Sigma_{\rm g,1\au}$ (gcm$^{-2}$) & 1265 & 2933 & \\
$\alpha$ & 1 & 1 & \\
$T_{\rm 1\au}$ (K) & 543 & 899 & \\
$\beta$ & 0.5 & 0.5 & \\

\end{tabular}
\caption{System and disc parameters.
References: $^1$ \citet{Doyle11,Triaud22}, $^2$ \citet{Welsh12}, $^3$ \citet{Everett13}.}
\label{tab:system_param}
\end{table}

\section{Basic Setup and disc evolution}
\label{sec:example_disc}

With the above sections describing the physical model and additions due to the binary stars, we now describe the initial conditions for the simulations and then present a fiducial model of the discs around both Kepler-16 and Kepler-34, and describe the evolution of the gas surface densities, temperatures and aspect ratios, as well as the migration behaviours of embedded planets.
We initialise the disc surface density and temperature profiles using, $\Sigma_{\rm g}(r) = \Sigma_{\rm g,1\au}(r/\au)^{-\alpha}$and $T(r) = T_{\rm 1\au}(r/\au)^{-\beta}$ respectively, where the values for $\Sigma_{\rm g,1\au}$, $T_{\rm 1\au}$, $\alpha$, and $\beta$ can be found in Table \ref{tab:system_param}.
The stellar and other disc parameters for Kepler-16 and -34 can also be found in Table \ref{tab:system_param}.

We include 37 and 47 planets in the simulations for Kepler-16 and Kepler-34 respectively, with their mass being set to $10^{-3}\me$.
We choose this mass as it is similar to the transition mass through the disc (with typical values between $10^{-4}$--$10^{-2}\me$ \citep{Bitsch15,Coleman21}, and also consistent with the most massive object that forms from the gravitational collapse of a pebble cloud yielding a distribution of planetesimals sizes that follows a power law plus an exponential tail \citep{Schafer17,Abod19,Liu20,Coleman21}.
In each disc the planets were placed with equidistant spacing in the interval 2.25--20$\au$ for Kepler-16 and between 3--25$\au$ for Kepler-34.
The reason for the increased number of planets and the wider initial semimajor axis range is due to the increased combined stellar mass of the Kepler-34 system, which reduces the formation and growth times of planets through initial planetesimal accretion and oligarchic growth.
This allows planetesimals that form further out in the disc to grow into planets and undergo substantial migration and pebble accretion.
We set the planet densities to either 1.5 or 3 $\rm gcm^{-2}$ depending on whether they form outside or inside the iceline respectively.
Planet eccentricities and inclinations were initialised randomly between 0--0.02, and 0--0.36$^{\rm o}$, respectively.
These non-zero values represent the initial formation process of the planets where they attain non-zero eccentricities through interactions with other planets and planetesimals, and through non-axisymmetric perturbations in the circumbinary discs.
The disc properties that we vary as part of our parameter study are limited to the initial disc mass and metallicity\footnote{We take the metallicity relative to Solar, which we assume is equal to 1$\%$.}, with our parameter choices shown in Table \ref{tab:sim_param}.
For each combination of parameters we run 10 realisations where we use a different random number seed to generate the initial planet positions and velocities.

\begin{figure}
\centering
\includegraphics[scale=0.6]{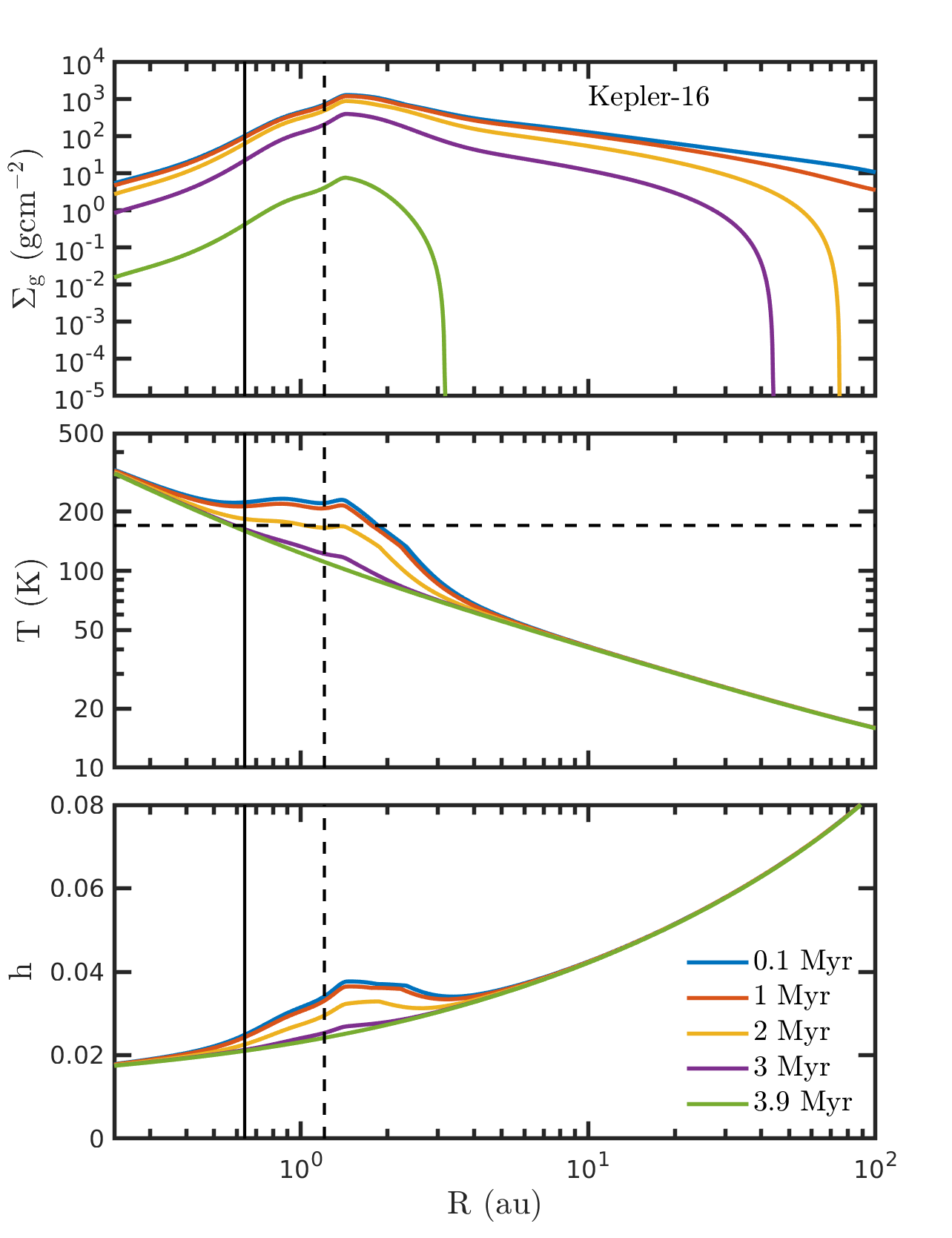}
\caption{Temporal evolution of the surface density (top panel), midplane temperature (middle panel) and aspect ratio (bottom panel) for a disc around Kepler-16. The vertical lines denote the outer edge of the zone of dynamical instability (solid line) and the cavity apocentre (dashed line), whilst the horizontal dashed line in the middle panel represents the water iceline.}
\label{fig:kep16_disc}
\end{figure}

\begin{figure}
\centering
\includegraphics[scale=0.6]{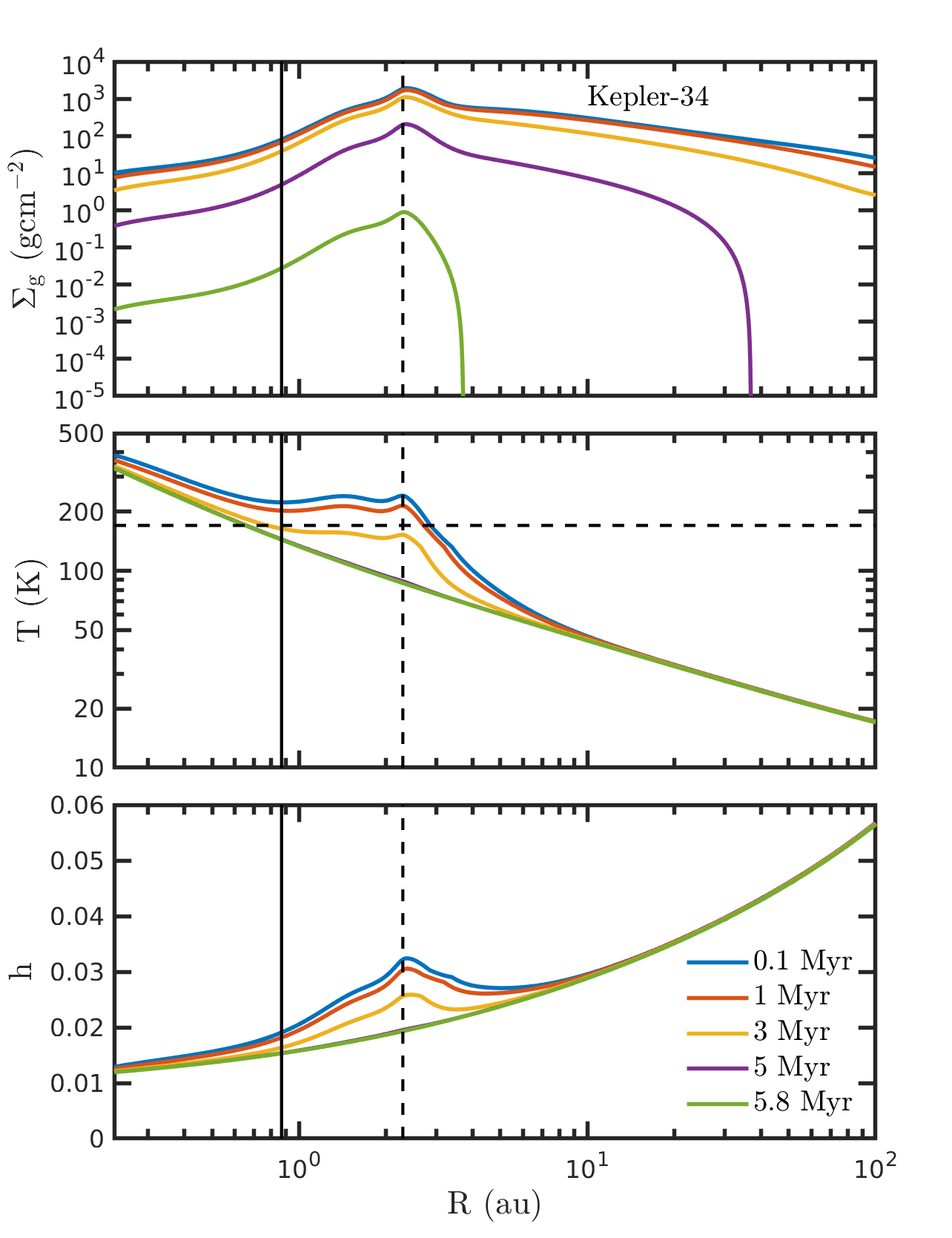}
\caption{Temporal evolution of the surface density (top panel), midplane temperature (middle panel) and aspect ratio (bottom panel) for a disc around Kepler-34. The vertical lines denote the outer edge of the zone of dynamical instability (solid line) and the cavity apocentre (dashed line), whilst the horizontal dashed line in the middle panel represents the water iceline.}
\label{fig:kep34_disc}
\end{figure}

\subsection{Disc Profiles}
Figure \ref{fig:kep16_disc} shows the temporal evolution of the gas surface density (top panel), temperature (middle panel), and aspect ratio (bottom panel), for our fiducial disc around Kepler-16.
The initial disc had a mass equal to 0.1$M_{\rm bin}$, and had a lifetime of 3.9 Myr which we define as when there is only $10^{-2}\me$ of gas remaining in the disc.
The black vertical lines denote the outer edge of the zone of dynamical instability (inner line), and the location of the cavity apocentre (outer line).
The effects of the binary stars on creating a central cavity can be easily seen in the inner regions of the disc (<2$\au$), where material concentrates at the apocentre of the cavity region, before the tidal torques from the binary stars carve out the central cavity.
This central cavity can be seen here in the left-hand most parts of fig. \ref{fig:kep16_disc} where the surface density decreases by three orders of magnitude.

The accumulation of material just outside the cavity is also clearly evident in all of the surface density profiles between 1--2$\au$.
This build-up of material, induces significant viscous heating at this location, increasing the midplane temperature above that of an irradiation dominated disc.
With the cavity being present just interior to this location, this leads to a decrease in the viscous heating due to the depletion of gas.
This is seen by the flat temperature profile, sometimes decreasing between the cavity apocentre, and just interior to the outer edge of the zone of dynamical instability, where the temperature becomes consistent with that calculated by irradiation alone.
Interestingly, for Kepler-16, the region where material accumulates also corresponds to the expected location of the water iceline for our disc models, where $T\sim170$~K.
Around this temperature, there are numerous transitions in the opacity, depending on the expected composition and structure of the dust grains.
These transitions can act as planet migration traps due to generating more favourable surface density and temperature gradients that allow corotation torques to balance Lindblad torques.

As the disc evolves, the effects of photoevaporation can be seen in the outer part of the disc after 2 Myr (yellow line), where external photoevaporation is beginning to truncate the disc.
This truncation continues in the purple line, whilst internal photoevaporation and accretion onto the central stars also continue to operate, removing material from the disc.
When looking at the temperature around the cavity, the reduction of gas as the disc evolves limits the amount of viscous heating, reducing the temperature in this region.
This can be seen in the middle panel of fig. \ref{fig:kep16_disc} as the midplane temperature gradually relaxes to the irradiation dominated temperature as the disc evolves.
After 3.9 Myr, photoevaporation has completely cleared the disc exterior to the cavity in an outside-in manner \citep{Coleman22}, leaving only a small amount of gas to finish accreting onto the central stars before the disc is fully dispersed.

\begin{figure*}
\centering
\includegraphics[scale=0.5]{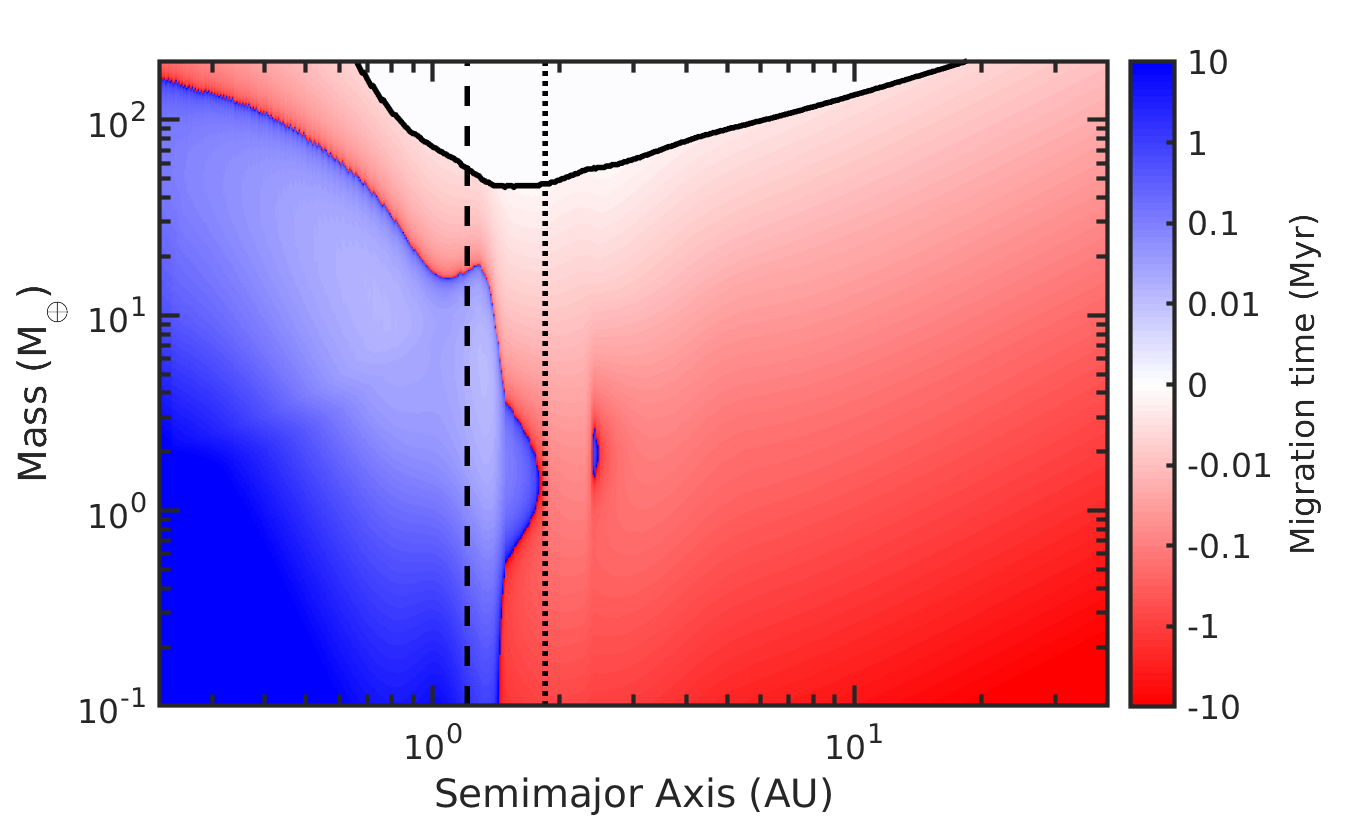}
\hspace{0.5cm}
\includegraphics[scale=0.5]{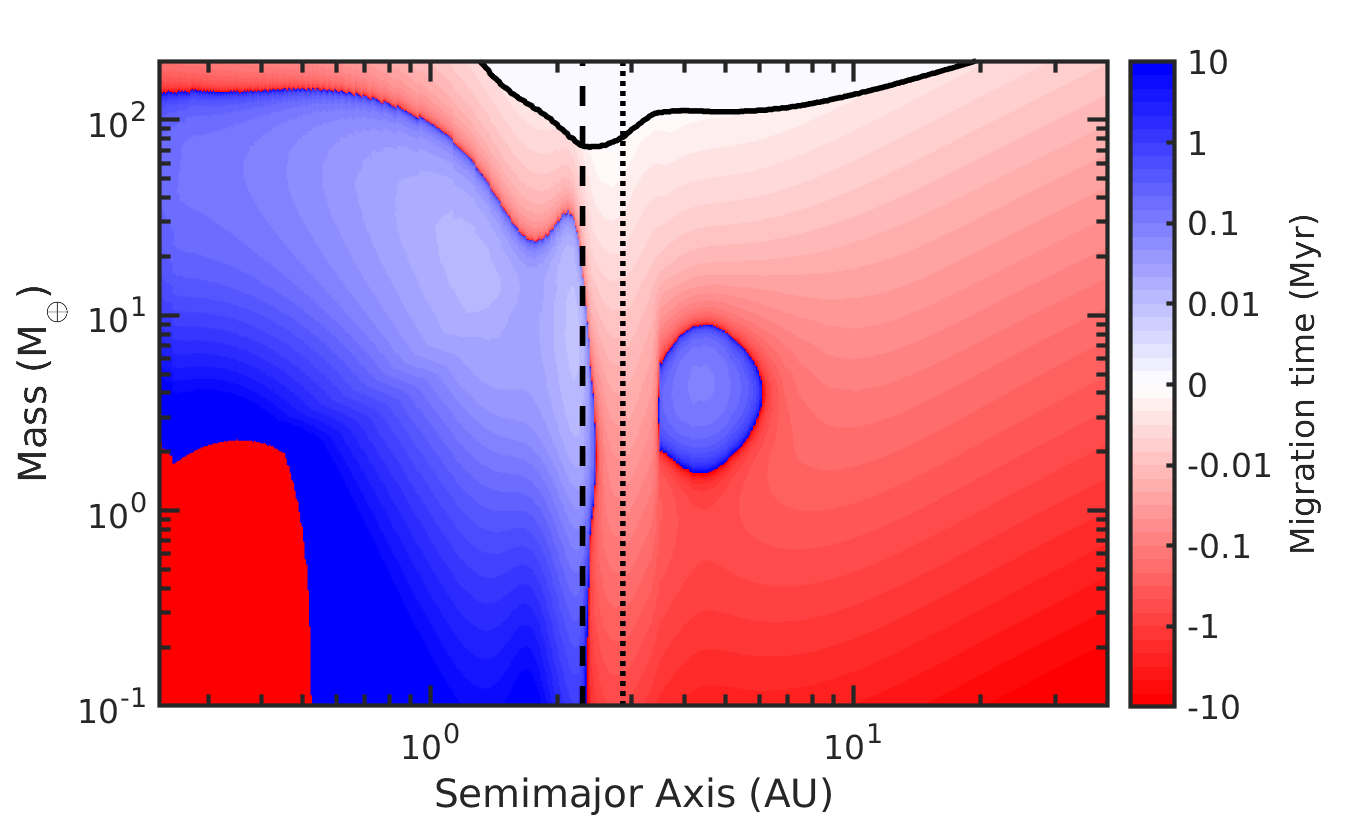}
\caption{Migration time-scales for discs around Kepler-16 (left panel) and Kepler-34 (right panel) with an initial disc mass of 0.1$\times M_{\rm bin}$ after a time of 0.1 Myr. Red regions show migration time-scales for inwardly migrating planets, whilst blue regions show outwardly migrating regions. The vertical dashed lines denote the outer edge of the zone of dynamical instability, and the vertical dotted lines show the water iceline. The upper black solid line shows the gap opening mass where migration switches from type-1 to type-2 migration.}
\label{fig:both_discs_migr}
\end{figure*}

Qualitatively, the evolution of the disc around Kepler-34, is similar to that of Kepler-16, in regards to the profiles in the disc, as well as the temporal evolution when including photoevaporation.
These similarities can be seen in fig. \ref{fig:kep34_disc} which again shows the gas surface density (top panel), temperature (middle panel) and aspect ratio (bottom panel), but for a disc with lifetime 5.8 Myr, around Kepler-34.
Like the disc for Kepler-16, the cavity can be easily seen in the left most parts of the disc, with the concentration of material at the cavity apocentre, before the surface density significantly drops in the cavity.
The longer disc lifetime arises due to the significantly larger initial disc mass, even though both examples had initial disc masses of 10 per cent of the combined binary mass, the Kepler-34 binary system has a combined mass that is  $\sim 2.3\times$ larger than that of Kepler-16.

\subsection{Migration behaviour}
In fig. \ref{fig:both_discs_migr} we show the migration time-scales for planets of different mass (y-axis) located at different locations of the disc (x-axis), after an evolution time of 0.1 Myr.
The colours show the migration time-scales in Myr, with red denoting inwards migration, and blue showing outwards migration.
The darker the colour, the longer the migration time-scale.
The vertical dashed line denotes the location of the cavity apocentre, since it can act as a significant migration trap, and the solid black line shows the gap opening mass (eq. \ref{eq:gapopening}) with the white region above showing that planets are in the type II migration regime.
In the left panel, we show the contours for a disc in the Kepler-16 system, with the right panel being for Kepler-34.

Features seen in previous works \citep[e.g.][]{ColemanNelson14,ColemanNelson16} such as the outward migration regions near opacity transitions (i.e. the iceline) are again seen in the circumbinary discs around Kepler-16 and -34.
This is seen for planets of masses 2--4$\me$ at around 2.5$\au$ around Kepler-16, and 2--10$\me$ between 4--7$\au$ for Kepler-34.
The outward migration region is larger in the discs for Kepler-34 since the the discs are slightly hotter due to the increased viscous heating and irradiation temperatures, as well as thermal diffusion time-scales being more comparable to horseshoe libration time-scales for lower mass planets.
The latter effect increases the strength of the entropy components of the horseshoe drag and linear corotation torques, allowing them to overcome Lindblad torques and create the regions of outwards migration.

The main difference in the migration maps to previous works, is the introduction of the inner cavity, carved by the binary stars.
The strong positive surface gradients associated with the cavity can be seen by the large region of outward migration around and interior to the cavity apocentre, shown by the vertical dashed lines.
Such strong positive surface density gradients act to substantially increase the strength of the corotation torque, for both the viscous and thermal components, so that it becomes stronger than planetary Lindblad torque and forms a migration trap.
This trap acts to halt the migration of low-mass planets undergoing type-I migration, until they reach masses in excess of 30$\me$, at which point those planets are about to enter the runaway gas accretion regime and open a gap in the disc, transitioning to type-II migration.

As the discs evolve, the migration trap located around the cavity region remains, but reduces in effectiveness and as such moves to slightly lower planet masses.
In regards to the outward migration region due to the changes in disc opacities, this region diminishes as the disc evolves, and the scale of viscous heating reduces and causes the disc midplane temperature to be similar to that derived from an irridated disc.
The reduction in temperature unfavourably alters the temperature gradients, making the corotation torque weaker than the Lindblad torque, and thus removing the region of outwards migration.

\begin{table}
\centering
\begin{tabular}{l|cc}
Parameter & Kepler-16 & Kepler-34\\
\hline
Initial disc mass ($M_{\rm bin}$) & [0.05,0.01,0.2] & [0.05,0.01,0.15]\\
Metallicity (Solar) & [0.5,1,2] & [0.5,0.75,1,2]\\
No. of realisations & 10 & 10\\
Initial $m_{\rm p}$ ($\me$) & $10^{-3}$ & $10^{-3}$\\
Initial $a_{\rm p}$ range ($\au$) & 2.25--20 & 3--25\\
No. of planets & 37 & 47\\
\end{tabular}
\caption{Simulation parameters for Kepler-16 and Kepler-34}
\label{tab:sim_param}
\end{table}

\section{Kepler 16}
\label{sec:kep16}

We now present the results of our simulations around Kepler-16.
Initially we will discuss the formation of a system similar to Kepler-16, before analysing the planet population produced by the simulations as a whole.

\begin{figure}
\centering
\includegraphics[scale=0.6]{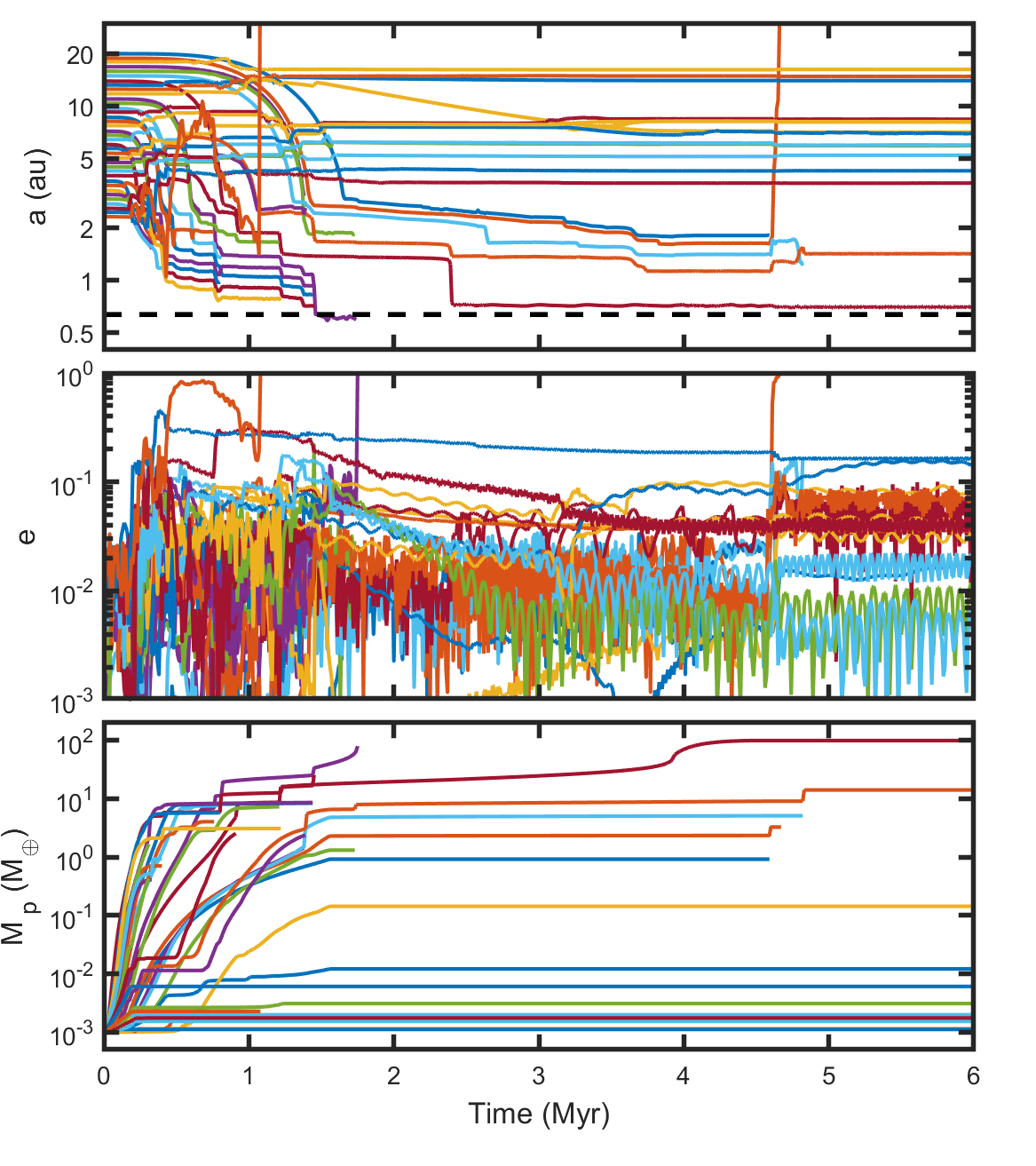}
\caption{Temporal evolution of planet semimajor axes (top), eccentricities (middle) and masses (bottom) for the example simulation described in sect. \ref{sec:kep16_sim}. The dashed horizontal black line denotes the outer edge of the zone of dynamical instability.}
\label{fig:kep16_exact_time}
\end{figure}

\begin{figure}
\centering
\includegraphics[scale=0.6]{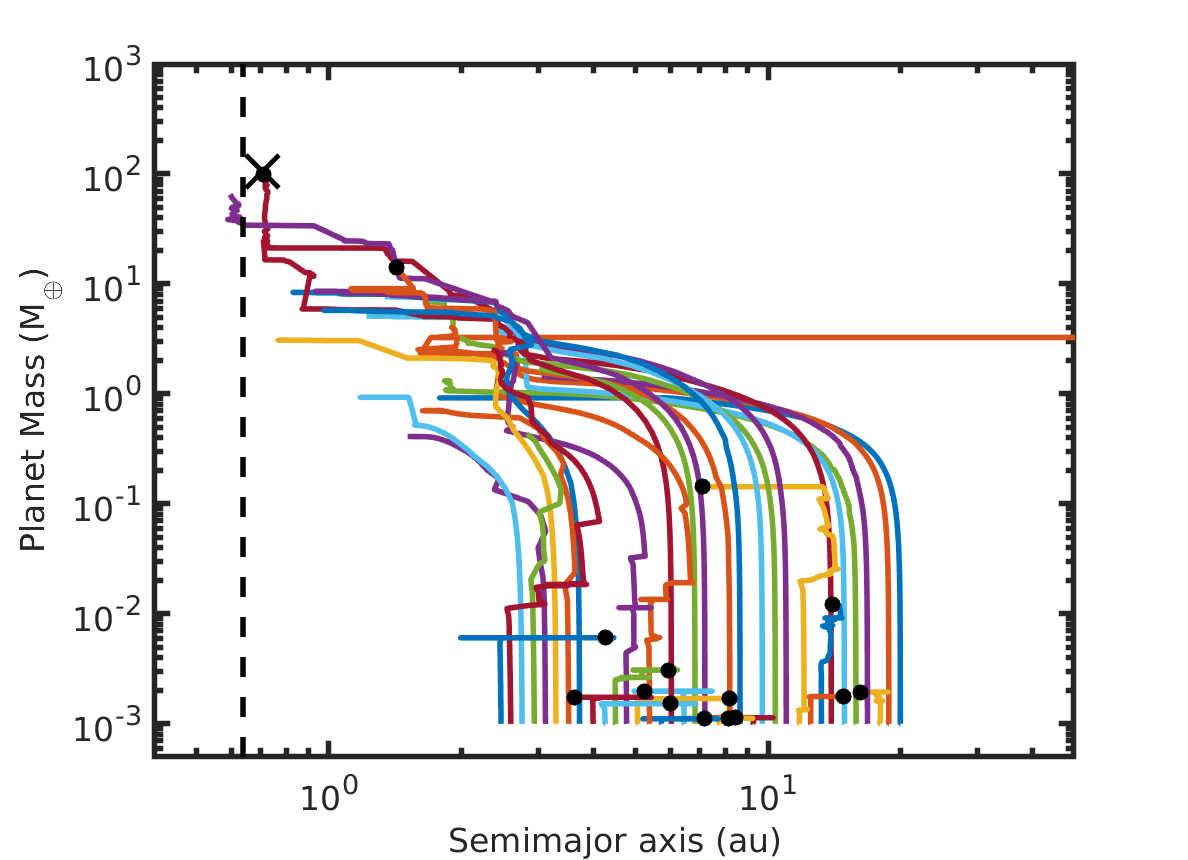}
\caption{Evolution of planet mass versus semimajor axis for the example simulation described in sect. \ref{sec:kep16_sim}. Filled black circles represent final masses and semimajor axes for surviving planets. The black cross shows the mass and semimajor axis of Kepler-16b, whilst the dashed vertical black line denotes the outer edge of the zone of dynamical instability.}
\label{fig:kep16_exact_mva}
\end{figure}

\subsection{Forming a Kepler-16b analogue}
\label{sec:kep16_sim}
Figure \ref{fig:kep16_exact_time} shows the temporal evolution of planet semimajor axes (top panel), eccentricities (middle panel) and masses (bottom panel) of one simulation that formed a system similar to Kepler-16b.
Figure~\ref{fig:kep16_exact_mva} shows the mass versus semimajor axis evolution of forming planets, where the black points represent the final planet masses and semimajor axes.
The dashed black line in figs. \ref{fig:kep16_exact_time} and \ref{fig:kep16_exact_mva} shows the outer edge of the zone of dynamical instability for Kepler-16 \citep{Holman99}, while the black cross represents Kepler-16b \citep{Doyle11}.
The initial disc mass for this simulation was equal to 16$\%$ of the combined stellar mass and the metallicity was equal to 0.5 $\times$ the Solar metallicity.

As the disc evolves in time, a pebble production front moves slowly outwards, converting dust into pebbles at the midplane that then drift in towards the central stars.
Planets are then able to accrete pebbles as they drift inwards.
As the amount of pebbles that are accreted by the planets depends on their eccentricities and inclinations (see eq. \ref{eq:peb_ecc_inc}), two populations of planets quickly appear, those that are on circular and coplanar orbits that can increase their mass, and those that are on slightly inclined and eccentric orbits which accrete very few pebbles.
After $\sim 0.3$ Myr, a number of planets around 3$\au$ grow to the super-Earth mass regime and begin to migrate in towards the inner cavity region.
Interactions within this group of planets lead to a number of collisions, allowing some of the planets to increase their core mass and accrete more pebbles, and reach masses around 10$\me$ that slowly accrete gas from the surrounding disc.
This pebble accretion can be seen in the rapid increase in mass of planets from $10^{-3}\me$ to $10\me$ in the far left part of the bottom panel of fig.~\ref{fig:kep16_exact_time}.
This group of planets forms a resonant convoy that migrates to the cavity region before being trapped there due to the positive surface density gradients allowing the corotation torques to balance the Lindblad torques, halting migration.

Over the next 0.25~Myr, three other planets accrete pebbles and migrate towards the cavity, joining the resonant chain.
This acts to slightly destabilise the chain of planets, leading to collisions between planetary cores.
Ultimately this period of instability reduces the number of planets in the chain from ten to seven after 1~Myr, with the most massive planet now being equal to 22$\me$.
This planet continues to accrete gas and reaches a mass of 25$\me$ after 1.45 Myr, where the resonant chain becomes unstable.
The instability in the chain leads to a 10$\me$ core colliding with the more massive planet, creating a 35$\me$ core.
This planet then migrates in past the cavity apocentre to halt its migration near the zone of dynamical instability.
As the planet migrates, it induces a number of other collisions, whilst also causing another planet to be ejected from the system.
The planet continues to orbit near the zone of dynamical instability, accreting gas, until 1.75 Myr where it interacted with the binary and was ejected from the system with a final mass of 80$\me$.
    
While the more massive planet orbits near the zone of dynamical instability and grows, a chain of five planets grows and migrates towards the cavity, becoming trapped there.
This chain of planets continues to grow, until the innermost planet reaches a mass of 21$\me$ and migrates past the cavity apocentre, into the cavity.
The other planets in the chain then migrate inwards to orbit near the cavity apocentre.
All of the planets continue to accrete gas, where the inner planet orbiting near the zone of dynamical instability begins to accrete at a significantly faster rate after 3.9 Myr when it reaches a mass of 40$\me$.
This allows the planet to quickly reach a mass of 70$\me$ where its torques were able to influence the disc and open a common gap with cavity.
The planet then transitioned to the slower viscous accretion regime.

After 4.55 Myr, the disc is fully dispersed, leaving the giant planet with a mass of 100$\me$ near the zone of dynamical instability, as well as a chain of four planets outside of where the cavity was.
A number of planets less than 0.2$\me$ are also orbiting further out in the system.
Whilst the giant planet with mass and period properties similar to Kepler-16b survives until the end of the simulation after 10 Myr, the resonant chain of planets quickly destabilises, and after 4.9 Myr, only one planet remains, having ejected one planet and after colliding with the remaining two.
This planet has a mass of 14$\me$, orbiting with an orbital period of 1.8 yrs.

\begin{figure}
\centering
\includegraphics[scale=0.6]{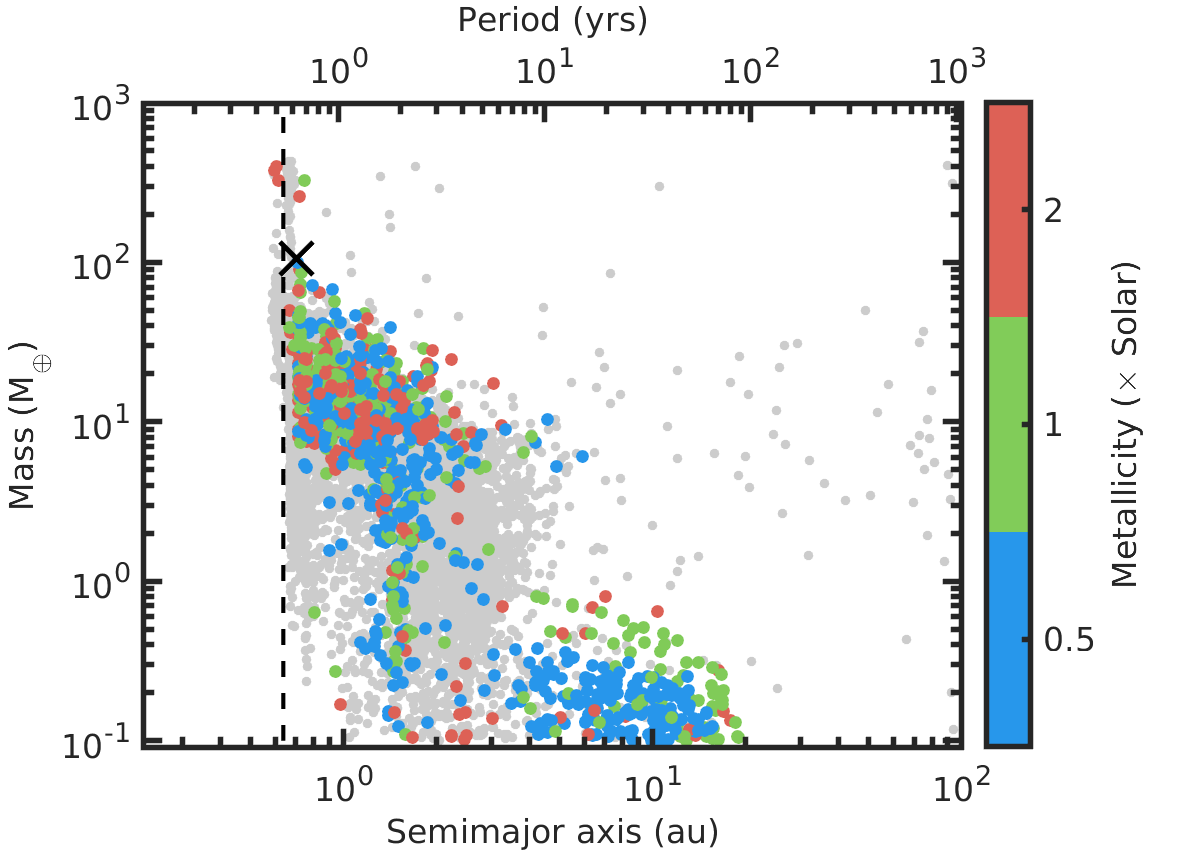}
\caption{Mass versus semimajor axis plot for all planets from the Kepler-16 simulations. Grey points show planets that have been lost from the simulations either by collisions or ejections. Coloured points show the surviving planets with different colours representing different initial disc metallicities: 0.5$\times$ Solar (blue), 1$\times$ Solar (green) and 2$\times$ Solar (red). The dashed vertical black line denotes the outer edge of the zone of dynamical instability, whilst the black cross shows the mass and semimajor axis of Kepler-16b.}
\label{fig:kep16_mva}
\end{figure}

\subsection{Overall population}

With the section above describing the formation of a system containing a planet similar to Kepler-16b, we now examine the population as a whole that arises from our suite of Kepler-16 simulations.
Whilst a number of simulations were not able to form a planet similar to Kepler-16b, the general formation pathways of the planets accreting pebbles and migrating to the edge of the cavity in resonant chains remained the same.
In fig.~\ref{fig:kep16_mva}, we show the masses versus semimajor axes for all planets in the simulations.
Planets denoted by grey points are those that have been lost from the simulations, either through collisions with other planets or via ejections following interactions with the central binary.
The colour coding for the surviving planets shows the metallicity of the circumbinary disc in which they formed, whilst the black cross shows the location of Kepler-16b.
The dashed black line denotes the outer edge of the zone of dynamical instability.

As can be seen in fig. \ref{fig:kep16_mva}, there are an abundance of planets with masses between 5--40 $\me$ with orbits outside the zone of dynamical instability, and extending to $\sim 2\au$, roughly the region around and immediately exterior to the cavity carved by the central binary.
Also apparent from fig.~\ref{fig:kep16_mva} is the trapping of planets at the cavity, near 1.4 $\au$, as well as the effects of the zone of dynamical instability where very few surviving planets are orbiting inside of the limit.
Those planets that are orbiting inside the stability are only just inside the limit, showing that there are orbital configurations that allow planets to orbit slightly closer than the empirically defined outer edge of the zone dynamical instability allows.

Interestingly, there are very few giants that form in the simulations, with only a handful of giant planets seen to be orbiting near the zone of dynamical instability.
The lack of giant planets has two main roots: migration into the cavity of cores that form in higher metallicity systems before they can undergo runaway gas accretion, and in lower metallicity systems the lack of sufficiently massive cores that form because of the abundance of solids being too small.
Whilst the migration affects all simulations, with the discs evolving similarly, the lack of mass in solids can be addressed by increasing the metallicity of the disc.
Increasing the metallicity boosts the amount of pebbles that can be accreted by the planets, allowing more massive cores to form, with some of them able to undergo runaway gas accretion within the disc lifetime.
This metallicity effect can be seen in fig. \ref{fig:kep16_mva} where as the metallicity increases (shown by blue points going to green and then red), the distribution of planets at higher masses also increases.
This can be particularly seen for planets above 100$\me$, where only those discs with metallicities of Solar or twice Solar are able to form such surviving planets.
For the giants that were formed in the discs and then subsequently ejected, 57 giant planets formed in discs with $2\times$ Solar metallicity, compared to 18 in discs with 0.5$\times$ Solar metallicity, again showing the effects of increased metallicity, whilst the average giant mass was 22$\%$ higher for those that formed in the metal-rich discs compared to those in metal-poor discs.

When comparing the resultant population to Kepler-16b, fig.~\ref{fig:kep16_mva} shows there are a few simulated planets that match the mass and period of the real system.
With the mass of Kepler-16b being roughly a Saturn mass, this places the planet amidst the runaway gas accretion regime, where cores can quickly bypass this mass by growing quickly into giant planets, as can be seen by the number of surviving giant planets above Kepler-16b in fig.~\ref{fig:kep16_mva}.
However some planets, similar to that described in sect. \ref{sec:kep16_sim}, were able to fortuitously time their formation so that they reached a mass similar to Kepler-16b late in the disc lifetime.
Interestingly, the simulations were able to produce Kepler-16b analogues in discs with all tested metallicities, however the more metal-rich discs formed decent analogues more frequently.

\begin{figure}
\centering
\includegraphics[scale=0.6]{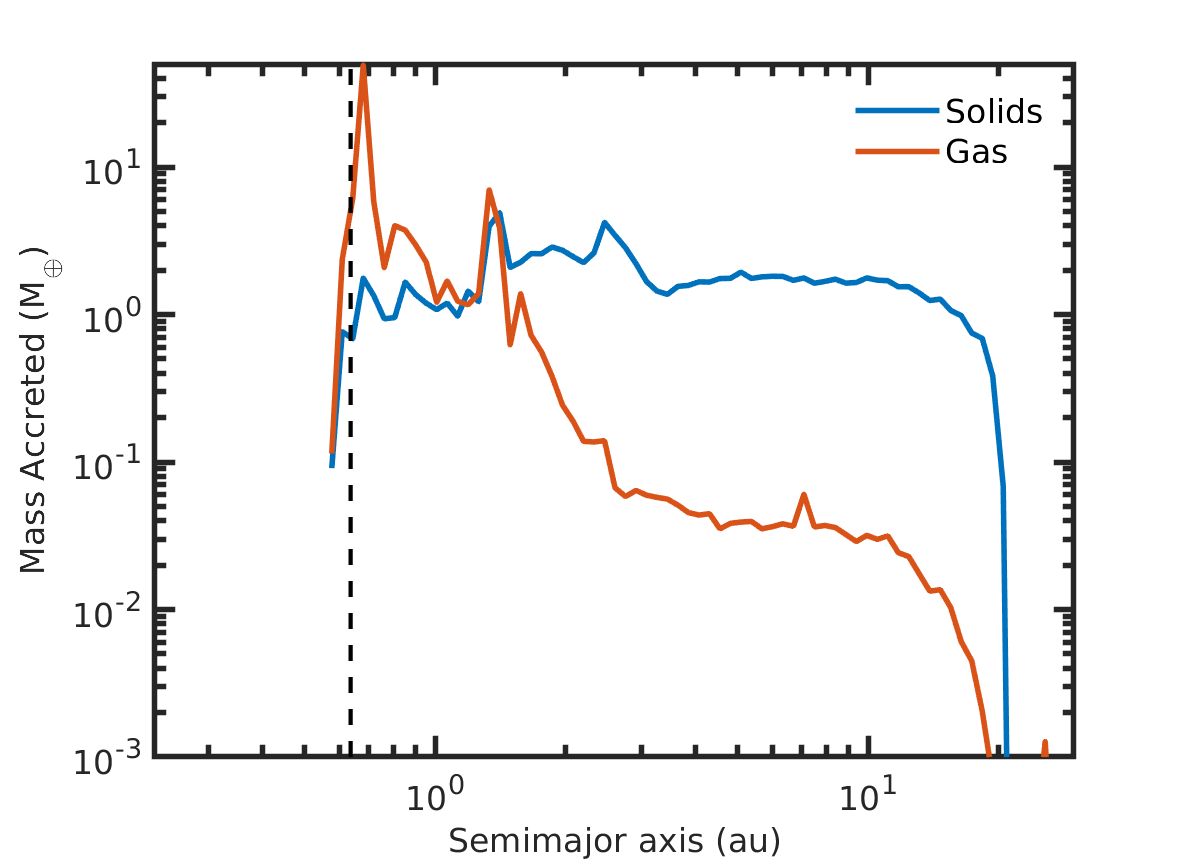}
\caption{Profiles showing the average radial distribution of mass accreted by planets througohut the circumbinary disc. Solid accretion is shown in blue whilst gas accretion is shown in red.}
\label{fig:kep16_accretion}
\end{figure}

\subsection{Where is mass accreted?}
Where planets accrete material can be important for the final composition of the planet \citep{Oberg11}.
Figure \ref{fig:kep16_accretion} shows the locations where mass is accreted in circumbinary discs averaged across all of the simulations.
The blue line shows the distribution for mass accreted in solids, the red line shows for gas, and the dashed line again shows the outer edge of the zone of dynamical instability.
Looking at the blue line, it is clear that solids are accreted throughout the discs where there are growing embryos.
This is especially true in the outer disc at distances $>3 \au$.
At around 2.5 $\au$, there is a spike in mass accreted, arising from an opacity transition acting to slow down migration for low mass planets (see the dark red region showing slow inward migration at this location in the left panel of fig. \ref{fig:both_discs_migr}).
An additional spike is also seen around the apocentre of the cavity region.
As discussed previously, the structure of the cavities act as efficient migration traps by driving enhanced positive corotation torques, that allows planets to congregate in these regions.
This allows planets more time to accrete pebbles at this location.
Collisions also occur here, due to the concentration of planets, again increasing accretion in this region.
Interestingly, the average amount of mass accreted drops interior to this location, since pebbles are unable to drift into this region as a result of trapping at the cavity outer edge.
However some planets are seen to grown in solid mass here, again a result of mutual collisions between planets.

Whilst the mass accreted in solids is spread across the majority of the circumbinary discs, the same cannot be said for gas.
With migration acting to bring giant planet cores close to the central stars, this gives such cores little time to accrete gas at large orbital distances.
The red line in fig.~\ref{fig:kep16_accretion} shows this, where there is little gas accreted in each simulation at distances larger than 3$\au$.
In fact, the majority of gas accreted by planets in the simulations is situated in and around the cavity region.
Spikes in the gas mass accreted can be seen at the cavity apocentre ($r_{\rm p}\sim1.4\au$), and close to the zone of dynamical instability.
These correspond to the locations where planet's become trapped as their migration stalls, allowing them to slowly accrete the surrounding gas and undergo runaway gas accretion.
With the majority of the accreted gas originating near the cavity, for planets around Kepler-16b, this corresponds thermally to temperatures just interior to the water iceline, assumed here to be where $T=150$~K.
This would result in the planets accreting gas with low C/O ratios since water would be in the gaseous phase, and accreted along with the gas, instead of as ices \citep{Oberg11}.
Should the atmospheres of observed circumbinary planets such as Kepler-16b be characterised \citep[see e.g.][ for a recent review]{Madhusudhan19}, it could in principle be determined whether or not the planets did indeed accrete their envelopes near the water iceline, and close to the cavity region.

\begin{figure}
\centering
\includegraphics[scale=0.6]{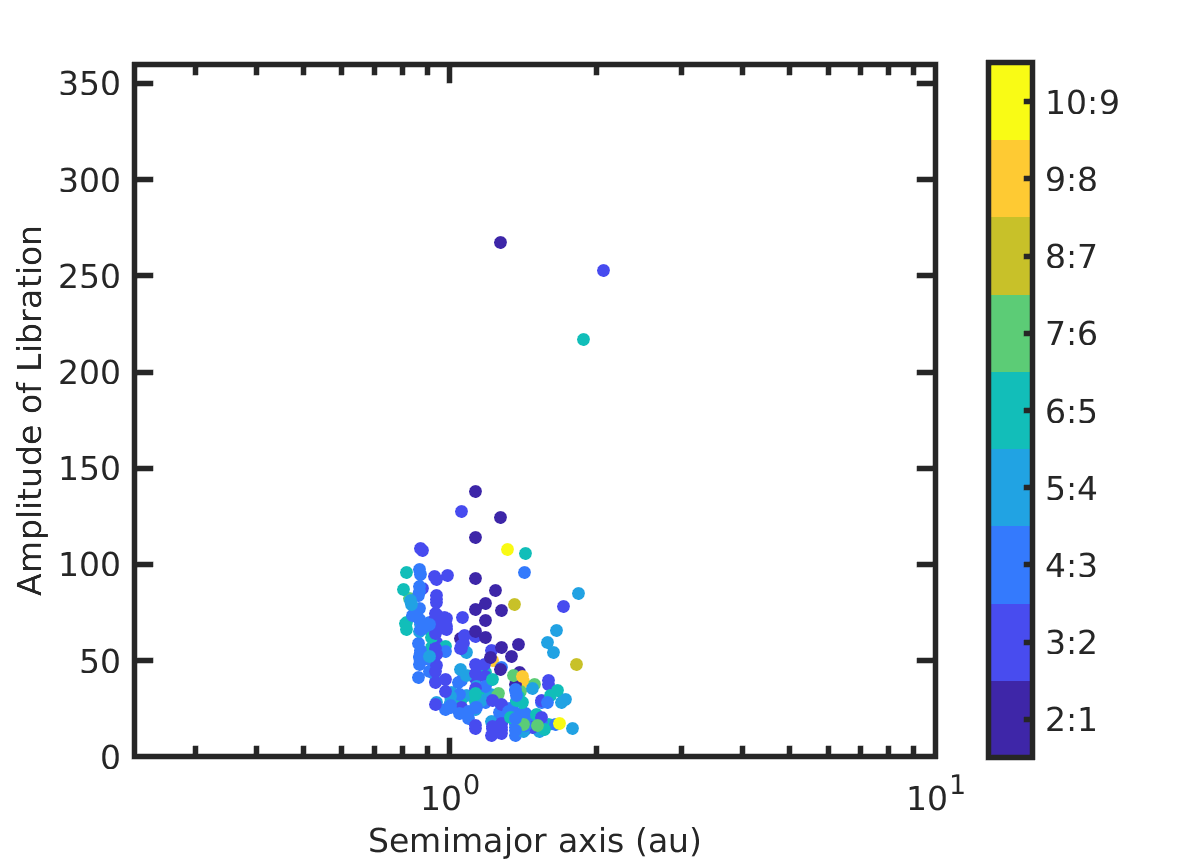}
\caption{Amplitudes of libration for resonant pairs of planets. Points are situated for the outer resonant planet. Colours show the degree of resonance, e.g. 2:1, 3:2, etc.}
\label{fig:kep16_res}
\end{figure}

\subsection{Systems in resonance}
\label{sec:kep16_res}
Whilst numerous single star systems containing multiple planets with integer period ratios have been observed (e.g. Kepler-223 \citep{Mills16}, Trappist-1 \citep{GillonTrappist17}, to name a couple), such systems have yet to be observed in circumbinary systems.
Resonant systems are also a natural outcome of planet formation scenarios \citep[see for example][]{ColemanNelson16,ColemanNelson16b,Coleman19} but how frequently they form around circumbinary stars has yet to be tested.
In sect. \ref{sec:kep16_sim} we described how resonant chains formed and migrated towards the central cavity, and we will now examine the prevalence of resonances throughout the entire population.

To account for N-body interactions after the end of the disc lifetime, we evolved all of the systems up to 100 Myr, and then checked over a 0.1 Myr time period whether the planets were in resonance.
We check all first-order resonances between 2:1 and 10:9, and define them as being in resonance if their libration amplitudes do not exceed 330 degrees.
Figure \ref{fig:kep16_res} shows the libration amplitudes for all planet pairs in resonance as a function of orbital distance, in which all planets have masses $m_{\rm p}>1\me$.
The colours denote the degree of resonance.
As can be seen a large number of planet pairs remain in resonance after 100 Myr, with $68\%$ of these being in either 2:1, 3:2 or 4:3.
The sculpting of the resonant population by the binary can also be seen in the left part of the distribution, where the planet pairs closer to the binary have larger amplitudes of libration on average, and the minimum libration amplitude increases the closer the planet pairs get to the binary.
There are also a number of planet pairs with larger amplitudes of libration that are close to falling out of resonance.

We now look at systems close to the central binary and of observable mass, and how these demographics change over time.
For this we limit our planet pairs to those with periods $P\leq 3$yrs, and masses $m_{\rm p}>1\me$.
After 10 Myr, 339 out of 480 systems contain planets within the observable region, with 133 being single planet systems and 206 multi-planet systems.
Of these 206 multi-planet systems, 141 contain resonant pairs of planets.
As the systems evolve up to 100 Myr, some of the resonant chains in the multiple planet systems become unstable leading to some collisions, but mainly ejections after planets interact with the central binary.
This reduces the number of systems with planets in the observable region to 305 out of 480, with single planet systems comprising 150 of these and 155 remaining as multiple planet systems, and only 123 systems with resonant pairs.

In terms of the total number of resonant pairs, interactions over the 100 Myr time period reduced the number of pairs from 303 to 257, such that roughly a sixth go unstable.
This is still a large number of resonant pairs compared to the number of systems, with a quarter of systems containing resonant pairs of planets.
Given that this outcome is similar to populations around single stars that over predict the number of resonances \citep[e.g.][albeit they only ran for 10 Myr]{ColemanNelson16}, this may imply that the number of resonant systems here are equally over predicted.
With the systems here evolved until 100 Myr, and with the extra perturbations arising from interactions with the binary stars, should the number of resonant systems be over predicted, this would imply that the resonant systems go unstable before the end of the disc lifetimes.
Such instabilities could arise out of 2D effects, such as stochastic migration \citep{Adams2008,ReinPapaloizou2009} or overstability in librations around resonant centres \citep{GoldreichSchlichting2014}.
Studies including two-dimensional effects will be investigated in future work.

\begin{figure}
\centering
\includegraphics[scale=0.6]{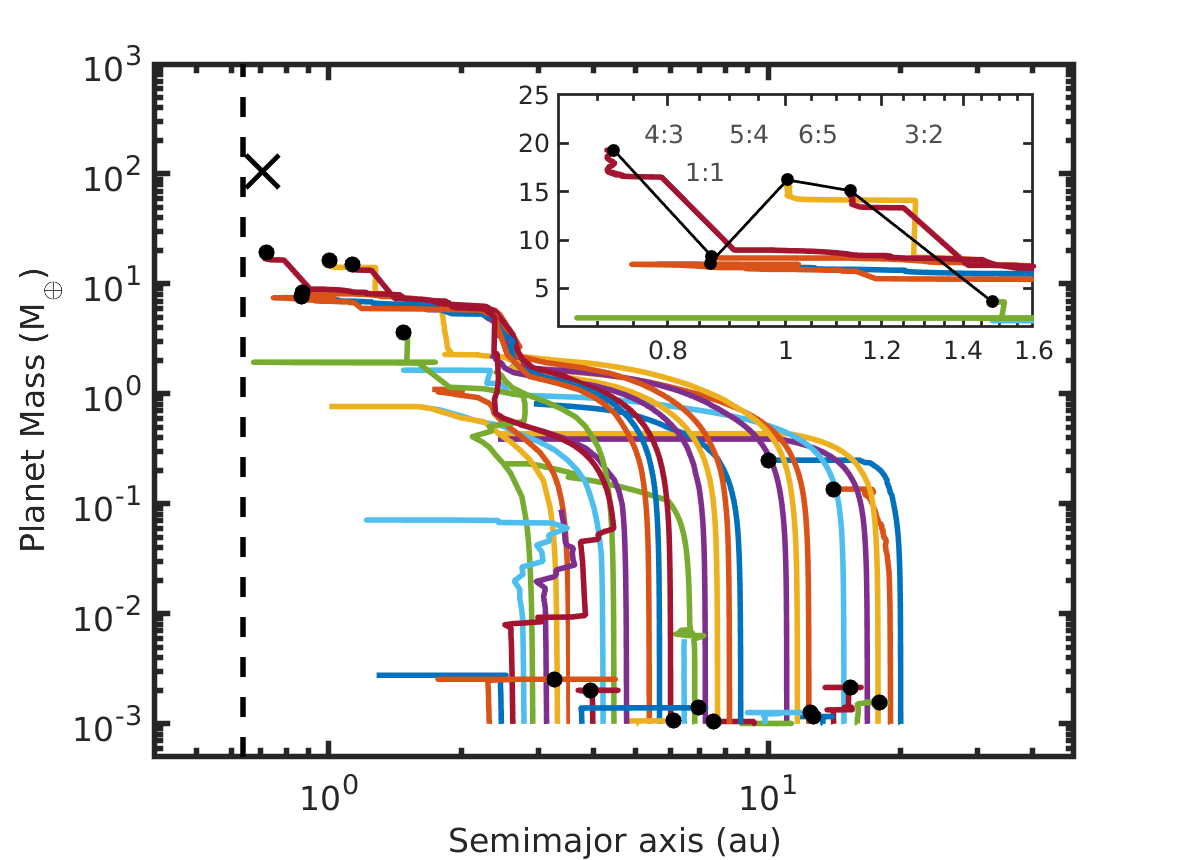}
\caption{Evolution of planet mass versus semimajor axis for an simulation resulting in resonant system. Filled black circles represent final masses and semimajor axes for surviving planets. The inset plot shows a zoom in of the inner system, with solid lines denoting resonant pairs of planets with the respective resonances denoted above the lines. The black cross shows the mass and semimajor axis of Kepler-16b, whilst the dashed vertical black line denotes the outer edge of the zone of dynamical instability.}
\label{fig:kep16_res_system}
\end{figure}

Figure \ref{fig:kep16_res_system} shows the mass versus semimajor axis evolution of an example resonant planetary system that forms in the simulations, and is stable for at least 100 Myr.
The system contains 6 planets with masses between 3--20 $\me$, which form further out in the disc and migrate in towards the central binary before getting trapped at the opacity transition and then at the cavity, with both traps allowing the planets to grow either through pebble accretion or collisions.
The inset of fig. \ref{fig:kep16_res_system} shows the final 6 planet system, along with the resonances between planet pairs.
Note that the second and third planet in the system form a coorbital pair, with libration angles indicating a trojan configuration.
A number of these coorbital planets appeared in other simulations as well, all in resonant chains, which have been found to aid in stabilising the coorbital resonance \citep{Leleu19}.
This system is representative of most resonant systems that formed in the simulations for Kepler-16.

\begin{figure}
\centering
\includegraphics[scale=0.6]{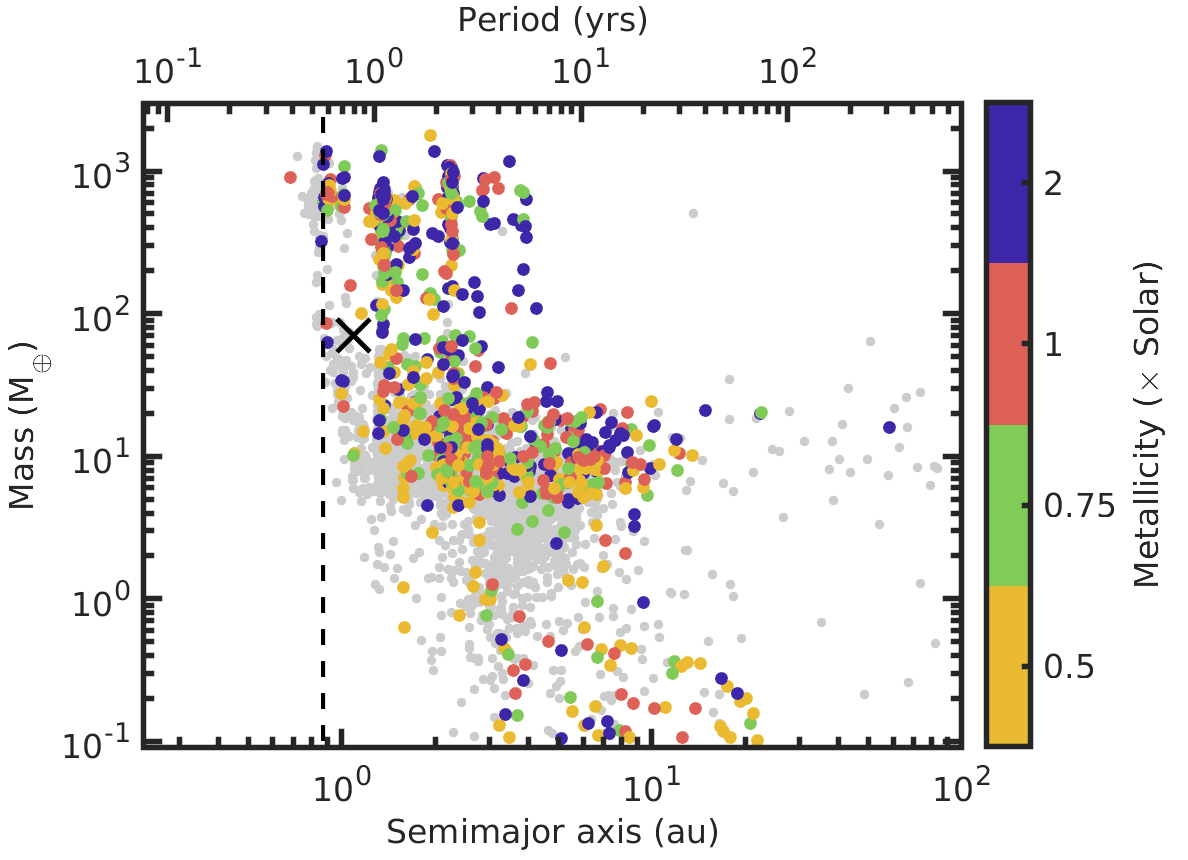}
\caption{Same as fig. \ref{fig:kep16_mva} but for Kepler-34, with the black cross representing Kepler-34b.}
\label{fig:kep34_mva}
\end{figure}

\section{Kepler 34}
\label{sec:kep34}

We now discuss the simulations for the Kepler-34 system.
Even though the mass of the central binary is over twice that of Kepler-16, the general formation processes are found to be qualitatively similar to that described in sect. \ref{sec:kep16_sim}.
As planetary embryos grow and accrete pebbles, they begin to migrate in towards the central binary, where they become trapped at either one of the opacity transitions (note from the right panel of fig. \ref{fig:both_discs_migr} the blue region denoting outwards migration at around 4--5 $\au$), or at the edge of the cavity carved by the binary.
Once at these locations, planets can then grow without undergoing much migration, which allows them to attain masses conducive to undergoing runaway gas accretion.
With the combined mass of the central binary $M_{\rm AB}\sim 2\msun$, the initial circumbinary discs were more massive, providing a greater abundance of pebbles, that allowed giant planet cores to more easily form than around Kepler-16.
Once the gas discs dispersed, the systems were again evolved for 100~Myr to determine final states of systems after undergoing N-body interactions in the absence of damping by a gaseous disc.

Figure~\ref{fig:kep34_mva} shows the mass versus semimajor axis for all planets in simulations around Kepler-34. The colours of the dots again denote the metallicity of the system, whilst the dashed line indicates the outer edge of the zone of dynamical instability, and the black cross shows the mass and semimajor axis for Kepler-34b.
Planets that were lost in the simulations, either through collisions or ejections, are shown by grey dots.
As can be seen, there are a large number of giant planets (those with $m_{\rm p}>100\me$) that have formed in the simulations.
These giant planets are mainly situated near the cavity at 1--2$\au$.
Some giant planets are orbiting further away, out at 5$\au$, though these planets are typically in multiple giant planet systems, with an inner giant orbiting near where the cavity would have been located.
The increase in the numbers of giant planets compared to Kepler-16 is not unsurprising given that the total mass of the binary system is more than double for our chosen parameters.
The scaling of disc mass with central binary mass allows many giant planet cores to form and undergo runaway gas accretion.

Whilst there are an abundance of giant planets forming around Kepler-34, fig. \ref{fig:kep34_mva} also shows a large number of super-Earth to Neptune mass planets (5--30$\me$) orbiting near the cavity, out to a semimajor axis $\sim 10 \au$.
With migration acting to bring planets in this mass range in towards the cavity region, the large population orbiting with semimajor axes less than 3$\au$ is unsurprising.
However, there is then a question of how so many super-Earths and Neptune mass planets are orbiting with semimajor axes out to 10$\au$, since migration should bring them to the cavity region.
Looking at the other planets in these systems reveals that there are actually two populations of super-Earth and Neptune mass planets.
The majority of the planets in the first population that are orbiting near the cavity are doing so in systems that do not contain giant planets, either due to giants not forming in those discs, or because the giant planets have been ejected from the system.
For the super-Earths and Neptune mass planets orbiting further away from the central stars, the second population, these systems contain at least one interior giant planet. The presence of inner giant planets can have multiple effects that explain the exterior population of planets. These include: scattering of the planets to larger semimajor axes, from where they can migrate back in whilst damping their eccentricities; forming chains of planets that dynamically act against the migration torques, stalling their migration; and forming gaps that act as migration traps.
Observing such low mass planets in these regions would therefore indicate that there should either be giant planets also orbiting in the system, or that giant planets did form in the system before later being lost, most likely through ejections.

\begin{figure}
\centering
\includegraphics[scale=0.6]{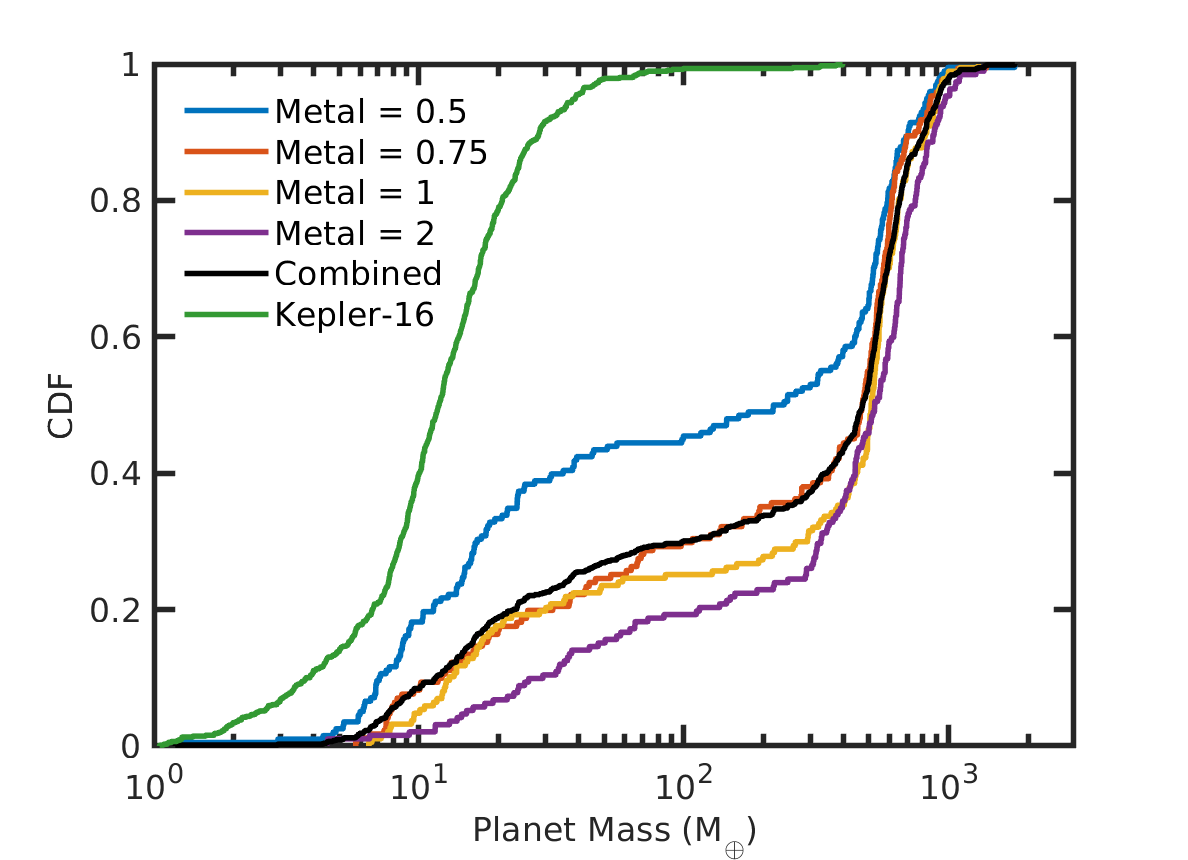}
\caption{Cumulative distribution functions of planet masses for planets with periods less than 3 years and masses greater than 1$\me$. The black line shows the combined distribution, whilst the coloured lines show the distributions for discs with initial metallicites equal to: 0.5$\times$ Solar (blue), 0.75$\times$ Solar (red), 1$\times$ Solar (yellow) and 2$\times$ Solar (purple). The green lines shows the combined mass distribution from the Kepler-16 simulations.}
\label{fig:kep34_mass_cdf}
\end{figure}

Having discussed the populations of giant planets and super-Earths/Neptunes, it is interesting to look at the distributions of planets that form around Kepler-34.
Figure~\ref{fig:kep34_mass_cdf} shows cumulative distribution functions for planet mass for all planets with periods $P<3$ yrs, and masses $m_{\rm p}>1\me$.
The black line shows the distribution for all simulations, whilst the coloured lines show the distributions for discs with different metallicities.
The green line shows the same combined distribution, but for the Kepler-16 simulations.
From the black line, the fraction of giant planets that orbit near the cavity is clear, with 70\% of planets with periods less than 3 years having masses greater than 100 $\me$.
The sharp gradient shows that a large fraction of the planets have masses between 1.5--3 $M_{\rm J}$.
The effect of higher metallicities can also easily be seen in fig. \ref{fig:kep34_mass_cdf} where giant planets comprised of 55\% of the planets in discs with metallicities of 0.5$\times$ Solar, compared to 80\% for discs with twice Solar metallicity.
This is not unexpected as the higher metallicities allow planets to accrete more mass in pebbles, resulting in greater numbers of giant planet cores that can undergo runaway gas accretion.
For the metal-rich discs, there are also very few planets with masses below 10$\me$.
Interestingly, there are very few planets with masses less than 4$\me$ across discs with all metallicities.
This is due to the efficiency of pebble accretion for Earth mass planets.

In comparing the combined distribution (black line) for Kepler-34 to that for Kepler-16 (green line), the populations of planets are clearly different.
The distribution for Kepler-16 is pushed to much lower masses, with only $\sim$1\% of the planets being giant planets, compared to 70\% for Kepler-34.
This is down to the increase in mass of the Kepler-34 binary, and thus the initial mass in solids within the circumbinary discs.
Interestingly, for both distributions, planets between 10--30$\me$ are relatively common, more so around Kepler-16 where they could not accrete sufficient amounts of solids to allow their cores to undergo runaway gas accretion, but also for Kepler-34 where they are the largest population of planets after the giant planet population.
It would therefore be expected that planets of this mass range, and with periods less than 3 years could be common around binary stars, and could possibly be observed with future missions such as PLATO.

Turning now to the issue of how well Kepler-34b is emulated in the population, fig.~\ref{fig:kep34_mva} shows that there are a few planets with masses similar to that inferred from the observations.
Generally though these planets are orbiting slightly further from the central stars, typically where the apocentre of the cavity was situated $\sim 2.5 \au$.
These planets stalled their migration there as giant planet cores, but were unable to undergo runaway gas accretion before the end of the disc lifetime, leaving them with masses similar to Kepler-34b.
In regards to also matching the period of Kepler-34b, those planets that were most similar also contained giant planets in their systems, which acted to push the Kepler-34b analogues through the trap at the apocentre of the cavity to their final locations.
This presents a problem for reproducing the observations, since to date there is no giant planet observed orbiting further out in the Kepler-34 system.
One effect that is not included in these simulations is partial gap opening which could allow the planet to affect the cavity structure, circularising it, and allowing the planets to migrate closer to the observed period of Kepler-34b \citep{Penzlin21}.

\section{Discussion}
\label{sec:discussions}

Sections \ref{sec:kep16} and \ref{sec:kep34} detailed example simulations and the resultant populations around Kepler-16 and Kepler-34 respectively.
We now simultaneously examine both populations, looking at possible observable characteristics.

\subsection{Ejections}
\label{sec:eject}
Around single stars, the majority of planets lost in N-body simulations is either through mutual collisions with other planets or collisions with the central star itself.
This is because in such populations, planet masses are not large enough to increase another planet's velocity to greater than the escape speed from the system during close encounters, and as such collisions normally occur between planets instead of ejection.
With Kepler-16 and Kepler-34 being binary systems, this is no longer the case, since the main perturbers in the system are the binary stars themselves.
Once planets become slightly eccentric near the cavity, they then interact with one or both of the central binary stars, increasing their velocity.
Eventually their velocity becomes larger than the escape velocity of the system, and they are ejected.
In total, of the planets lost around Kepler-16 and Kepler-34, ejections from the systems accounted for 40\% and 56\%, respectively.
It is understandable that there were more ejections in the Kepler-34 simulations, as more massive planets were able to form, increasing the relative velocities that planets would attain before interacting with the central binaries, that are also more massive and on more eccentric orbits.

Whilst it is interesting to know the percentage of planets ejected in the systems, understanding the mass distribution of ejected planets could be tested by future observations.
Assuming that the ejected planets do not collide with other objects or become captured by other stars, such free-floating objects can be detected through microlensing surveys (e.g. Nancy Grace Roman Telescope \citep{Spergel15,Bennett18} or the Large Synoptic Survey Telescope \citep{LSST_2019}).
Looking at the number of planets ejected from the simulations, we find that Kepler-16 like systems ejected 6.3 planets on average, whilst Kepler-34 like systems ejected 9.3 planets per system.
This increase in the number of planets is due to more giant planets forming in discs around Kepler-34 increasing the strength of the perturbations felt by other smaller planets.
In terms of the number of giant planets ejected, systems around Kepler-16 ejected 0.26 giant planets on average whilst those around Kepler-34 ejected 0.33 giants per system.
This shows that typically for binaries similar to Kepler-16 and Kepler-34, in terms of combined central mass and binary separation or eccentricity, 1 in 3 to 1 in 4 systems will eject a giant planet.
Naturally this work has not investigated populations of binary systems, but rather two specific examples, and as such we are unable to accurately comment on the total number of free-floating giant planets that originated in circumbinary discs.

\begin{figure}
\centering
\includegraphics[scale=0.6]{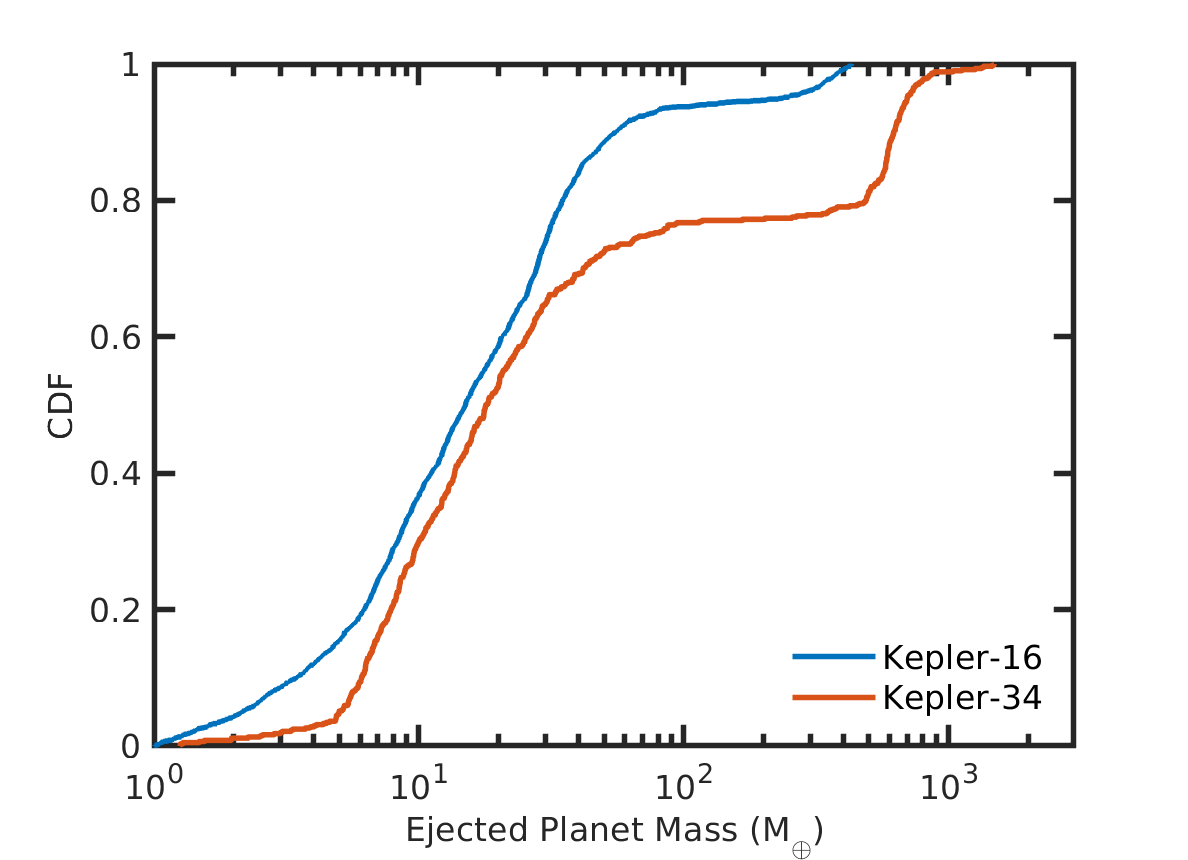}
\caption{Cumulative distribution functions of planet masses for planets that were ejected in the Kepler-16 (blue line) and Kepler-34 (red line) simulations. We only show planets with masses $m_{\rm p}\ge1\me$.}
\label{fig:ejected_cdf}
\end{figure}

Whilst the number of planets and giant planets ejected can yield insights into the frequency of free floating planets, the mass distribution of ejected planets is also relevant.
Figure \ref{fig:ejected_cdf} shows the cumulative distribution function for all ejected planets above $1\me$ for Kepler-16 (blue line) and Kepler-34 (red line).
Interestingly it is clear that a greater fraction of planets ejected around Kepler-34 are giant planets, approximately 23\%, compared to only 6\% from Kepler-16.
Note that the ratio between these percentages is different to the ratio of giant planets ejected from each system, since ejections from systems around Kepler-34 involved a large number of planets with sub-terrestrial masses.
This was due to more massive planets forming earlier in those discs, exciting the eccentricities of neighbouring planets that resulted in them accreting fewer pebbles due to larger relative velocities between the planets and pebbles.
Whilst the fraction of giant planets ejected around Kepler-34 were higher, it is interesting to see that fig. \ref{fig:ejected_cdf} shows that super-earths and Neptune mass planets are most commonly ejected, that is of all planets with $m_{\rm p}\ge 1\me$.
For both Kepler-16 and Kepler-34, $\sim60\%$ of planets with masses greater than 1$\me$ were in the mass range 5--30$\me$, showing that for detectable free floating planets, super-Earths and Neptune mass planets are the most common.

\begin{figure}
\centering
\includegraphics[scale=0.6]{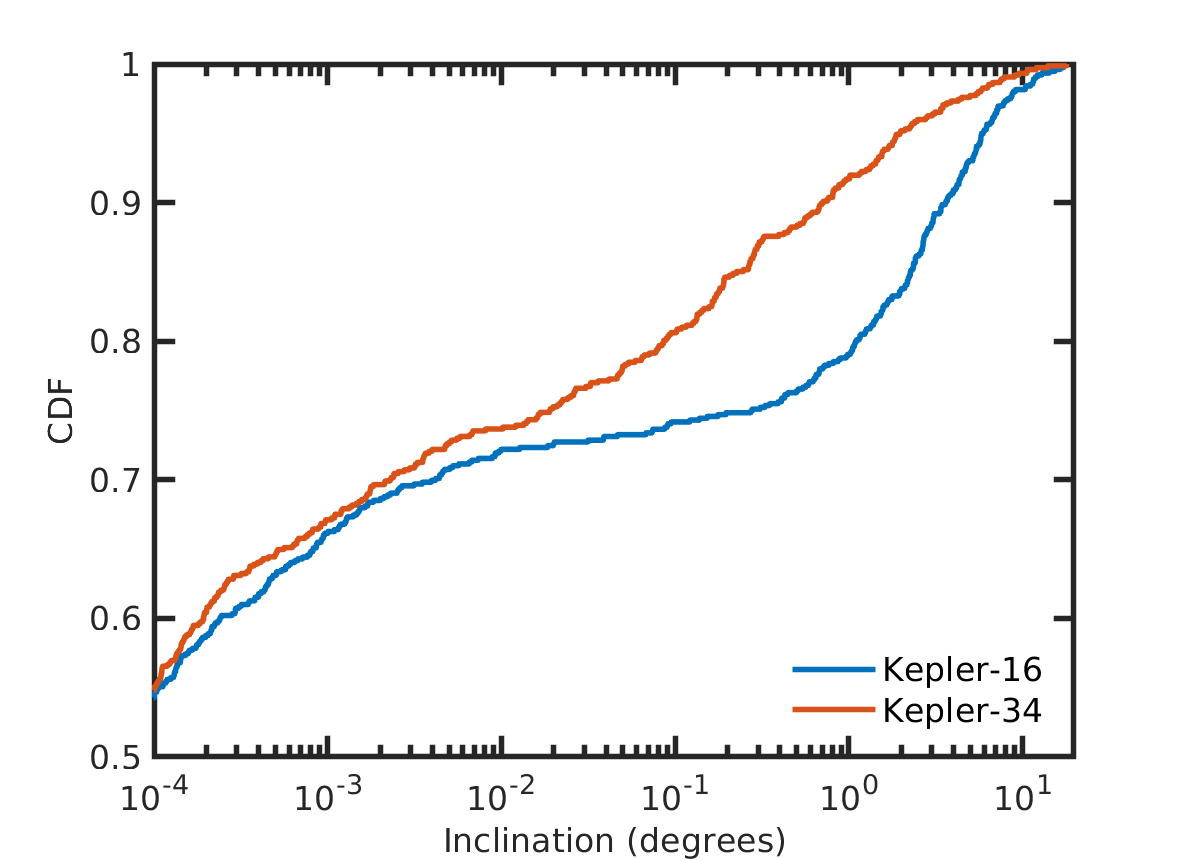}
\caption{Cumulative distribution functions of planet inclinations for surviving planets with periods less than 3 years and masses greater than 1$\me$ for Kepler-16 (blue line) and Kepler-34 (red line).}
\label{fig:inclination_cdf}
\end{figure}

\subsection{Transiting Probabilities}
\label{sec:transit}

As Kepler-16b and Kepler-34b were both found with transit surveys \citep{Doyle11,Welsh12}, and with transit surveys discovering the majority of currently known circumbinary planets, it is useful to judge the transit observability for planets formed in our simulations.
Since the transit probability depends on the mutual inclination of the planets compared to the binary stars, fig.~\ref{fig:inclination_cdf} shows the cumulative distribution function of planet inclinations for planets with periods less than 3 years and masses greater than 1$\me$.
Interestingly, the coplanarity of the formed planets is significant with 55\% of all selected planets in both the Kepler-16 and Kepler-34 models having inclinations less than $10^{-4}$ degrees.
This is due to the damping from the gas disc acting on the planets, and with few vertical perturbations due to there only being a small number of planets, their inclinations remained low.
Such perturbations could arise through interactions with nearby planets and planetesimals on inclined orbits or through density perturbations in the vertical and azimuthal plane in a 3D gas disc.

One main difference between the inclination distributions between Kepler-16 and Kepler-34, is that Kepler-16 appears to have two populations of planets based on their inclinations.
This can be seen by the blue line in fig. \ref{fig:inclination_cdf} for Kepler-16 where few planets have inclinations between $10^{-2}$ and 1 degree (7\%), but 21\% of planets have inclinations greater than 1 degree.
The main reason for these populations arising in the Kepler-16 systems is that in the systems where mass growth was limited, i.e. only super-Earths were able to form, planet inclinations were able to remain high since inclination damping due to the disc was not able to overcome excitation through mutual planet interactions.
This allowed planet inclinations to remain non-negligible.
For systems where planets were able to reach masses similar to Neptune, damping from the circumbinary discs reduced planet inclinations to less than $10^{-2}$, of which it was then difficult to excite them back to meaningful values.

For planets in the Kepler-34 systems on the other hand, there appears to only be a continuous distribution.
This distribution arises, because in many systems around Kepler-34, many planets were able to reach larger masses and reduce their inclinations.
However, as many giant planets formed in the Kepler-34 models, they were able to significantly perturb other planets in the system, resulting in those planets attaining some inclination.

Assuming that a potential observer is located in the plane of the binary system, circumbinary planets will always transit their parent stars when their maximum height above the plane of the binary stars is less than the radius of the stars themselves, which we take to be equal to the minimum of the two stars present day values ($0.22 \rsun$ for Kepler-16  and $1.09 \rsun$ for Kepler-34).
When looking at the fraction of planets that are always transiting, we find that for Kepler-16 like systems, 73\% of planets with periods less than 3 years and masses greater than 1$\me$ have low inclinations allowing them to always transit.
This fraction of transiting planets increases to 83.6\% for planets that formed around Kepler-34.
The difference in the fraction of transiting planets is a result of a number of effects.
Planets around Kepler-34 are more massive than those around Kepler-16, allowing them to more effectively damp their inclinations closer to the midplane of the disc, and thus the binary orbital plane.
Other planets in the systems also excite planetary inclinations through mutual interactions, and when examining the multiplicity of planets around Kepler-16 and Kepler-34, we find that systems around Kepler-16 are more plentiful.
On average systems around Kepler-16 contain 2 planets with periods less than 3 years, whilst for Kepler-34, there are only 1.77 planets on average.
Systems around Kepler-16 are also more likely to host larger numbers of planets, with 13\% of systems containing 4 or more planets, compared to 0.5\% for Kepler-34.
This is a result of the the cavity being more extended around Kepler-34, as well as the more massive planets decreasing the stability of large chains of planets.

\subsection{Other Circumbinary Systems}
Whilst this work focuses on the Kepler-16 and Kepler-34, a number of other discovered circumbinary planetary systems have binary parameters similar to those explored here, and it is therefore possible that the planets that form around those systems should be similar to the results of our simulations.

\subsubsection{Kepler-47}

The Kepler-47 system was the first multi-planet circumbinary system to be found, with three planets of masses between 2--19$\me$, orbiting binary stars with properties similar to Kepler-16 \citep{Orosz12_k47,Orosz19}.
As discussed in sect. \ref{sec:kep16}, super-Earth and Neptune mass planets were the most common to form around Kepler-16, since they were not massive enough to undergo runaway gas accretion.
They also typically formed in resonant chains that migrated in to the cavity region, before some underwent dynamical instabilities.
Given that current estimates of the planetary compositions indicate they are icy bodies \citep{Orosz19}, this lends weight to the scenario of formation and migration from outside of the water snowline, as seen in the multi-planet systems discussed in sect. \ref{sec:kep16}.
The planetary system around Kepler-47 fits in well with the results of the Kepler-16 simulations presented here, and as such is a good example of those types of systems that are seen to form.

\subsubsection{Kepler-1647}

The more recent discovery of Kepler-1647b, a $\sim1.5 M_{\rm J}$ planet orbiting a 1.2 and 0.97 $\msun$ binary pair, far from where the cavity would have been\citep{Kostov16}, raises questions as to how it formed and why it is not orbiting near the zone of dynamical instability, similar to previously discovered circumbinary planets.
With the planet also currently being the most massive of the known circumbinary planets, this also raises questions as to how it formed and survived in the system, since previous hydrodynamical simulations of giant planets showed that they should be migrate towards the central cavity, where interactions with the binary stars could lead to their ejection \citep{Pierens08b}.
Given that the combined central mass of the system is similar to that for Kepler-34, we compare the planet to our results in sect. \ref{sec:kep34}, finding that it's observed semi-major axis and mass is compatible with the more distant giant planets formed in those simulations.
As can be seen, those planets are not found at the location of the cavity.
This arises due to their formation pathways, where they either formed in multiple giant planet systems and so torques from more interior giant planets acted to prevent them from migrating closer to the cavity, or the planets did migrate near to the cavity, but interactions with other planets scattered them outwards where they could then damp their eccentricities through interactions with the surrounding gas disc.
In both of these scenarios, scattering events could have ejected the other giant planets, leaving only a single giant planet on a longer period orbit, similar to Kepler-1647b.

\subsubsection{TOI-1338/BEBOP-1}

Whilst most circumbinary planets have been discovered by \textit{Kepler}, recently \textit{TESS} has begun finding transiting circumbinary planets.
TOI-1338/BEBOP-1 is the first such system found with \textit{TESS} where a $7R_{\oplus}$ planet is orbiting near the zone of dynamical instability in a binary system with a similar mass ratio and binary eccentricity to Kepler-16, although the combined binary mass is $\sim$50\% larger than Kepler-16 \citep{Kostov20}.
Radial velocity observations from the BEBOP program \citep{Martin19} have recently discovered an additional $\sim 65 \me$ planet on a longer period orbit than the transiting planet \citep{Standing23}.
The transiting planet seen by \textit{TESS} was not seen in the radial velocity data yielding an upper mass limit of $22\me$.
Given the similarity in the binary properties between Kepler-16 and TOI-1338/BEBOP-1c, we can compare the population to the observed planets, finding that the inner planet is well placed amongst the population of super-Earth--Neptune mass planets orbiting just exterior to the zone of dynamical instability, with the outer planet being situated near the edge of surviving planets.
Typically such planets that formed were part of a multiple planet system, indicating that there could be additional planets in the system, as has been found by \citet{Standing23}.

\section{Conclusions}
\label{sec:conclusions}

In this work, we have explored the formation of circumbinary planets and planetary systems around systems akin to Kepler-16 and Kepler-34.
We used an updated version of N-body code \textsc{mercury6} including the effects of a central binary, and coupled to this a self-consistent 1D viscously evolving disc model containing prescriptions for planet migration, accretion of gaseous envelopes, pebble accretion and disc removal through photoevaporative winds.
To account for the eccentric precessing cavity carved by the binary stars, we adjusted the viscous $\alpha$ parameter to evacuate the inner disc of gas similar to that seen in 2D hydrodynamical models.
We also included prescriptions for the gravitational torque exerted by the precessing, eccentric circumbinary disc, and the role of non-Keplerian gas/dust velocities on pebble accretion rates.
Whilst we mainly focused on the Kepler-16 and Kepler-34 systems, we also compared our results to other observed circumbinary systems.
The main results from our study can be summarised as follows.

(1) Planets are found to accrete pebbles mainly outside of the water snowline, far from the cavity.
This allows them to significantly increase their mass up to super-Earth--Neptune masses, where they begin to migrate in resonant chains towards the binary stars, where they eventually become trapped at the outer edge of the eccentric cavity, carved by the binaries.
Whilst trapped around the cavity, the more massive planets are able to accrete sufficient quantities of gas, and undergo runaway gas accretion forming giant planets.
Some of the resonant chains go unstable leading to collisions, and quite often, ejections from the system.

(2) For Kepler-16, the initial solid mass is insufficient to consistently grow cores massive enough to undergo runaway gas accretion.
Only the more metal rich discs were able to form giant planets around the cavity region.
Typically, most systems forming around Kepler-16 contained multiple super-Earth--Neptune mass planets, orbiting near and exterior to the cavity.
In comparing Kepler-16b to the model results, we find that some systems were able to form similar planets, where a resonant chain went unstable leaving a more massive core that was able to effectively accrete gas, allowing it to reach a final mass similar to Kepler-16b once the disc fully dispersed.

(3) With Kepler-34 being a much more massive central binary than Kepler-16, this allows more planets to attain core masses amenable to undergoing runaway gas accretion, resulting in a larger number of giant planets forming around the cavity region.
These giant planets are then able to survive until after the disc is fully dispersed by settling into stable orbits, either in multiple planet systems or singularly around the central binary.
The effects of increasing metallicity is also evident, as more giant planets form in systems with super-Solar metallicity compared to sub-Solar.
Examining the planets in the inner region of the disc, with periods of up to 3 years, the fraction of giant planets also increases with metallicity, rising from 55\% to 80\% when increasing the metallicity from 0.5 to 2 $\times$ Solar.
In regards to Kepler-34b, few systems contained planets similar to that observed.
This was due to the mass of Kepler-34b, $\sim60\me$, being amidst the mass range that runaway gas accretion tends to quickly bypass, forming giant planets instead, as seen in fig. \ref{fig:kep34_mva}.
Of the planets that could match the mass of Kepler-34b, they typically orbit with slightly longer periods, and other mechanisms, such as partial gap opening, may be required for the models to better match both the observed mass and period of Kepler-34b.

(4) An interesting outcome of the simulations is the number of resonant systems that form, especially around Kepler-16 analogues, where fewer giant planets were able to form, resulting in many systems containing multiple super-Earth--Neptune like planets.
Whilst no resonant circumbinary systems have been found to date, the large number that formed and survived for 100~Myr in the simulations indicates that they could be observed by future missions.
The main hurdle to observing such planets would be the long orbital periods of planets in such long chains, as seen in fig. \ref{fig:kep16_res_system}, which can be much longer than typical discovery missions.

(5) With the central systems being two binary stars, interactions with either of the stars can easily lead to ejection from the circumbinary systems.
Our suite of simulations show that a large number of planets are ejected from both the Kepler-16 and Kepler-34 analogues, including both terrestrial planets and gas giants.
For Kepler-16, each system on average ejected 6.3 planets, with 0.26 of these being giant planets.
These values changed to 9.3 and 0.33 per system respectively for Kepler-34.
The simulations therefore show that around 1-in-3 to 1-in-4 binary systems eject giant planets, yielding a large number of giant free floating planets that can be detected by microlensing surveys \citep{Spergel15,LSST_2019}.

(6) Whilst this work mainly focused on the Kepler-16 and Kepler-34 systems, other circumbinary planets orbit stars with similar binary properties.
In comparing those planets to the simulation results, it is clear that the models can adequately reproduce the currently observed circumbinary planets, including multiple planetary systems such as Kepler-47.
The models can also place predictions on the presence of other planets in the systems, for example the TOI-1338/BEBOP-1 system, as well as the formation history of others, such as the possibility of giant planets being ejected from the Kepler-1647 system.

The simulations we have presented here show that circumbinary planets akin to Kepler-16b and Kepler-34b can easily form through the pebble accretion scenario in circumbinary discs.
However it has yet to be investigated whether other flavours of planet formation, e.g. via planetesimal accretion, or a hybrid
pebble/planetesimal scenario, yield similar results.
Whilst such a question is beyond the scope of this study, other works have compared such scenarios around single stars.
\citet{Coleman19} compared the two scenarios for systems forming around low mass stars similar to Trappist-1, and found that both scenarios consistently formed planetary systems similar to those observed.
More recently, \citet{Brugger20} compared pebble accretion to planetesimal accretion scenarios for single planet populations around single solar-mass stars, and found that both scenarios could form a wide diversity of planets from a wide range of initial parameters. However the time that such planets formed and the initial disc parameters they formed from differed across the two scenarios, which would have implications on the properties of those planets, i.e. composition.

In future work, we will compute a planetesimal accretion scenario population for circumbinary planets, and compare to the results from the pebble accretion scenario presented here.
If the results of the two scenarios are materially different, then comparing them to observations could yield insights into the formation of circumbinary planets, and into which accretion scenario may be dominant for planet formation around circumbinary systems and around single stars.
Comparing simulated circumbinary populations to single star populations could also give hints into planet formation processes as observable properties that are similar or different can be tested with observations of circumbinary planets, as well as the much larger exoplanet population around single stars.

\section*{Acknowledgements}
We thank the anonymous referee for useful comments on the paper.
We thank John Chambers for providing an updated version of {\sc{mercury6}} including the integrators for circumbinary systems.
GALC was funded by the Leverhulme Trust through grant RPG-2018-418. RPN acknowledges support from STFC through grants ST/P000592/1 and ST/T000341/1.
This research utilised Queen Mary's Apocrita HPC facility, supported by QMUL Research-IT (http://doi.org/10.5281/zenodo.438045).
This work was performed using the DiRAC Data Intensive service at Leicester, operated by the University of Leicester IT Services, which forms part of the STFC DiRAC HPC Facility (www.dirac.ac.uk). The equipment was funded by BEIS capital funding via STFC capital grants ST/K000373/1 and ST/R002363/1 and STFC DiRAC Operations grant ST/R001014/1. DiRAC is part of the National e-Infrastructure.
This research received funding from the European Research Council (ERC) under the European Union's Horizon 2020 research and innovation programme (grant agreement n$^\circ$ 803193/BEBOP)

\section*{Data Availability}
The data underlying this article will be shared on reasonable request to the corresponding author.

\bibliographystyle{mnras}
\bibliography{references}{}

\begin{thebibliography}{}
\makeatletter
\relax
\def\mn@urlcharsother{\let\do\@makeother \do\$\do\&\do\#\do\^\do\_\do\%\do\~}
\def\mn@doi{\begingroup\mn@urlcharsother \@ifnextchar [ {\mn@doi@}
  {\mn@doi@[]}}
\def\mn@doi@[#1]#2{\def\@tempa{#1}\ifx\@tempa\@empty \href
  {http://dx.doi.org/#2} {doi:#2}\else \href {http://dx.doi.org/#2} {#1}\fi
  \endgroup}
\def\mn@eprint#1#2{\mn@eprint@#1:#2::\@nil}
\def\mn@eprint@arXiv#1{\href {http://arxiv.org/abs/#1} {{\tt arXiv:#1}}}
\def\mn@eprint@dblp#1{\href {http://dblp.uni-trier.de/rec/bibtex/#1.xml}
  {dblp:#1}}
\def\mn@eprint@#1:#2:#3:#4\@nil{\def\@tempa {#1}\def\@tempb {#2}\def\@tempc
  {#3}\ifx \@tempc \@empty \let \@tempc \@tempb \let \@tempb \@tempa \fi \ifx
  \@tempb \@empty \def\@tempb {arXiv}\fi \@ifundefined
  {mn@eprint@\@tempb}{\@tempb:\@tempc}{\expandafter \expandafter \csname
  mn@eprint@\@tempb\endcsname \expandafter{\@tempc}}}

\bibitem[\protect\citeauthoryear{{Abod}, {Simon}, {Li}, {Armitage}, {Youdin}
  \& {Kretke}}{{Abod} et~al.}{2019}]{Abod19}
{Abod} C.~P.,  {Simon} J.~B.,  {Li} R.,  {Armitage} P.~J.,  {Youdin} A.~N.,
  {Kretke} K.~A.,  2019, \mn@doi [\apj] {10.3847/1538-4357/ab40a3}, \href
  {https://ui.adsabs.harvard.edu/abs/2019ApJ...883..192A} {883, 192}

\bibitem[\protect\citeauthoryear{{Adams}, {Laughlin}  \& {Bloch}}{{Adams}
  et~al.}{2008}]{Adams2008}
{Adams} F.~C.,  {Laughlin} G.,   {Bloch} A.~M.,  2008, \mn@doi [\apj]
  {10.1086/589986}, \href {http://adsabs.harvard.edu/abs/2008ApJ...683.1117A}
  {683, 1117}

\bibitem[\protect\citeauthoryear{{Alexander}}{{Alexander}}{2012}]{Alexander12}
{Alexander} R.,  2012, \mn@doi [\apjl] {10.1088/2041-8205/757/2/L29}, \href
  {https://ui.adsabs.harvard.edu/abs/2012ApJ...757L..29A} {757, L29}

\bibitem[\protect\citeauthoryear{{Alexander} \& {Armitage}}{{Alexander} \&
  {Armitage}}{2007}]{Alexander07}
{Alexander} R.~D.,  {Armitage} P.~J.,  2007, \mn@doi [\mnras]
  {10.1111/j.1365-2966.2006.11341.x}, \href
  {http://adsabs.harvard.edu/abs/2007MNRAS.375..500A} {375, 500}

\bibitem[\protect\citeauthoryear{{Alexander} \& {Armitage}}{{Alexander} \&
  {Armitage}}{2009}]{Alexander09}
{Alexander} R.~D.,  {Armitage} P.~J.,  2009, \mn@doi [\apj]
  {10.1088/0004-637X/704/2/989}, \href
  {http://adsabs.harvard.edu/abs/2009ApJ...704..989A} {704, 989}

\bibitem[\protect\citeauthoryear{{Alexander} \& {Pascucci}}{{Alexander} \&
  {Pascucci}}{2012}]{AlexanderPascucci12}
{Alexander} R.~D.,  {Pascucci} I.,  2012, \mn@doi [\mnras]
  {10.1111/j.1745-3933.2012.01243.x}, \href
  {http://adsabs.harvard.edu/abs/2012MNRAS.422L..82A} {422, 82}

\bibitem[\protect\citeauthoryear{{Artymowicz} \& {Lubow}}{{Artymowicz} \&
  {Lubow}}{1994}]{Artymowicz94}
{Artymowicz} P.,  {Lubow} S.~H.,  1994, \mn@doi [\apj] {10.1086/173679}, \href
  {https://ui.adsabs.harvard.edu/abs/1994ApJ...421..651A} {421, 651}

\bibitem[\protect\citeauthoryear{{Ataiee}, {Baruteau}, {Alibert}  \&
  {Benz}}{{Ataiee} et~al.}{2018}]{Ataiee18}
{Ataiee} S.,  {Baruteau} C.,  {Alibert} Y.,   {Benz} W.,  2018, \mn@doi [\aap]
  {10.1051/0004-6361/201732026}, \href
  {https://ui.adsabs.harvard.edu/abs/2018A&A...615A.110A} {615, A110}

\bibitem[\protect\citeauthoryear{{Bell} \& {Lin}}{{Bell} \&
  {Lin}}{1994}]{Bell94}
{Bell} K.~R.,  {Lin} D.~N.~C.,  1994, \mn@doi [\apj] {10.1086/174206}, \href
  {http://adsabs.harvard.edu/abs/1994ApJ...427..987B} {427, 987}

\bibitem[\protect\citeauthoryear{{Bell}, {Cassen}, {Klahr}  \&
  {Henning}}{{Bell} et~al.}{1997}]{Bell97}
{Bell} K.~R.,  {Cassen} P.~M.,  {Klahr} H.~H.,   {Henning} T.,  1997, \apj,
  \href {http://adsabs.harvard.edu/abs/1997ApJ...486..372B} {486, 372}

\bibitem[\protect\citeauthoryear{{Ben{\'\i}tez-Llambay} \&
  {Masset}}{{Ben{\'\i}tez-Llambay} \& {Masset}}{2016}]{FARGO-3D-2016}
{Ben{\'\i}tez-Llambay} P.,  {Masset} F.~S.,  2016, \mn@doi [\apjs]
  {10.3847/0067-0049/223/1/11}, \href
  {https://ui.adsabs.harvard.edu/abs/2016ApJS..223...11B} {223, 11}

\bibitem[\protect\citeauthoryear{{Bennett} et~al.,}{{Bennett}
  et~al.}{2018}]{Bennett18}
{Bennett} D.~P.,  et~al., 2018, arXiv e-prints, \href
  {https://ui.adsabs.harvard.edu/abs/2018arXiv180308564B} {p. arXiv:1803.08564}

\bibitem[\protect\citeauthoryear{{Birnstiel}, {Klahr}  \&
  {Ercolano}}{{Birnstiel} et~al.}{2012}]{Birnstiel12}
{Birnstiel} T.,  {Klahr} H.,   {Ercolano} B.,  2012, \mn@doi [\aap]
  {10.1051/0004-6361/201118136}, \href
  {https://ui.adsabs.harvard.edu/abs/2012A&A...539A.148B} {539, A148}

\bibitem[\protect\citeauthoryear{{Bitsch} \& {Kley}}{{Bitsch} \&
  {Kley}}{2010}]{Bitsch}
{Bitsch} B.,  {Kley} W.,  2010, \mn@doi [\aap] {10.1051/0004-6361/201014414},
  \href {http://adsabs.harvard.edu/abs/2010A%26A...523A..30B} {523, A30}

\bibitem[\protect\citeauthoryear{{Bitsch}, {Lambrechts}  \&
  {Johansen}}{{Bitsch} et~al.}{2015}]{Bitsch15}
{Bitsch} B.,  {Lambrechts} M.,   {Johansen} A.,  2015, \mn@doi [\aap]
  {10.1051/0004-6361/201526463}, \href
  {http://adsabs.harvard.edu/abs/2015A%26A...582A.112B} {582, A112}

\bibitem[\protect\citeauthoryear{{Bitsch}, {Morbidelli}, {Johansen}, {Lega},
  {Lambrechts}  \& {Crida}}{{Bitsch} et~al.}{2018}]{Bitsch18}
{Bitsch} B.,  {Morbidelli} A.,  {Johansen} A.,  {Lega} E.,  {Lambrechts} M.,
  {Crida} A.,  2018, \mn@doi [\aap] {10.1051/0004-6361/201731931}, \href
  {https://ui.adsabs.harvard.edu/abs/2018A&A...612A..30B} {612, A30}

\bibitem[\protect\citeauthoryear{{Bromley} \& {Kenyon}}{{Bromley} \&
  {Kenyon}}{2015}]{Bromley15}
{Bromley} B.~C.,  {Kenyon} S.~J.,  2015, \mn@doi [\apj]
  {10.1088/0004-637X/806/1/98}, \href
  {https://ui.adsabs.harvard.edu/abs/2015ApJ...806...98B} {806, 98}

\bibitem[\protect\citeauthoryear{{Br{\"u}gger}, {Burn}, {Coleman}, {Alibert}
  \& {Benz}}{{Br{\"u}gger} et~al.}{2020}]{Brugger20}
{Br{\"u}gger} N.,  {Burn} R.,  {Coleman} G.~A.~L.,  {Alibert} Y.,   {Benz} W.,
  2020, \mn@doi [\aap] {10.1051/0004-6361/202038042}, \href
  {https://ui.adsabs.harvard.edu/abs/2020A&A...640A..21B} {640, A21}

\bibitem[\protect\citeauthoryear{{Chambers}}{{Chambers}}{1999}]{Chambers}
{Chambers} J.~E.,  1999, \mn@doi [\mnras] {10.1046/j.1365-8711.1999.02379.x},
  \href {http://adsabs.harvard.edu/abs/1999MNRAS.304..793C} {304, 793}

\bibitem[\protect\citeauthoryear{{Chambers}, {Quintana}, {Duncan}  \&
  {Lissauer}}{{Chambers} et~al.}{2002}]{ChambersBinary}
{Chambers} J.~E.,  {Quintana} E.~V.,  {Duncan} M.~J.,   {Lissauer} J.~J.,
  2002, \mn@doi [\aj] {10.1086/340074}, \href
  {https://ui.adsabs.harvard.edu/abs/2002AJ....123.2884C} {123, 2884}

\bibitem[\protect\citeauthoryear{{Chametla}, {Masset}, {Baruteau}  \&
  {Bitsch}}{{Chametla} et~al.}{2022}]{Chametla22}
{Chametla} R.~O.,  {Masset} F.~S.,  {Baruteau} C.,   {Bitsch} B.,  2022,
  \mn@doi [\mnras] {10.1093/mnras/stab3753}, \href
  {https://ui.adsabs.harvard.edu/abs/2022MNRAS.510.3867C} {510, 3867}

\bibitem[\protect\citeauthoryear{{Clarke}, {Gendrin}  \& {Sotomayor}}{{Clarke}
  et~al.}{2001}]{Clarke2001}
{Clarke} C.~J.,  {Gendrin} A.,   {Sotomayor} M.,  2001, \mn@doi [\mnras]
  {10.1046/j.1365-8711.2001.04891.x}, \href
  {http://adsabs.harvard.edu/abs/2001MNRAS.328..485C} {328, 485}

\bibitem[\protect\citeauthoryear{{Coleman}}{{Coleman}}{2021}]{Coleman21}
{Coleman} G. A.~L.,  2021, \mn@doi [\mnras] {10.1093/mnras/stab1904}, \href
  {https://ui.adsabs.harvard.edu/abs/2021MNRAS.506.3596C} {506, 3596}

\bibitem[\protect\citeauthoryear{{Coleman} \& {Haworth}}{{Coleman} \&
  {Haworth}}{2022}]{Coleman22}
{Coleman} G. A.~L.,  {Haworth} T.~J.,  2022, \mn@doi [MNRAS]
  {10.1093/mnras/stac1513}, \href
  {https://ui.adsabs.harvard.edu/abs/2022MNRAS.514.2315C} {514, 2315}

\bibitem[\protect\citeauthoryear{{Coleman} \& {Nelson}}{{Coleman} \&
  {Nelson}}{2014}]{ColemanNelson14}
{Coleman} G.~A.~L.,  {Nelson} R.~P.,  2014, \mn@doi [\mnras]
  {10.1093/mnras/stu1715}, \href
  {http://adsabs.harvard.edu/abs/2014MNRAS.445..479C} {445, 479}

\bibitem[\protect\citeauthoryear{{Coleman} \& {Nelson}}{{Coleman} \&
  {Nelson}}{2016a}]{ColemanNelson16}
{Coleman} G.~A.~L.,  {Nelson} R.~P.,  2016a, \mn@doi [\mnras]
  {10.1093/mnras/stw149}, \href
  {http://adsabs.harvard.edu/abs/2016MNRAS.457.2480C} {457, 2480}

\bibitem[\protect\citeauthoryear{{Coleman} \& {Nelson}}{{Coleman} \&
  {Nelson}}{2016b}]{ColemanNelson16b}
{Coleman} G.~A.~L.,  {Nelson} R.~P.,  2016b, \mn@doi [\mnras]
  {10.1093/mnras/stw1177}, \href
  {http://adsabs.harvard.edu/abs/2016MNRAS.460.2779C} {460, 2779}

\bibitem[\protect\citeauthoryear{{Coleman}, {Nelson}, {Paardekooper},
  {Dreizler}, {Giesers}  \& {Anglada-Escud{\'e}}}{{Coleman}
  et~al.}{2017a}]{ColemanProxima17}
{Coleman} G.~A.~L.,  {Nelson} R.~P.,  {Paardekooper} S.~J.,  {Dreizler} S.,
  {Giesers} B.,   {Anglada-Escud{\'e}} G.,  2017a, \mn@doi [\mnras]
  {10.1093/mnras/stx169}, \href
  {http://adsabs.harvard.edu/abs/2017MNRAS.467..996C} {467, 996}

\bibitem[\protect\citeauthoryear{{Coleman}, {Papaloizou}  \&
  {Nelson}}{{Coleman} et~al.}{2017b}]{CPN17}
{Coleman} G.~A.~L.,  {Papaloizou} J.~C.~B.,   {Nelson} R.~P.,  2017b, \mn@doi
  [\mnras] {10.1093/mnras/stx1297}, \href
  {http://adsabs.harvard.edu/abs/2017MNRAS.470.3206C} {470, 3206}

\bibitem[\protect\citeauthoryear{{Coleman}, {Leleu}, {Alibert}  \&
  {Benz}}{{Coleman} et~al.}{2019}]{Coleman19}
{Coleman} G.~A.~L.,  {Leleu} A.,  {Alibert} Y.,   {Benz} W.,  2019, \mn@doi
  [\aap] {10.1051/0004-6361/201935922}, \href
  {https://ui.adsabs.harvard.edu/abs/2019A&A...631A...7C} {631, A7}

\bibitem[\protect\citeauthoryear{{Coleman}, {Nelson}  \& {Triaud}}{{Coleman}
  et~al.}{2022}]{Coleman22b}
{Coleman} G. A.~L.,  {Nelson} R.~P.,   {Triaud} A. H.~M.~J.,  2022, \mn@doi
  [\mnras] {10.1093/mnras/stac1029}, \href
  {https://ui.adsabs.harvard.edu/abs/2022MNRAS.513.2563C} {513, 2563}

\bibitem[\protect\citeauthoryear{{Cox}}{{Cox}}{2000}]{Cox}
{Cox} A.~N.,  2000, {Allen's astrophysical quantities}

\bibitem[\protect\citeauthoryear{{Cresswell} \& {Nelson}}{{Cresswell} \&
  {Nelson}}{2008}]{cressnels}
{Cresswell} P.,  {Nelson} R.~P.,  2008, \mn@doi [\aap]
  {10.1051/0004-6361:20079178}, \href
  {http://adsabs.harvard.edu/abs/2008A%26A...482..677C} {482, 677}

\bibitem[\protect\citeauthoryear{{Crida}, {Morbidelli}  \& {Masset}}{{Crida}
  et~al.}{2006}]{Crida}
{Crida} A.,  {Morbidelli} A.,   {Masset} F.,  2006, \mn@doi [\icarus]
  {10.1016/j.icarus.2005.10.007}, \href
  {http://adsabs.harvard.edu/abs/2006Icar..181..587C} {181, 587}

\bibitem[\protect\citeauthoryear{{D'Angelo} \& {Marzari}}{{D'Angelo} \&
  {Marzari}}{2012}]{Dangelo12}
{D'Angelo} G.,  {Marzari} F.,  2012, \mn@doi [\apj]
  {10.1088/0004-637X/757/1/50}, \href
  {http://adsabs.harvard.edu/abs/2012ApJ...757...50D} {757, 50}

\bibitem[\protect\citeauthoryear{{Doyle} et~al.,}{{Doyle}
  et~al.}{2011}]{Doyle11}
{Doyle} L.~R.,  et~al., 2011, \mn@doi [Science] {10.1126/science.1210923},
  \href {https://ui.adsabs.harvard.edu/abs/2011Sci...333.1602D} {333, 1602}

\bibitem[\protect\citeauthoryear{{Dullemond}, {Hollenbach}, {Kamp}  \&
  {D'Alessio}}{{Dullemond} et~al.}{2007}]{Dullemond}
{Dullemond} C.~P.,  {Hollenbach} D.,  {Kamp} I.,   {D'Alessio} P.,  2007,
  Protostars and Planets V, \href
  {http://adsabs.harvard.edu/abs/2007prpl.conf..555D} {pp 555--572}

\bibitem[\protect\citeauthoryear{{Dunhill} \& {Alexander}}{{Dunhill} \&
  {Alexander}}{2013}]{Dunhill13}
{Dunhill} A.~C.,  {Alexander} R.~D.,  2013, \mn@doi [\mnras]
  {10.1093/mnras/stt1456}, \href
  {https://ui.adsabs.harvard.edu/abs/2013MNRAS.435.2328D} {435, 2328}

\bibitem[\protect\citeauthoryear{{Dutrey}, {Guilloteau}  \& {Simon}}{{Dutrey}
  et~al.}{1994}]{Dutrey94}
{Dutrey} A.,  {Guilloteau} S.,   {Simon} M.,  1994, \aap, \href
  {https://ui.adsabs.harvard.edu/abs/1994A&A...286..149D} {286, 149}

\bibitem[\protect\citeauthoryear{{Everett}, {Howell}, {Silva}  \&
  {Szkody}}{{Everett} et~al.}{2013}]{Everett13}
{Everett} M.~E.,  {Howell} S.~B.,  {Silva} D.~R.,   {Szkody} P.,  2013, \mn@doi
  [\apj] {10.1088/0004-637X/771/2/107}, \href
  {https://ui.adsabs.harvard.edu/abs/2013ApJ...771..107E} {771, 107}

\bibitem[\protect\citeauthoryear{{Fendyke} \& {Nelson}}{{Fendyke} \&
  {Nelson}}{2014}]{Fendyke}
{Fendyke} S.~M.,  {Nelson} R.~P.,  2014, \mn@doi [\mnras]
  {10.1093/mnras/stt1867}, \href
  {http://adsabs.harvard.edu/abs/2014MNRAS.437...96F} {437, 96}

\bibitem[\protect\citeauthoryear{{Gillon} et~al.,}{{Gillon}
  et~al.}{2017}]{GillonTrappist17}
{Gillon} M.,  et~al., 2017, \mn@doi [\nat] {10.1038/nature21360}, \href
  {http://adsabs.harvard.edu/abs/2017Natur.542..456G} {542, 456}

\bibitem[\protect\citeauthoryear{{Goldreich} \& {Schlichting}}{{Goldreich} \&
  {Schlichting}}{2014}]{GoldreichSchlichting2014}
{Goldreich} P.,  {Schlichting} H.~E.,  2014, \mn@doi [\aj]
  {10.1088/0004-6256/147/2/32}, \href
  {http://adsabs.harvard.edu/abs/2014AJ....147...32G} {147, 32}

\bibitem[\protect\citeauthoryear{{Gorti} \& {Hollenbach}}{{Gorti} \&
  {Hollenbach}}{2009}]{Gorti09a}
{Gorti} U.,  {Hollenbach} D.,  2009, \mn@doi [\apj]
  {10.1088/0004-637X/690/2/1539}, \href
  {https://ui.adsabs.harvard.edu/abs/2009ApJ...690.1539G} {690, 1539}

\bibitem[\protect\citeauthoryear{{Gorti}, {Dullemond}  \& {Hollenbach}}{{Gorti}
  et~al.}{2009}]{Gorti09}
{Gorti} U.,  {Dullemond} C.~P.,   {Hollenbach} D.,  2009, \mn@doi [\apj]
  {10.1088/0004-637X/705/2/1237}, \href
  {https://ui.adsabs.harvard.edu/abs/2009ApJ...705.1237G} {705, 1237}

\bibitem[\protect\citeauthoryear{{Gorti}, {Hollenbach}  \& {Dullemond}}{{Gorti}
  et~al.}{2015}]{Gorti15}
{Gorti} U.,  {Hollenbach} D.,   {Dullemond} C.~P.,  2015, \mn@doi [\apj]
  {10.1088/0004-637X/804/1/29}, \href
  {https://ui.adsabs.harvard.edu/abs/2015ApJ...804...29G} {804, 29}

\bibitem[\protect\citeauthoryear{{Gundlach} \& {Blum}}{{Gundlach} \&
  {Blum}}{2015}]{Gundlach15}
{Gundlach} B.,  {Blum} J.,  2015, \mn@doi [\apj] {10.1088/0004-637X/798/1/34},
  \href {https://ui.adsabs.harvard.edu/abs/2015ApJ...798...34G} {798, 34}

\bibitem[\protect\citeauthoryear{{G{\"u}ttler}, {Blum}, {Zsom}, {Ormel}  \&
  {Dullemond}}{{G{\"u}ttler} et~al.}{2010}]{Guttler10}
{G{\"u}ttler} C.,  {Blum} J.,  {Zsom} A.,  {Ormel} C.~W.,   {Dullemond} C.~P.,
  2010, \mn@doi [\aap] {10.1051/0004-6361/200912852}, \href
  {https://ui.adsabs.harvard.edu/abs/2010A&A...513A..56G} {513, A56}

\bibitem[\protect\citeauthoryear{{Habing}}{{Habing}}{1968}]{Habing68}
{Habing} H.~J.,  1968, \bain, \href
  {https://ui.adsabs.harvard.edu/abs/1968BAN....19..421H} {19, 421}

\bibitem[\protect\citeauthoryear{{Haworth}, {Clarke}, {Rahman}, {Winter}  \&
  {Facchini}}{{Haworth} et~al.}{2018}]{Haworth18}
{Haworth} T.~J.,  {Clarke} C.~J.,  {Rahman} W.,  {Winter} A.~J.,   {Facchini}
  S.,  2018, \mn@doi [\mnras] {10.1093/mnras/sty2323}, \href
  {https://ui.adsabs.harvard.edu/abs/2018MNRAS.481..452H} {481, 452}

\bibitem[\protect\citeauthoryear{{Hellary} \& {Nelson}}{{Hellary} \&
  {Nelson}}{2012}]{Hellary}
{Hellary} P.,  {Nelson} R.~P.,  2012, \mn@doi [\mnras]
  {10.1111/j.1365-2966.2011.19815.x}, \href
  {http://adsabs.harvard.edu/abs/2012MNRAS.419.2737H} {419, 2737}

\bibitem[\protect\citeauthoryear{{Holman} \& {Wiegert}}{{Holman} \&
  {Wiegert}}{1999}]{Holman99}
{Holman} M.~J.,  {Wiegert} P.~A.,  1999, \mn@doi [\aj] {10.1086/300695}, \href
  {https://ui.adsabs.harvard.edu/abs/1999AJ....117..621H} {117, 621}

\bibitem[\protect\citeauthoryear{{Ivezi{\'c}} et~al.,}{{Ivezi{\'c}}
  et~al.}{2019}]{LSST_2019}
{Ivezi{\'c}} {\v{Z}}.,  et~al., 2019, \mn@doi [\apj]
  {10.3847/1538-4357/ab042c}, \href
  {https://ui.adsabs.harvard.edu/abs/2019ApJ...873..111I} {873, 111}

\bibitem[\protect\citeauthoryear{{Johansen} \& {Lambrechts}}{{Johansen} \&
  {Lambrechts}}{2017}]{Johansen17}
{Johansen} A.,  {Lambrechts} M.,  2017, \mn@doi [Annual Review of Earth and
  Planetary Sciences] {10.1146/annurev-earth-063016-020226}, \href
  {http://adsabs.harvard.edu/abs/2017AREPS..45..359J} {45, 359}

\bibitem[\protect\citeauthoryear{{Kley} \& {Nelson}}{{Kley} \&
  {Nelson}}{2010}]{Kley10}
{Kley} W.,  {Nelson} R.~P.,  2010, {Early Evolution of Planets in Binaries:
  Planet-Disk Interaction}.
p.~135, \mn@doi{10.1007/978-90-481-8687-7\_6}

\bibitem[\protect\citeauthoryear{{Kley}, {Thun}  \& {Penzlin}}{{Kley}
  et~al.}{2019}]{Kley19}
{Kley} W.,  {Thun} D.,   {Penzlin} A. B.~T.,  2019, \mn@doi [\aap]
  {10.1051/0004-6361/201935503}, \href
  {https://ui.adsabs.harvard.edu/abs/2019A&A...627A..91K} {627, A91}

\bibitem[\protect\citeauthoryear{{Kostov} et~al.,}{{Kostov}
  et~al.}{2016}]{Kostov16}
{Kostov} V.~B.,  et~al., 2016, \mn@doi [\apj] {10.3847/0004-637X/827/1/86},
  \href {https://ui.adsabs.harvard.edu/abs/2016ApJ...827...86K} {827, 86}

\bibitem[\protect\citeauthoryear{{Kostov} et~al.,}{{Kostov}
  et~al.}{2020}]{Kostov20}
{Kostov} V.~B.,  et~al., 2020, \mn@doi [\aj] {10.3847/1538-3881/ab8a48}, \href
  {https://ui.adsabs.harvard.edu/abs/2020AJ....159..253K} {159, 253}

\bibitem[\protect\citeauthoryear{{Lambrechts} \& {Johansen}}{{Lambrechts} \&
  {Johansen}}{2012}]{Lambrechts12}
{Lambrechts} M.,  {Johansen} A.,  2012, \mn@doi [\aap]
  {10.1051/0004-6361/201219127}, \href
  {http://adsabs.harvard.edu/abs/2012A%26A...544A..32L} {544, A32}

\bibitem[\protect\citeauthoryear{{Lambrechts} \& {Johansen}}{{Lambrechts} \&
  {Johansen}}{2014}]{Lambrechts14}
{Lambrechts} M.,  {Johansen} A.,  2014, \mn@doi [\aap]
  {10.1051/0004-6361/201424343}, \href
  {http://adsabs.harvard.edu/abs/2014A%26A...572A.107L} {572, A107}

\bibitem[\protect\citeauthoryear{{Langford} \& {Weiss}}{{Langford} \&
  {Weiss}}{2023}]{Langford23}
{Langford} A.,  {Weiss} L.~M.,  2023, arXiv e-prints, \href
  {https://ui.adsabs.harvard.edu/abs/2023arXiv230200580L} {p. arXiv:2302.00580}

\bibitem[\protect\citeauthoryear{{Leleu}, {Coleman}  \& {Ataiee}}{{Leleu}
  et~al.}{2019}]{Leleu19}
{Leleu} A.,  {Coleman} G. A.~L.,   {Ataiee} S.,  2019, \mn@doi [\aap]
  {10.1051/0004-6361/201834486}, \href
  {https://ui.adsabs.harvard.edu/abs/2019A&A...631A...6L} {631, A6}

\bibitem[\protect\citeauthoryear{{Lin} \& {Papaloizou}}{{Lin} \&
  {Papaloizou}}{1986}]{LinPapaloizou86}
{Lin} D.~N.~C.,  {Papaloizou} J.,  1986, \mn@doi [\apj] {10.1086/164653}, \href
  {http://adsabs.harvard.edu/abs/1986ApJ...309..846L} {309, 846}

\bibitem[\protect\citeauthoryear{{Lines}, {Leinhardt}, {Paardekooper},
  {Baruteau}  \& {Thebault}}{{Lines} et~al.}{2014}]{Lines14}
{Lines} S.,  {Leinhardt} Z.~M.,  {Paardekooper} S.,  {Baruteau} C.,
  {Thebault} P.,  2014, \mn@doi [\apjl] {10.1088/2041-8205/782/1/L11}, \href
  {https://ui.adsabs.harvard.edu/abs/2014ApJ...782L..11L} {782, L11}

\bibitem[\protect\citeauthoryear{{Liu}, {Lambrechts}, {Johansen}, {Pascucci}
  \& {Henning}}{{Liu} et~al.}{2020}]{Liu20}
{Liu} B.,  {Lambrechts} M.,  {Johansen} A.,  {Pascucci} I.,   {Henning} T.,
  2020, \mn@doi [\aap] {10.1051/0004-6361/202037720}, \href
  {https://ui.adsabs.harvard.edu/abs/2020A&A...638A..88L} {638, A88}

\bibitem[\protect\citeauthoryear{{Madhusudhan}}{{Madhusudhan}}{2019}]{Madhusudhan19}
{Madhusudhan} N.,  2019, \mn@doi [\araa] {10.1146/annurev-astro-081817-051846},
  \href {https://ui.adsabs.harvard.edu/abs/2019ARA&A..57..617M} {57, 617}

\bibitem[\protect\citeauthoryear{{Martin}}{{Martin}}{2018}]{Martin18}
{Martin} D.~V.,  2018, in {Deeg} H.~J.,  {Belmonte} J.~A.,  eds, , Handbook of
  Exoplanets.
p.~156, \mn@doi{10.1007/978-3-319-55333-7\_156}

\bibitem[\protect\citeauthoryear{{Martin} \& {Fitzmaurice}}{{Martin} \&
  {Fitzmaurice}}{2022}]{Martin22}
{Martin} D.~V.,  {Fitzmaurice} E.,  2022, \mn@doi [\mnras]
  {10.1093/mnras/stac090}, \href
  {https://ui.adsabs.harvard.edu/abs/2022MNRAS.512..602M} {512, 602}

\bibitem[\protect\citeauthoryear{{Martin}, {Armitage}  \& {Alexander}}{{Martin}
  et~al.}{2013}]{Martin13}
{Martin} R.~G.,  {Armitage} P.~J.,   {Alexander} R.~D.,  2013, \mn@doi [\apj]
  {10.1088/0004-637X/773/1/74}, \href
  {https://ui.adsabs.harvard.edu/abs/2013ApJ...773...74M} {773, 74}

\bibitem[\protect\citeauthoryear{{Martin} et~al.,}{{Martin}
  et~al.}{2019}]{Martin19}
{Martin} D.~V.,  et~al., 2019, \mn@doi [\aap] {10.1051/0004-6361/201833669},
  \href {https://ui.adsabs.harvard.edu/abs/2019A&A...624A..68M} {624, A68}

\bibitem[\protect\citeauthoryear{{Marzari} \& {Scholl}}{{Marzari} \&
  {Scholl}}{2000}]{Marzari00}
{Marzari} F.,  {Scholl} H.,  2000, \mn@doi [\apj] {10.1086/317091}, \href
  {https://ui.adsabs.harvard.edu/abs/2000ApJ...543..328M} {543, 328}

\bibitem[\protect\citeauthoryear{{Marzari}, {Th{\'e}bault}  \&
  {Scholl}}{{Marzari} et~al.}{2008}]{Marzari08}
{Marzari} F.,  {Th{\'e}bault} P.,   {Scholl} H.,  2008, \mn@doi [\apj]
  {10.1086/588423}, \href
  {https://ui.adsabs.harvard.edu/abs/2008ApJ...681.1599M} {681, 1599}

\bibitem[\protect\citeauthoryear{{Marzari}, {Thebault}, {Scholl}, {Picogna}  \&
  {Baruteau}}{{Marzari} et~al.}{2013}]{Marzari13}
{Marzari} F.,  {Thebault} P.,  {Scholl} H.,  {Picogna} G.,   {Baruteau} C.,
  2013, \mn@doi [\aap] {10.1051/0004-6361/201220893}, \href
  {https://ui.adsabs.harvard.edu/abs/2013A&A...553A..71M} {553, A71}

\bibitem[\protect\citeauthoryear{{Matsuyama}, {Johnstone}  \&
  {Hartmann}}{{Matsuyama} et~al.}{2003}]{Matsuyama03}
{Matsuyama} I.,  {Johnstone} D.,   {Hartmann} L.,  2003, \mn@doi [\apj]
  {10.1086/344638}, \href
  {https://ui.adsabs.harvard.edu/abs/2003ApJ...582..893M} {582, 893}

\bibitem[\protect\citeauthoryear{{Menou} \& {Goodman}}{{Menou} \&
  {Goodman}}{2004}]{Menou}
{Menou} K.,  {Goodman} J.,  2004, \mn@doi [\apj] {10.1086/382947}, \href
  {http://adsabs.harvard.edu/abs/2004ApJ...606..520M} {606, 520}

\bibitem[\protect\citeauthoryear{{Meschiari}}{{Meschiari}}{2012a}]{Meschiari12a}
{Meschiari} S.,  2012a, \mn@doi [\apj] {10.1088/0004-637X/752/1/71}, \href
  {https://ui.adsabs.harvard.edu/abs/2012ApJ...752...71M} {752, 71}

\bibitem[\protect\citeauthoryear{{Meschiari}}{{Meschiari}}{2012b}]{Meschiari12b}
{Meschiari} S.,  2012b, \mn@doi [\apjl] {10.1088/2041-8205/761/1/L7}, \href
  {https://ui.adsabs.harvard.edu/abs/2012ApJ...761L...7M} {761, L7}

\bibitem[\protect\citeauthoryear{{Meschiari}}{{Meschiari}}{2014}]{Meschiari14}
{Meschiari} S.,  2014, \mn@doi [\apj] {10.1088/0004-637X/790/1/41}, \href
  {https://ui.adsabs.harvard.edu/abs/2014ApJ...790...41M} {790, 41}

\bibitem[\protect\citeauthoryear{{Mihalas} \& {Mihalas}}{{Mihalas} \&
  {Mihalas}}{1984}]{Mihalas}
{Mihalas} D.,  {Mihalas} B.~W.,  1984, {Foundations of radiation hydrodynamics}

\bibitem[\protect\citeauthoryear{{Mills}, {Fabrycky}, {Migaszewski}, {Ford},
  {Petigura}  \& {Isaacson}}{{Mills} et~al.}{2016}]{Mills16}
{Mills} S.~M.,  {Fabrycky} D.~C.,  {Migaszewski} C.,  {Ford} E.~B.,  {Petigura}
  E.,   {Isaacson} H.,  2016, \mn@doi [\nat] {10.1038/nature17445}, \href
  {http://adsabs.harvard.edu/abs/2016Natur.533..509M} {533, 509}

\bibitem[\protect\citeauthoryear{{Musiolik} \& {Wurm}}{{Musiolik} \&
  {Wurm}}{2019}]{Musiolik19}
{Musiolik} G.,  {Wurm} G.,  2019, \mn@doi [\apj] {10.3847/1538-4357/ab0428},
  \href {https://ui.adsabs.harvard.edu/abs/2019ApJ...873...58M} {873, 58}

\bibitem[\protect\citeauthoryear{{Mutter}, {Pierens}  \& {Nelson}}{{Mutter}
  et~al.}{2017a}]{Mutter17D}
{Mutter} M.~M.,  {Pierens} A.,   {Nelson} R.~P.,  2017a, \mn@doi [\mnras]
  {10.1093/mnras/stw2768}, \href
  {https://ui.adsabs.harvard.edu/abs/2017MNRAS.465.4735M} {465, 4735}

\bibitem[\protect\citeauthoryear{{Mutter}, {Pierens}  \& {Nelson}}{{Mutter}
  et~al.}{2017b}]{Mutter17P}
{Mutter} M.~M.,  {Pierens} A.,   {Nelson} R.~P.,  2017b, \mn@doi [\mnras]
  {10.1093/mnras/stx1113}, \href
  {https://ui.adsabs.harvard.edu/abs/2017MNRAS.469.4504M} {469, 4504}

\bibitem[\protect\citeauthoryear{{Nakagawa}, {Sekiya}  \& {Hayashi}}{{Nakagawa}
  et~al.}{1986}]{Nakagawa86}
{Nakagawa} Y.,  {Sekiya} M.,   {Hayashi} C.,  1986, \mn@doi [\icarus]
  {10.1016/0019-1035(86)90121-1}, \href
  {https://ui.adsabs.harvard.edu/abs/1986Icar...67..375N} {67, 375}

\bibitem[\protect\citeauthoryear{{Nelson}}{{Nelson}}{2003}]{Nelson03}
{Nelson} R.~P.,  2003, \mn@doi [\mnras] {10.1046/j.1365-8711.2003.06929.x},
  \href {https://ui.adsabs.harvard.edu/abs/2003MNRAS.345..233N} {345, 233}

\bibitem[\protect\citeauthoryear{{{\"O}berg}, {Murray-Clay}  \&
  {Bergin}}{{{\"O}berg} et~al.}{2011}]{Oberg11}
{{\"O}berg} K.~I.,  {Murray-Clay} R.,   {Bergin} E.~A.,  2011, \mn@doi [\apjl]
  {10.1088/2041-8205/743/1/L16}, \href
  {https://ui.adsabs.harvard.edu/abs/2011ApJ...743L..16O} {743, L16}

\bibitem[\protect\citeauthoryear{{Ormel} \& {Cuzzi}}{{Ormel} \&
  {Cuzzi}}{2007}]{Ormel07}
{Ormel} C.~W.,  {Cuzzi} J.~N.,  2007, \mn@doi [\aap]
  {10.1051/0004-6361:20066899}, \href
  {https://ui.adsabs.harvard.edu/abs/2007A&A...466..413O} {466, 413}

\bibitem[\protect\citeauthoryear{{Ormel} \& {Klahr}}{{Ormel} \&
  {Klahr}}{2010}]{OrmelKlahr2010}
{Ormel} C.~W.,  {Klahr} H.~H.,  2010, \mn@doi [\aap]
  {10.1051/0004-6361/201014903}, \href
  {http://adsabs.harvard.edu/abs/2010A%26A...520A..43O} {520, A43}

\bibitem[\protect\citeauthoryear{{Orosz} et~al.,}{{Orosz}
  et~al.}{2012}]{Orosz12_k47}
{Orosz} J.~A.,  et~al., 2012, \mn@doi [Science] {10.1126/science.1228380},
  \href {https://ui.adsabs.harvard.edu/abs/2012Sci...337.1511O} {337, 1511}

\bibitem[\protect\citeauthoryear{{Orosz} et~al.,}{{Orosz}
  et~al.}{2019}]{Orosz19}
{Orosz} J.~A.,  et~al., 2019, \mn@doi [\aj] {10.3847/1538-3881/ab0ca0}, \href
  {https://ui.adsabs.harvard.edu/abs/2019AJ....157..174O} {157, 174}

\bibitem[\protect\citeauthoryear{{Owen}, {Clarke}  \& {Ercolano}}{{Owen}
  et~al.}{2012}]{Owen12}
{Owen} J.~E.,  {Clarke} C.~J.,   {Ercolano} B.,  2012, \mn@doi [\mnras]
  {10.1111/j.1365-2966.2011.20337.x}, \href
  {https://ui.adsabs.harvard.edu/abs/2012MNRAS.422.1880O} {422, 1880}

\bibitem[\protect\citeauthoryear{{Paardekooper} \& {Leinhardt}}{{Paardekooper}
  \& {Leinhardt}}{2010}]{Paardekooper10}
{Paardekooper} S.~J.,  {Leinhardt} Z.~M.,  2010, \mn@doi [\mnras]
  {10.1111/j.1745-3933.2010.00816.x}, \href
  {https://ui.adsabs.harvard.edu/abs/2010MNRAS.403L..64P} {403, L64}

\bibitem[\protect\citeauthoryear{{Paardekooper} \& {Mellema}}{{Paardekooper} \&
  {Mellema}}{2006}]{PaardekooperMellema06}
{Paardekooper} S.-J.,  {Mellema} G.,  2006, \mn@doi [\aap]
  {10.1051/0004-6361:20066304}, \href
  {http://adsabs.harvard.edu/abs/2006A%26A...459L..17P} {459, L17}

\bibitem[\protect\citeauthoryear{{Paardekooper}, {Baruteau}, {Crida}  \&
  {Kley}}{{Paardekooper} et~al.}{2010}]{pdk10}
{Paardekooper} S.-J.,  {Baruteau} C.,  {Crida} A.,   {Kley} W.,  2010, \mn@doi
  [\mnras] {10.1111/j.1365-2966.2009.15782.x}, \href
  {http://adsabs.harvard.edu/abs/2010MNRAS.401.1950P} {401, 1950}

\bibitem[\protect\citeauthoryear{{Paardekooper}, {Baruteau}  \&
  {Kley}}{{Paardekooper} et~al.}{2011}]{pdk11}
{Paardekooper} S.-J.,  {Baruteau} C.,   {Kley} W.,  2011, \mn@doi [\mnras]
  {10.1111/j.1365-2966.2010.17442.x}, \href
  {http://adsabs.harvard.edu/abs/2011MNRAS.410..293P} {410, 293}

\bibitem[\protect\citeauthoryear{{Paardekooper}, {Leinhardt}, {Th{\'e}bault}
  \& {Baruteau}}{{Paardekooper} et~al.}{2012}]{Paardekooper12}
{Paardekooper} S.-J.,  {Leinhardt} Z.~M.,  {Th{\'e}bault} P.,   {Baruteau} C.,
  2012, \mn@doi [\apjl] {10.1088/2041-8205/754/1/L16}, \href
  {https://ui.adsabs.harvard.edu/abs/2012ApJ...754L..16P} {754, L16}

\bibitem[\protect\citeauthoryear{{Papaloizou} \& {Larwood}}{{Papaloizou} \&
  {Larwood}}{2000}]{Papaloizou2000}
{Papaloizou} J.~C.~B.,  {Larwood} J.~D.,  2000, \mn@doi [\mnras]
  {10.1046/j.1365-8711.2000.03466.x}, \href
  {https://ui.adsabs.harvard.edu/abs/2000MNRAS.315..823P} {315, 823}

\bibitem[\protect\citeauthoryear{{Papaloizou} \& {Nelson}}{{Papaloizou} \&
  {Nelson}}{2005}]{PapNelson2005}
{Papaloizou} J.~C.~B.,  {Nelson} R.~P.,  2005, \mn@doi [\aap]
  {10.1051/0004-6361:20042029}, \href
  {http://adsabs.harvard.edu/abs/2005A%26A...433..247P} {433, 247}

\bibitem[\protect\citeauthoryear{{Papaloizou} \& {Terquem}}{{Papaloizou} \&
  {Terquem}}{1999}]{Pap-Terquem-envelopes}
{Papaloizou} J.~C.~B.,  {Terquem} C.,  1999, \mn@doi [\apj] {10.1086/307581},
  \href {http://adsabs.harvard.edu/abs/1999ApJ...521..823P} {521, 823}

\bibitem[\protect\citeauthoryear{{Penzlin}, {Kley}  \& {Nelson}}{{Penzlin}
  et~al.}{2021}]{Penzlin21}
{Penzlin} A. B.~T.,  {Kley} W.,   {Nelson} R.~P.,  2021, \mn@doi [\aap]
  {10.1051/0004-6361/202039319}, \href
  {https://ui.adsabs.harvard.edu/abs/2021A&A...645A..68P} {645, A68}

\bibitem[\protect\citeauthoryear{{Penzlin}, {Kley}, {Audiffren}  \&
  {Sch{\"a}fer}}{{Penzlin} et~al.}{2022}]{Penzlin22}
{Penzlin} A. B.~T.,  {Kley} W.,  {Audiffren} H.,   {Sch{\"a}fer} C.~M.,  2022,
  \mn@doi [\aap] {10.1051/0004-6361/202141399}, \href
  {https://ui.adsabs.harvard.edu/abs/2022A&A...660A.101P} {660, A101}

\bibitem[\protect\citeauthoryear{{Picogna}, {Ercolano}, {Owen}  \&
  {Weber}}{{Picogna} et~al.}{2019}]{Picogna19}
{Picogna} G.,  {Ercolano} B.,  {Owen} J.~E.,   {Weber} M.~L.,  2019, \mn@doi
  [\mnras] {10.1093/mnras/stz1166}, \href
  {https://ui.adsabs.harvard.edu/abs/2019MNRAS.487..691P} {487, 691}

\bibitem[\protect\citeauthoryear{{Pierens} \& {Nelson}}{{Pierens} \&
  {Nelson}}{2007}]{Pierens07}
{Pierens} A.,  {Nelson} R.~P.,  2007, \mn@doi [\aap]
  {10.1051/0004-6361:20077659}, \href
  {https://ui.adsabs.harvard.edu/abs/2007A&A...472..993P} {472, 993}

\bibitem[\protect\citeauthoryear{{Pierens} \& {Nelson}}{{Pierens} \&
  {Nelson}}{2008a}]{Pierens08a}
{Pierens} A.,  {Nelson} R.~P.,  2008a, \mn@doi [\aap]
  {10.1051/0004-6361:20078844}, \href
  {https://ui.adsabs.harvard.edu/abs/2008A&A...478..939P} {478, 939}

\bibitem[\protect\citeauthoryear{{Pierens} \& {Nelson}}{{Pierens} \&
  {Nelson}}{2008b}]{Pierens08b}
{Pierens} A.,  {Nelson} R.~P.,  2008b, \mn@doi [\aap]
  {10.1051/0004-6361:200809453}, \href
  {https://ui.adsabs.harvard.edu/abs/2008A&A...483..633P} {483, 633}

\bibitem[\protect\citeauthoryear{{Pierens} \& {Nelson}}{{Pierens} \&
  {Nelson}}{2013}]{Pierens13}
{Pierens} A.,  {Nelson} R.~P.,  2013, \mn@doi [\aap]
  {10.1051/0004-6361/201321777}, \href
  {https://ui.adsabs.harvard.edu/abs/2013A&A...556A.134P} {556, A134}

\bibitem[\protect\citeauthoryear{{Pierens}, {McNally}  \& {Nelson}}{{Pierens}
  et~al.}{2020}]{Pierens20}
{Pierens} A.,  {McNally} C.~P.,   {Nelson} R.~P.,  2020, \mn@doi [\mnras]
  {10.1093/mnras/staa1550}, \href
  {https://ui.adsabs.harvard.edu/abs/2020MNRAS.496.2849P} {496, 2849}

\bibitem[\protect\citeauthoryear{{Pierens}, {Nelson}  \& {McNally}}{{Pierens}
  et~al.}{2021}]{Pierens21}
{Pierens} A.,  {Nelson} R.~P.,   {McNally} C.~P.,  2021, \mn@doi [\mnras]
  {10.1093/mnras/stab2853}, \href
  {https://ui.adsabs.harvard.edu/abs/2021MNRAS.508.4806P} {508, 4806}

\bibitem[\protect\citeauthoryear{{Poon}, {Nelson}  \& {Coleman}}{{Poon}
  et~al.}{2021}]{Poon21}
{Poon} S. T.~S.,  {Nelson} R.~P.,   {Coleman} G. A.~L.,  2021, \mn@doi [\mnras]
  {10.1093/mnras/stab1466}, \href
  {https://ui.adsabs.harvard.edu/abs/2021MNRAS.505.2500P} {505, 2500}

\bibitem[\protect\citeauthoryear{Press, Teukolsky, Vetterling  \&
  Flannery}{Press et~al.}{2007}]{Press07}
Press W.~H.,  Teukolsky S.~A.,  Vetterling W.~T.,   Flannery B.~P.,  2007,
  Numerical Recipes 3rd Edition: The Art of Scientific Computing, 3 edn.
Cambridge University Press, New York, NY, USA

\bibitem[\protect\citeauthoryear{{Rafikov} \& {Silsbee}}{{Rafikov} \&
  {Silsbee}}{2015}]{Rafikov15}
{Rafikov} R.~R.,  {Silsbee} K.,  2015, \mn@doi [\apj]
  {10.1088/0004-637X/798/2/69}, \href
  {https://ui.adsabs.harvard.edu/abs/2015ApJ...798...69R} {798, 69}

\bibitem[\protect\citeauthoryear{{Rein} \& {Papaloizou}}{{Rein} \&
  {Papaloizou}}{2009}]{ReinPapaloizou2009}
{Rein} H.,  {Papaloizou} J.~C.~B.,  2009, \mn@doi [\aap]
  {10.1051/0004-6361/200811330}, \href
  {http://adsabs.harvard.edu/abs/2009A%26A...497..595R} {497, 595}

\bibitem[\protect\citeauthoryear{{Rice}, {Armitage}, {Wood}  \&
  {Lodato}}{{Rice} et~al.}{2006}]{Rice06}
{Rice} W.~K.~M.,  {Armitage} P.~J.,  {Wood} K.,   {Lodato} G.,  2006, \mn@doi
  [\mnras] {10.1111/j.1365-2966.2006.11113.x}, \href
  {https://ui.adsabs.harvard.edu/abs/2006MNRAS.373.1619R} {373, 1619}

\bibitem[\protect\citeauthoryear{{Sch{\"a}fer}, {Yang}  \&
  {Johansen}}{{Sch{\"a}fer} et~al.}{2017}]{Schafer17}
{Sch{\"a}fer} U.,  {Yang} C.-C.,   {Johansen} A.,  2017, \mn@doi [\aap]
  {10.1051/0004-6361/201629561}, \href
  {https://ui.adsabs.harvard.edu/abs/2017A&A...597A..69S} {597, A69}

\bibitem[\protect\citeauthoryear{{Scholl}, {Marzari}  \&
  {Th{\'e}bault}}{{Scholl} et~al.}{2007}]{Scholl07}
{Scholl} H.,  {Marzari} F.,   {Th{\'e}bault} P.,  2007, \mn@doi [\mnras]
  {10.1111/j.1365-2966.2007.12145.x}, \href
  {https://ui.adsabs.harvard.edu/abs/2007MNRAS.380.1119S} {380, 1119}

\bibitem[\protect\citeauthoryear{{Shakura} \& {Sunyaev}}{{Shakura} \&
  {Sunyaev}}{1973}]{Shak}
{Shakura} N.~I.,  {Sunyaev} R.~A.,  1973, \aap, \href
  {http://adsabs.harvard.edu/abs/1973A%26A....24..337S} {24, 337}

\bibitem[\protect\citeauthoryear{{Spergel} et~al.,}{{Spergel}
  et~al.}{2015}]{Spergel15}
{Spergel} D.,  et~al., 2015, arXiv e-prints, \href
  {https://ui.adsabs.harvard.edu/abs/2015arXiv150303757S} {p. arXiv:1503.03757}

\bibitem[\protect\citeauthoryear{{Standing} et~al.,}{{Standing}
  et~al.}{2023}]{Standing23}
{Standing} M.~R.,  et~al., 2023, arXiv e-prints, \href
  {https://ui.adsabs.harvard.edu/abs/2023arXiv230110794S} {p. arXiv:2301.10794}

\bibitem[\protect\citeauthoryear{{Tanaka} \& {Ward}}{{Tanaka} \&
  {Ward}}{2004}]{Tanaka04}
{Tanaka} H.,  {Ward} W.~R.,  2004, \mn@doi [\apj] {10.1086/380992}, \href
  {https://ui.adsabs.harvard.edu/abs/2004ApJ...602..388T} {602, 388}

\bibitem[\protect\citeauthoryear{{Thun} \& {Kley}}{{Thun} \&
  {Kley}}{2018}]{Thun18}
{Thun} D.,  {Kley} W.,  2018, \mn@doi [\aap] {10.1051/0004-6361/201832804},
  \href {https://ui.adsabs.harvard.edu/abs/2018A&A...616A..47T} {616, A47}

\bibitem[\protect\citeauthoryear{{Thun}, {Kley}  \& {Picogna}}{{Thun}
  et~al.}{2017}]{Thun17}
{Thun} D.,  {Kley} W.,   {Picogna} G.,  2017, \mn@doi [\aap]
  {10.1051/0004-6361/201730666}, \href
  {https://ui.adsabs.harvard.edu/abs/2017A&A...604A.102T} {604, A102}

\bibitem[\protect\citeauthoryear{{Triaud} et~al.,}{{Triaud}
  et~al.}{2022}]{Triaud22}
{Triaud} A. H.~M.~J.,  et~al., 2022, \mn@doi [\mnras] {10.1093/mnras/stab3712},
  \href {https://ui.adsabs.harvard.edu/abs/2022MNRAS.511.3561T} {511, 3561}

\bibitem[\protect\citeauthoryear{{Weidenschilling}}{{Weidenschilling}}{1977}]{Weidenschilling_77}
{Weidenschilling} S.~J.,  1977, \mn@doi [\mnras] {10.1093/mnras/180.1.57},
  \href {http://adsabs.harvard.edu/abs/1977MNRAS.180...57W} {180, 57}

\bibitem[\protect\citeauthoryear{{Welsh} et~al.,}{{Welsh}
  et~al.}{2012}]{Welsh12}
{Welsh} W.~F.,  et~al., 2012, \mn@doi [\nat] {10.1038/nature10768}, \href
  {https://ui.adsabs.harvard.edu/abs/2012Natur.481..475W} {481, 475}

\bibitem[\protect\citeauthoryear{{Wilson}, {Walker}  \& {Thornley}}{{Wilson}
  et~al.}{1997}]{Wilson97}
{Wilson} C.~D.,  {Walker} C.~E.,   {Thornley} M.~D.,  1997, \mn@doi [\apj]
  {10.1086/304216}, \href
  {https://ui.adsabs.harvard.edu/abs/1997ApJ...483..210W} {483, 210}

\makeatother
\end{thebibliography}

\appendix
\section{Opacity table}
\label{sec:opacity}

The opacity $\kappa$ is calculated using the temperature and density dependant formulae from \citet{Bell97} for temperatures below 3730 K, and by \citet{Bell94} above 3730 K:
\begin{equation}
\kappa[{\rm cm}^2/{\rm g}] = \left\{ \begin{array}{ll}
10^{-4}\rm T^{2.1} & \rm T<132 \, {\rm K} \\
3\rm T^{-0.01} & 132\le \rm T<170 \, {\rm K} \\
0.01\rm T^{-1.1} & 170\le \rm T<375 \, {\rm K} \\
5\text{x}10^{4}\rm T^{-1.5} & 375\le \rm T<390 \, {\rm K} \\
0.1\rm T^{0.7} & 390\le \rm T<580 \, {\rm K} \\
2\text{x}10^{15}\rm T^{-5.2} & 580\le \rm \rm T<680 \, {\rm K} \\
0.02\rm T^{0.8} & 680\le \rm T<960^1 \, {\rm K} \\
2\text{x}10^{81}\rho \rm T^{-24} & 960\le \rm T<1570^1 \, {\rm K} \\
10^{-8}\rho^{2/3}\rm T^{3} & 1570\le \rm T<3730^1 \, {\rm K} \\
10^{-36}\rho^{1/3}\rm T^{10} & 3730\le \rm T<10000^1 \, {\rm K}
\end{array}\right.
\end{equation}
For the purpose of the above equations, where opacity is dependant on the local gas density, a density of $10^{-9}\rm gcm^{-3}$ is used to calculate the temperature ranges where that opacity law is appropriate.

\label{lastpage}
\end{document}